\documentclass[aps,prb,onecolumn,nofootinbib,citeautoscript,10pt]{revtex4-2}  
\usepackage{amsmath,amssymb,bm} 
\usepackage{graphicx,comment}

\usepackage[tight]{subfigure} 

\usepackage[dvipsnames]{xcolor} 
\usepackage[papersize={8.5in,11in}]{geometry}
\usepackage[colorlinks=true]{hyperref}
\hypersetup{
    bookmarks=true,         
    unicode=false,          
    pdftoolbar=true,        
    pdfmenubar=true,        
    pdffitwindow=false,     
    pdfstartview={FitH},    
    pdfkeywords={keyword1} {key2} {key3}, 
    pdfnewwindow=true,      
    colorlinks=true,       
    linkcolor=magenta, 
    citecolor=blue,        
    filecolor=magenta,      
    urlcolor=blue           
} 

\usepackage{graphicx}
\usepackage{dcolumn}
\usepackage{color}
\usepackage{amssymb,amsmath}
\usepackage{tabularx,graphicx}
\usepackage{epstopdf}
\usepackage{latexsym}
\usepackage{colortbl}
\usepackage{psfrag}
\usepackage{bbm,bm,array,physics}
\usepackage{dsfont}
\usepackage{float}

\def \nn{\nonumber \\}

\def\*#1{\mathbf{#1}} 

\begin{document}

\title{Electric, thermal, and thermoelectric magnetoconductivity for Weyl/multi-Weyl semimetals in planar Hall set-ups induced by the combined effects of topology and strain}

\author{Leonardo Medel$^{1}$}

\author{Rahul Ghosh$^{2}$}

\author{Alberto Martín-Ruiz$^{1}$}

\author{Ipsita Mandal$^{2,3}$}
\email{ipsita.mandal@snu.edu.in}

\affiliation{$^1$Instituto de Ciencias Nucleares, Universidad Nacional Aut\'{o}noma de M\'{e}xico, 04510 Ciudad de M\'{e}xico, M\'{e}xico}
\affiliation{$^2$Department of Physics, Shiv Nadar Institution of Eminence (SNIoE), Gautam Buddha Nagar, Uttar Pradesh 201314, India}
\affiliation{$^3$
Freiburg Institute for Advanced Studies (FRIAS), University of Freiburg, D-79104 Freiburg, Germany}

\begin{abstract} 
We continue our investigation of the response tensors in planar Hall (or planar thermal Hall) configurations where a three-dimensional Weyl/multi-Weyl semimetal is subjected to the combined influence of an electric field $\mathbf E $ (and/or temperature gradient $\nabla_{\mathbf r } T$) and an effective magnetic field $\mathbf B_\chi $, generalizing the considerations of Phys. Rev. B 108 (2023) 155132 and Physica E 159 (2024) 115914. The electromagnetic fields are oriented at a generic angle with respect to each other, thus leading to the possibility of having collinear components, which do not arise in a Hall set-up. The net effective magnetic field $\mathbf B_\chi $ consists of two parts --- (a) an actual/physical magnetic field $\mathbf B $ applied externally; and (b) an emergent magnetic field $\mathbf B_5 $ which quantifies the elastic deformations of the sample. $\mathbf B_5 $ is an axial pseudomagnetic field because it couples to conjugate nodal points with opposite chiralities with opposite signs. Using a semiclassical Boltzmann formalism, we derive the generic expressions for the response tensors, including the effects of the Berry curvature (BC) and the orbital magnetic moment (OMM), which arise due to a nontrivial topology of the bandstructures. We elucidate the interplay of the BC-only and the OMM-dependent parts in the longitudinal and transverse (or Hall) components of the electric, thermal, and thermoelectric response tensors. Especially, for the co-planar transverse components of the response tensors, the OMM part acts exclusively in opposition (sync) with the BC-only part for the Weyl (multi-Weyl) semimetals.
\end{abstract}

\maketitle


\section{Introduction}

There has been an incredible amount of research work focussing on the investigations of the transport properties of semimetals, which are systems harbouring band-crossing points in the Brillouin zone (BZ). Two or more bands cross at the nodal points where the densities of states go to zero. Among the three-dimensional (3d) semimetals with twofold nodal points, the well-known examples include the Weyl semimetals (WSMs) \cite{burkov11_weyl,yan17_topological} and the multi-Weyl semimetals (mWSMs) \cite{bernevig,bernevig2,dantas18_magnetotransport}, whose bandstructures exhibit nontrivial topology quantified by the Berry phase.
The nodal points for both the WSMs and the mWSMs are protected by the point-group symmetries of the crystal lattice \cite{bernevig2}. In the language of the Berry curvature (BC) flux, each nodal point acts as a source or sink in the momentum space and, thus, acts as an analogue of the elusive magnetic monopole. The value of the monopole charge is equal to the Chern number arising from the Berry connection. A consequence of the Nielson-Ninomiya theorem \cite{nielsen}, applicable for systems in an odd number of spatial dimensions, the nodal points exist in pairs, with each pair carrying Chern numbers $\pm J$. Thus, the pair acts as a source and a sink of the BC flux. Since the sign of the monopole charge is called the chirality $\chi $ of the associated node, the two nodes in a pair are of opposite chiralities (i.e., $\chi = \pm 1$). The values of $J$ for Weyl (e.g., TaAs \cite{huang15_observation, lv_Weyl, yang_Weyl} and HgTe-class materials~\cite{ruan_Weyl}), double-Weyl (e.g., $\mathrm{HgCr_2Se_4}$~\cite{Gang2011} and $\mathrm{SrSi_2}$~\cite{hasan_mweyl16, singh18_tunable}), and triple-Weyl nodes (e.g., transition-metal monochalcogenides~\cite{liu2017predicted}) are equal to one, two, and three, respectively.

\begin{figure*}[t]
	\includegraphics[width=0.5 \linewidth]{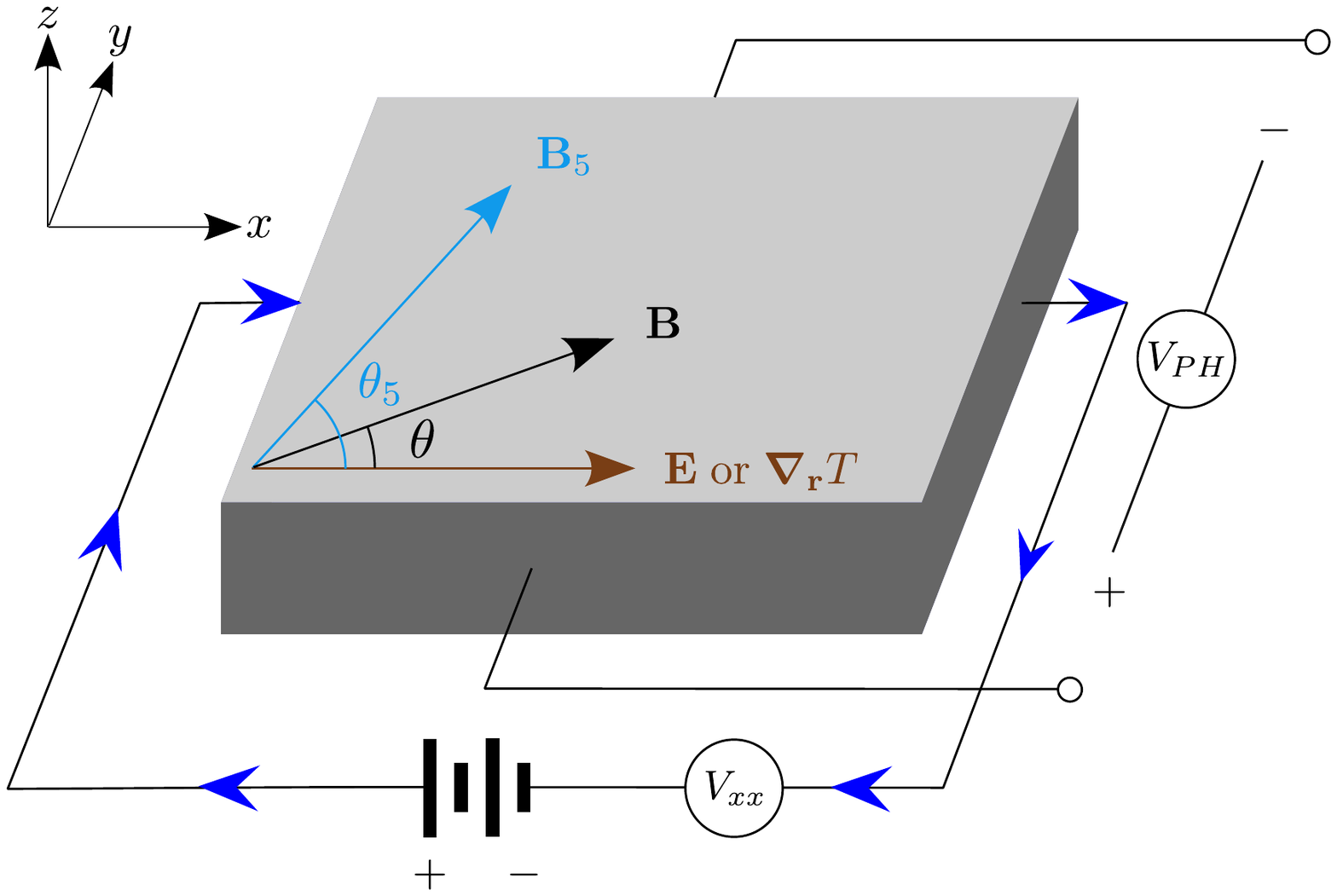}
	\caption{\label{figsetup}
Schematics showing the planar Hall (or planar thermal Hall) experimental set-up, where the sample is subjected to an external electric field $ E\, {\mathbf{\hat x}} $ (and/or a temperature gradient $\partial_x T\, {\mathbf{\hat x}}$). An external magnetic field $\mathbf B $ is applied such that it makes an angle $\theta $ with the existing electric field (and/or the temperature gradient). In addition, the sample is considered to be under the influence of a mechanical strain, whose effect is incorporated via an artificial chiral magnetic field $\mathbf B_5 $, making an angle $\theta_5$ with the $x$-axis. The resulting planar Hall (or planar thermal Hall) voltage, generated along the $y$-axis, is indicated by the symbol $V_{PH}$.
}
\end{figure*}

Suppose we consider an experimental set-up with a WSM/mWSM subjected to an external uniform electric field $ \mathbf E $ along the $x$-axis and a uniform external magnetic field $ \mathbf B $ along the $y$-axis. Since $ \mathbf B $ is perpendicular to $ \mathbf E $, a potential difference (known as the Hall voltage) will be generated along the $z$-axis. This phenomenon is the well-known Hall effect. However, if we apply $ \mathbf B $ making an angle $ \theta $ with $ \mathbf E $, where $  \theta  \neq \pi/2 \text{ or } 3\pi/2$, the conventional Hall voltage induced from the Lorentz force is zero along the $y$-axis. Nonetheless, due to nontrivial Chern numbers, a voltage difference $V_{PH}$ appears along this direction, as shown in Fig.~\ref{figsetup}. This is known as the planar Hall effect (PHE), arising due to the chiral anomaly~\cite{son13_chiral, burkov17_giant, li_nmr17, nandy_2017_chiral, nandy18_Berry, Nag_2020, ips-serena}.
The chiral anomaly refers to the phenomenon of charge pumping from one node to its conjugate when $\mathbf E \cdot \mathbf B \neq 0 $, originating from a local non-conservation of electric charge in the vicinity of an individual node, with the rate of change of the number density of chiral quasiparticles being proportional to $ J \left( \mathbf E \cdot \mathbf  B \right) $ \cite{chiral_ABJ, chiral_ano_mWSM}. The associated transport coefficients, related to this set-up, are referred to as the longitudinal magnetoconductivity (LMC) and the planar Hall conductivity (PHC), which depend on the value of $ \theta $. In an analogous set-up, we observe the planar thermal Hall effect (PTHE), where we replace the external electric field by (or add) an external temperature gradient $ \mathbf \nabla_{\mathbf r} T$. In this scenario, too, a potential difference is induced along the $y$-axis due to the chiral anomaly~\cite{girish1, ips-serena}  [cf. Fig.~\ref{figsetup}], with the response coefficients known as the longitudinal thermoelectric coefficient (LTEC) and transverse thermoelectric coefficient (TTEC). There has been a tremendous amount of efforts to determine the behaviour of these response tensors~\cite{zhang16_linear, chen16_thermoelectric, das19_linear, das19_linear2, das20_thermal,das22_nonlinear, pal22a_berry,pal22b_berry, fu22_thermoelectric, araki20_magnetic, mizuta14_contribution, timm_omm, onofre, ips-rahul-ph, ips-rahul-tilt, nodal_ph}.

In a planar Hall (or thermal Hall) set-up, if a semimetal is subjected to a mechanical strain, it induces elastic deformations of the material. The elastic deformations couple to the electronic degrees of freedom (i.e., quasiparticles) in such a way that they can be modelled as pseudogauge fields in the semimetals \cite{guinea10_energy, guinea10_generating,low10_strain, landsteiner_gaguge,liu_gauge, pikulin_gauge, arjona18_rotational, onofre}. The form of these elastic gauge fields shows that they couple to the quasiparticles with opposite chiralities with opposite signs \cite{landsteiner_gaguge,liu_gauge,pikulin_gauge,ghosh20_chirality,girish2023,onofre}. Due to the chiral nature of the coupling between the emergent vector fields and the itinerant fermionic carriers, this provides an example of axial gauge fields in three dimensions. This is to be contrasted with the actual electromagnetic fields, which couple to all the nodes with the same sign. While a uniform pseudomagnetic field $ \mathbf B_5$ can be generated when a WSM/mWSM nanowire is put under torsion, a pseudoelectric field $ \mathbf E_5$ appears on dynamically stretching and compressing the crystal along an axis (which can be achieved, for example, by driving longitudinal sound waves) \cite{pikulin_gauge}. Direct evidence of the generation of such pseudoelectromagnetic fields in doped semimetals has been obtained in experiments \cite{exp_gauge}. In an earlier work by the two of us \cite{ips-rahul-ph}, the response tensors have been computed for WSMs and mWSMs, but neglecting the orbital magnetic moment (OMM) \cite{xiao_review, sundaram99_wavepacket}, which is another artifact of a nontrivial topology in the bandstructures. Although the effects of OMM were included in the computations of Ref.~\cite{onofre}, only the electric conductivity at zero temperature was studied. Hence, in this paper, we derive all the relevant electric, thermal, and thermoelectric response tensors, associated with the planar Hall and planar thermal set-ups, which constitute a complete description incorporating both the BC and the OMM.

As shown in Fig.~\ref{figsetup}, the co-planar $\mathbf E$ and $\mathbf B_\chi =  \mathbf B + \chi \,\mathbf B_5 $ set-ups considered here consist of a nonzero $ \mathbf B_5$ and a nonzero $\mathbf B $ in the $xy$-plane. These two parts of the effective magnetic field $\mathbf B_\chi$ are oriented at the angles $ \theta$ and $\theta_5 $, respectively, with respect to $ \mathbf E $ (or $\nabla_{\mathbf r} T$) applied along the $x$-axis. In other words, $\mathbf B =  B \left( \cos \theta, \sin \theta, 0 \right) $ and $\mathbf B_5 =  B_5 \left( \cos \theta_5, \sin \theta_5, 0 \right) $, where $B \equiv |\mathbf B|$ and $B_5 \equiv |\mathbf B_5|$. We consider the weak magnetic field limit with low values of $ B$ and $B_5 $, such that the formation of the Landau levels can be ignored, and the magnetoelectric, magnetothermal, and magnetothermoelectric response can be derived using the semiclassical Boltzmann formalism. The paper is organized as follows: In Sec.~\ref{secmodel}, we describe the low-energy effective Hamiltonians for the WSMs and mWSMs. In Secs.~\ref{secsigma}, \ref{secalpha}, and \ref{secell}, we show the explicit expressions of the in-plane components of the response tensors, and discuss their behaviour in some relevant parameter regimes. Finally, we conclude with a summary and outlook in Sec.~\ref{secsum}. The appendices are devoted to explaining the details of the intermediate steps used to derive the final expressions in the main text.

\section{Model}
\label{secmodel}

\begin{figure*}[t]
\subfigure[]{\includegraphics[width=0.2\linewidth]{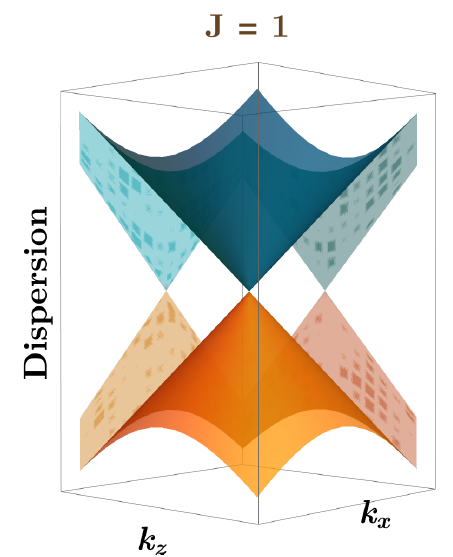}} \hspace{ 1.5 cm}
\subfigure[]{\includegraphics[width=0.2 \linewidth]{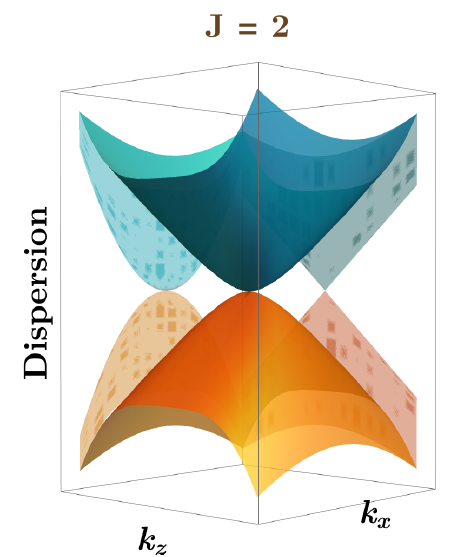}}\hspace{ 1.5 cm}
\subfigure[]{\includegraphics[width=0.2 \linewidth]{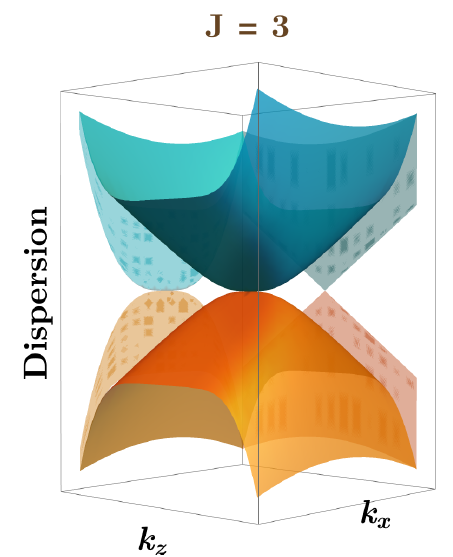}}
\caption{\label{figdis}Schematic dispersion of a single node in a (a) Weyl, (b) double-Weyl, and (c) triple-Weyl semimetal, plotted against the $k_z k_x $-plane. The double(triple)-Weyl node shows an anisotropic hybrid dispersion with a quadratic(cubic)-in-momentum dependence along the $k_x$-direction. In order to pinpoint the direction-dependent features, the projections of the dispersion along the respective momentum axes are also shown.}
\end{figure*}

In the vicinity of a nodal point with chirality $\chi$ and Berry monopole charge of magnitude $J$, the low-energy effective continuum Hamiltonian is given by \cite{liu2017predicted,bernevig,bernevig2}
\begin{align} 
\label{eqHweyl}
\mathcal{H}_\chi ( \mathbf k) & = 
\mathbf d_\chi( \mathbf k) \cdot \boldsymbol{\sigma} \,,
\quad k_\perp=\sqrt{k_x^2 + k_y^2}\,, \quad
\phi_k=\arctan({\frac{k_y}{k_x}})\,,
\quad \alpha_J=\frac{v_\perp}{k_0^{J-1}} \,, \nn
\mathbf d_\chi( \mathbf k) &=
\left \lbrace
\alpha_J \, k_\perp^J \cos(J\phi_k), \,
\alpha_J \, k_\perp^J \sin(J\phi_k), \,
\chi \, v_z \, k_z \right \rbrace,
\end{align}
where $ \boldsymbol{\sigma} = \lbrace \sigma_x, \, \sigma_y, \, \sigma_z \rbrace $ is the vector operator consisting of the three Pauli matrices, $\sigma_0$ is the $2 \times 2$ identity matrix, $\chi \in \lbrace 1, -1 \rbrace $ denotes the chirality of the node, and $v_z$ ($v_\perp$) is the Fermi velocity along the $z$-direction ($xy$-plane). The parameter $k_0$ has the dimension of momentum, whose value depends on the microscopic details of the material in consideration.
The eigenvalues of the Hamiltonian are given by
\begin{align} 
\label{eigenvalues_kperp_kz_phik}
\varepsilon_{\chi, s} ({ \mathbf k})= 
 (-1)^{s+1} \, \epsilon_{\mathbf k} \,, \quad
s \in \lbrace 1,2 \rbrace ,
\quad 
\epsilon_{\mathbf k}
= \sqrt{\alpha_J^2 \, k_\perp^{2J} + v_z^2 \, k_z^2}\,,
\end{align}
where the value $1$ ($2$) for $s$ represents the conduction (valence) band [cf. Fig.~\ref{figdis}]. 
We note that we recover the linear and isotropic nature of a WSM by setting $J=1$ and $\alpha_1= v_z$.

The band velocity of the chiral quasiparticles is given by
\begin{align}
{\boldsymbol v}^{(0)}_{\chi, s}( \mathbf k) 
\equiv \nabla_{\mathbf k} \varepsilon_{\chi, s}   (\mathbf k)
= - \frac{(-1) ^{s}}
{  \epsilon_{\mathbf k} }  \left\lbrace J\,  \alpha_J^2 \,  k_\perp^{2J-2} \,  k_{x} , \, J \,  \alpha_J^2 \,  k_\perp^{2J-2} \,  k_{y}  , \, v_z^2\,  k_z \right \rbrace\,.
\end{align}
The Berry curvature (BC) and the orbital magnetic moment (OMM), associated with the $s^{\rm{th}}$ band, are expressed by  \cite{xiao_review,xiao07_valley,konye21_microscopic}
\begin{align} 
\label{eqomm}
& {\mathbf \Omega}_{\chi, s}( \mathbf k)  = 
    i \, \langle  \nabla_{ \mathbf k}  \psi_s^\chi ({ \mathbf k})| \, 
    \cross  \, | \nabla_{ \mathbf k}  \psi_s^\chi ({ \mathbf k})\rangle
\Rightarrow
\Omega^i_{\chi, s}( \mathbf k)  =
 \frac{  (-1)^s \,  
\epsilon^i_{\,\,\,jl}}
 {4\,| \mathbf d_\chi (\mathbf k) |^3} \, 
 \mathbf d_\chi (\mathbf k) \cdot
 \left[   \partial_{k_j} \mathbf d_\chi (\mathbf k) \cross  
 \partial_{k_l } \mathbf d_\chi (\mathbf k) \right ] ,   
\nn & \text{and } 
{\boldsymbol{m}}_{\chi,s} ( \mathbf k) = 
-  \frac{ i \, e} {2 } \,
\langle  \mathbf \nabla_{ \mathbf k} \psi_s({ \mathbf k})| \cross
\left [\,
\left \lbrace \mathcal{H}({ \mathbf k}) -\mathcal{E}_{\chi, s}
({ \mathbf k}) 
\right \rbrace
| \mathbf \nabla_{ \mathbf k} \psi_s({ \mathbf k})\rangle \right ]
\Rightarrow
m^i_{\chi,s} ( \mathbf k) =
  \frac{e \, \epsilon^i_{\,\,\,jl}
  } 
{4 \, | \mathbf d_\chi (\mathbf k) |^2} \,  
\mathbf d_\chi (\mathbf k) \cdot
 \left[   \partial_{k_j} \mathbf d_\chi (\mathbf k) \cross  \partial_{k_l} \mathbf d_\chi (\mathbf k) \right ],
\end{align}
respectively, where the indices $i$, $j$, and $l$ $ \in \lbrace x, y, z \rbrace $, and are used to denote the Cartesian components of the 3d vectors and tensors. The symbol $ |  \psi_s^\chi ({ \mathbf k}) \rangle $ denotes the normalized eigenvector corresponding to the band labelled by $s$, with $ \lbrace |  \psi_1^\chi \rangle, \,  \lbrace |  \psi_2^\chi \rangle \rbrace $ forming an orthonornomal set for each node.

On evaluating the expressions in Eq.~\eqref{eqomm}, using Eq.~\eqref{eqHweyl}, we get
\begin{align}
\mathbf \Omega_{\chi, s }({ \mathbf k})= 
 \frac{ \chi \,(-1)^s
J \,v_z \, \alpha_J^2 \, k_\perp^{2J-2} }
{2 \,\epsilon^3_{\mathbf k}
} 
\left
\lbrace k_x, \, k_y, \, J\, k_z \right \rbrace , \quad 
   {\boldsymbol{m}}_{\chi, s }({ \mathbf k}) 
=  -   \frac{ \chi \, e\, J \,v_z \, \alpha_J^2 \, k_\perp^{2\,J-2} }
{2 \, \epsilon^2_{\mathbf k}} 
 \left \lbrace k_x, \, k_y, \, J \, k_z \right \rbrace .
\end{align}
From these expressions, we immediately observe the identity
\begin{align}
{\boldsymbol{m}}_{\chi, s }({ \mathbf k}) 
=  - (-1)^s \,  e \, \epsilon_{\mathbf k} \, \mathbf \Omega_{\chi, s }({ \mathbf k}) \, . 
\end{align}
While the BC changes sign with $s$, the OMM does not.

In this paper, we will take a positive value of the chemical potential $\mu$, such that it cuts the conduction bands with $s=1$. Henceforth, we will use the notations $ \varepsilon_{\chi, 1}  = \varepsilon_{\chi} $,
$ {\boldsymbol{v}}^{(0)}_{\chi, 1} =  {\boldsymbol{v}}^{(0)} $ (since it is independent of $\chi$),
$  {\mathbf  \Omega}_{\chi,  1 } =  {\mathbf  \Omega}_{\chi}$, and $ {\boldsymbol{m}}_{\chi,  1 } =  {\boldsymbol{m}}_{\chi}$, in order to avoid cluttering. 
The ranges of the values of the parameters that we will use in our computations are shown in Table~\ref{table_params}.

\section{Response tensors using the Boltzmann formalism}

Using the semiclassical Boltzmann formalism \cite{mermin, ips-kush-review}, the transport coefficients can be determined in the weak $|\mathbf B_\chi |$ limit, which applies to the regime of small cyclotron frequency, implying that the Landau-level quantization can be ignored. The detailed steps can be found in Appendix A of Ref.~\cite{ips-rahul-ph}, which we do not repeat here for the sake of brevity. Moreover, we work with the relation-time approximation for the collision integral, taking a simplistic momentum-independent relaxation time $\tau$. Furthermore, we assume that only intranode scatterings are dominant over the internode scattering processes, such that $\tau$ corresponds only to the former.

\begin{table}[t]
\begin{tabular}{|c|c|c|}
\hline
Parameter &   SI Units &   Natural Units  \\ \hline
$v_z$ from Ref.~\cite{Nag_floquet_2020} & $ 15 \times10^{5} $ m~s$^{-1} $ & $0.005$  \\ \hline
$\tau$ from Ref.~\cite{watzman18_dirac} & $ 10^{-13} \, \text{s} $ & $152 $ eV$^{-1}$  \\ \hline
$ T $ from Ref.~\cite{Nag_2020} & $ 10 - 100 $ K & $ 8.617 \cross 10^{-4} - 8.617 \cross 10^{-3}  $ eV \\ \hline
$ B $ and $B_5$ from Ref.~\cite{ghosh20_chirality} 
& $ 0 - 10  $ Tesla 
& $ 0 - 2000  $ eV$^{2}$
\\ \hline
$\mu$ from Refs.~\cite{Nag_2020, Nag_floquet_2020}
& $  1.6\times 10^{-21} -  1.6\times 10^{-20} $ J & $  0.01 - 0.1$ eV \\ \hline
\end{tabular}
\caption{\label{table_params}
The values of the various parameters which we have used in plotting the transport coefficients are tabulated here. Since $\alpha_J = {v_{\perp}}  /{k_0^{J-1}}$, we get
$\alpha_{1} = v_z$, $\alpha_{2}= {v_{\perp}} / {k_{0}}$, and $ \alpha_{3} = {v_{\perp}} / {k_{0}^2} = {\alpha_{2}^2} / {\alpha_{1}} $. In terms of the natural units, we need to set $\hbar=c=k_{B}=1$ and $4 \, \pi \,\epsilon_{0} = 137$. In our plots, we have used $v_\perp = v_z$ (from the table entry), leading to $\alpha_{2}=3.9 \times 10^{-5}$ eV$^{-1}$ and $\alpha_{3}=2.298 \times 10^{-6}$ eV$^{-2}$. 
For $ J=2$ and $J=3$, $v_\perp$ has been set equal to $ v_z$ for the sake of simplicity, while the isotropic dispersion for $J =1$ has $v_{\perp} = v_z$ automatically.
}
\end{table}

Let the contributions to the average electric and thermal current densities from the quasiparticles, associated with the node of chirality $\chi$, be ${\mathbf J}^\chi$ and ${\mathbf J}^{{\rm th},\chi}$, respectively. The response matrix, which relates the resulting generalized currents to the driving electric potential gradient or temperature gradient, can be expressed as
\begin{align}
\label{eqcur1}
\begin{pmatrix}
J_i^\chi \vspace{0.2 cm} \\
{J}_i^{{\rm th},\chi} 
\end{pmatrix} & = \sum \limits_j
\begin{pmatrix}
\sigma_{ij}^\chi & \alpha^\chi_{ i j }
\vspace{0.2 cm}  \\
T\, \alpha^\chi_{ij}  & \ell^\chi_{ i j }
\end{pmatrix}
\begin{pmatrix}
E_j
\vspace{0.2 cm}  \\
-{ \partial_{j} T } 
\end{pmatrix} .
\end{align}
Here, $\sigma^\chi_{ij} $ and $\alpha^\chi_{ij}$ represent the components of the magnetoelectric conductivity tensor ($\sigma^\chi$) and the magnetothermoelectric conductivity tensor ($\alpha^\chi $), respectively. While $\alpha^\chi$ determines the Peltier ($\Pi^\chi$), Seebeck ($S^\chi $), and Nernst coefficients, $\ell^\chi $ is the linear response tensor relating the heat current density to the temperature gradient, at a vanishing electric field. $ S^\chi $, $\Pi^\chi $, and the magnetothermal conductivity tensor $\kappa^\chi $ (which provides the coefficients between the heat current density and the temperature gradient at vanishing electric current) can be extracted from the coefficients on the right-hand-side of Eq.~\eqref{eqcur1}, via the following relations \cite{mermin, ips-kush-review}:
\begin{align}
\label{eq:kappa}
S_{ ij}^\chi = \sum \limits_{ i^\prime}
\left( \sigma^\chi \right)^{-1}_{ i i^\prime }
\alpha^\chi_{ i^\prime j} \, , \quad
\Pi_{ i j} ^\chi= T \sum \limits_{ i^\prime}
\alpha^\chi_{ i   i^\prime}   
\left(\sigma^\chi\right)^{-1}_{ i^\prime j} \,,\quad 
\kappa_{ij}^\chi =
\ell^\chi_{ ij }- T \sum \limits_{ i^\prime, \, j^\prime }
\alpha^\chi_{ i  i^\prime }
\left( \sigma^\chi \right)^{-1}_{  i^\prime  j^\prime }
\alpha^\chi_{ j^\prime j}  \,.
\end{align}

In order to include the effects from the OMM and the BC, we first define the quantitites
\begin{align}
& \mathcal{E}_{\chi} (\mathbf k) 
= \varepsilon_{\chi}  (\mathbf k) + \varepsilon_{\chi}^{ (m) }  (\mathbf k) \, ,
\quad 
\varepsilon_{\chi}^{(m)}   (\mathbf k) 
= - \,{\mathbf B}_{\chi} \cdot \mathbf{m }_{\chi}  (\mathbf k) \,, \quad
{\boldsymbol  v}_{\chi}({\mathbf k} ) \equiv 
 \nabla_{{\mathbf k}}   \mathcal{E}_{\chi} ({\mathbf k})
 = {\boldsymbol  v}^{(0)} ({\mathbf k} ) + {\boldsymbol  v}^{(m)}_\chi ({\mathbf k} ) \,,\nn
& {\boldsymbol  v}^{(m)}_\chi ({\mathbf k} )
= \nabla_{{\mathbf k}} \varepsilon_{\chi}^{(m)}   (\mathbf k) \,,
\quad  D_{\chi} = \left [1 
+ e \,  \left \lbrace 
{\mathbf B}_{\chi} \cdot \mathbf{\Omega }_{\chi}  (\mathbf k)
\right \rbrace  \right ]^{-1} ,
\end{align}
where $ \varepsilon_\chi^{ (m) } ({\mathbf k})$ is the Zeeman-like correction to the energy due to the OMM, $ {\boldsymbol  v}_{\chi}({\mathbf k} )  $ is the modified band velocity of the Bloch electrons after including $ \varepsilon_\chi^{ (m) } ({\mathbf k})$, and $ D_{\chi} $ is the modification factor of the phase space volume element due to a nonzero BC. The modification of the effective Fermi surface, on including the correction $ \varepsilon_\chi^{ (m) } ({\mathbf k})$, is shown schematically in Fig.~\ref{figfs}.


Our weak-magnetic-field limit implies that 
\begin{align}
 e \,| {\mathbf B}_{\chi} \cdot \mathbf{\Omega }_{\chi} |  \ll 1 . \label{Condition}
\end{align}
In our calculations, we keep terms upto $\order{ |{ \mathbf B}_{\chi}|^2}$ and, thus, use
\begin{align}
D_{\chi} &=
 1 - e \,  \left( {\mathbf B}_{\chi} \cdot \mathbf{\Omega }_{\chi}  \right) 
+   e^2  \,  \left( {\mathbf B}_{\chi} \cdot \mathbf{\Omega }_{\chi}  \right)^2  
+  \order{ |{ \mathbf B}_{\chi}|^3} \,. 
 \label{Exp_D}
\end{align}
Also, the condition in Eq.~(\ref{Condition}) implies that $ |\varepsilon_\chi ^{ (m) } (\mathbf k) | $ is small compared to $\varepsilon_\chi (\mathbf k)$:
\begin{align}
\vert {\mathbf B}_{\chi} \cdot  \boldsymbol{ m }_{\chi} \vert = \varepsilon_\chi \,  e
 \left | {\mathbf B}_{\chi} \cdot  \mathbf{\Omega}_{\chi} \right |
 \ll  \varepsilon_\chi   \,.
\end{align} 
This means that the Fermi-Dirac distribution can also be power expanded up to quadratic order in the magnetic field, as follows:
\begin{align}
f_{0} ( \mathcal{E}_{\chi}   ) = f_{0} ( \varepsilon_\chi) 
+  \varepsilon_\chi ^{ (m) } \,  
f'_{0} (\varepsilon_\chi) + \frac{1}{2} \left(  \varepsilon_\chi ^{(m)} \right)^2  \,
  f^{\prime \prime}_{0} ( \varepsilon_\chi )  +  \order{ |{ \mathbf B}_{\chi}|^3} \,,
\label{Exp_f}
\end{align}
where the prime indicates derivative with respect to the energy argument of $f_0$.

The general expression for the magnetoelectric conductivity tensor for an isolated node of chirality $\chi$, contributed by the conduction band, is given by
\begin{align}
\sigma ^{\chi}_{ij} &= - e^2 \,  \tau   \int \frac{  d^3 \mathbf k } { (2\, \pi )^3  } \,  D_{\chi}  
 \left[  v_{\chi i } + e  \,  (  {\boldsymbol  v}_{\chi} \cdot \mathbf{\Omega }_{\chi} ) \,  B_{\chi i} \right] 
 \left[  v_{\chi j } + e  \,  (  {\boldsymbol  v}_{\chi} \cdot \mathbf{\Omega }_{\chi} ) \, 
  B_{\chi j} \right] \,  \frac{\partial f_{0} (\mathcal{E}_{\chi}) }{ \partial \mathcal{E}_{\chi}  }  ,
\label{eq_elec}
\end{align}
where we do not include the  parts coming from the ``intrinsic anomalous Hall'' effect and the so-called Lorentz-force contribution. This is because of the following reasons:
\begin{enumerate}

\begin{figure*}[t]
\subfigure[]{\includegraphics[width=0.2 \linewidth]{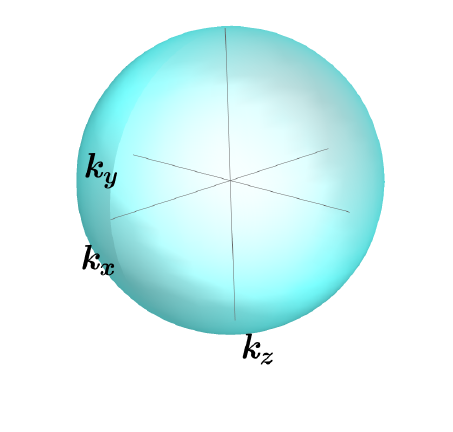} 
\includegraphics[width=0.2 \linewidth]{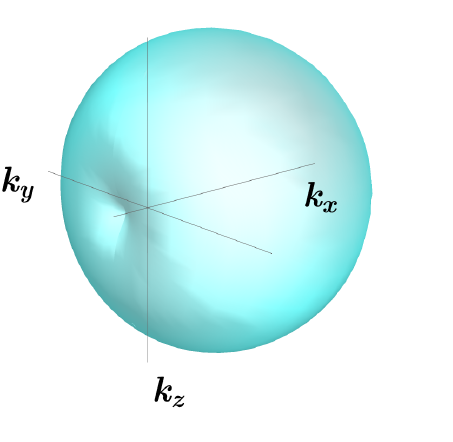}}\hspace{ 1.5 cm}
\subfigure[]{\includegraphics[width=0.22 \linewidth]{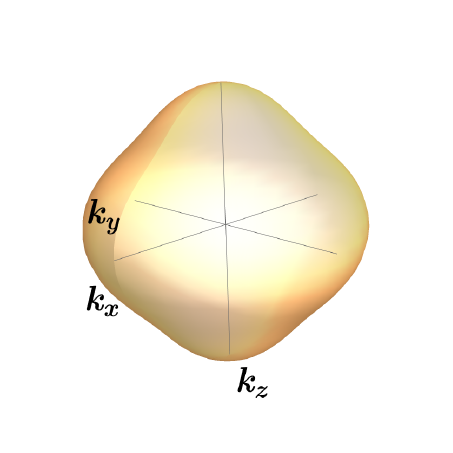}, \includegraphics[width=0.22 \linewidth]{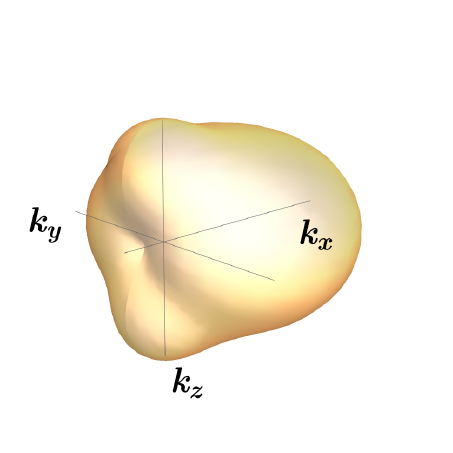}}
\caption{\label{figfs}Schematics of the Fermi surfaces for one node of a (a) WSM and (b) double-Weyl semimetal, without and with the OMM-correction for the effective energy dispersion. Here we have taken the effective magnetic field to be directed purely along the $x$-axis.
}
\end{figure*}

\item The intrinsic anomalous Hall term is given by
\begin{align}
\sigma_{ij}^{\mbox{\scriptsize AH},\chi} & = - \, e^2 \,\epsilon _{ijl}
\int \frac{  d^3 {\mathbf k} }{(2 \,\pi ) ^{3} } \, 
     \Omega _\chi^l ({\mathbf k}) \, f_{0} (\mathcal{E} _{\chi})
=  \sigma_{ij}^{\mbox{\scriptsize AH(0)},\chi}
+ \sigma_{ij}^{\mbox{\scriptsize AH(1)},\chi}
+\sigma_{ij}^{\mbox{\scriptsize AH(2)},\chi}
+ \order{ |{ \mathbf B}_{\chi}|^3}\,, \nn
\sigma_{ij}^{\mbox{\scriptsize AH(0)},\chi}
& = - \, e^2 \,\epsilon _{ijl} \int  \frac{  d^3 {\mathbf k} }{(2 \,\pi )^{3} } \, 
\Omega _\chi^l ({\mathbf k}) \,f_0  (\varepsilon_{\chi} )\,, 
\quad \sigma_{ij}^{\mbox{\scriptsize AH(1)},\chi}
 = -\,e^2 \,\epsilon _{ijl} \int  \frac{  d^3 {\mathbf k} }{(2 \,\pi )^{3} } \, 
\Omega _\chi^l ({\mathbf k}) \, \varepsilon_{\chi} ^{(m)} \, f_0^\prime (\varepsilon_{\chi} )
\,,\nn
\sigma_{ij}^{\mbox{\scriptsize AH(2)},\chi}
& = - \,\frac{e^2 \,\epsilon _{ijl} } {2}
 \int  \frac{  d^3 {\mathbf k} }{(2 \,\pi )^{3} } \, 
\Omega _\chi^l ({\mathbf k}) \,
\left ({\varepsilon_{\chi} ^{(m)}}\right )^2 \,
  f_0^{ \prime \prime} (\varepsilon_{\chi} )  \,,
\end{align}
whose diagonal components (i.e., $\sigma_{ii}^{\mbox{\scriptsize AH},\chi}$) are automatically zero.
The first term, $\sigma_{ij}^{\mbox{\scriptsize AH(0)},\chi}$, is $\mathbf{B}_\chi$-independent and vanishes identically. The nonzero OMM generates $\mathbf{B}_\chi$-dependent terms. However, for our configuration consisting of ${\mathbf E}$- and $ {\mathbf B_\chi} $-components lying in the $xy$-plane, we have
\begin{align}
 \sigma_{yx} ^{\mbox{\scriptsize AH}(1), \chi} 
 =  \sigma_{xy} ^{\mbox{\scriptsize AH}(1), \chi} 
\propto
\int\limits_{-\infty}^{\infty}k_z\, dk_z
\int\limits_0^{\infty} dk_\perp \, 
\frac{k_{\perp}^{4J-2}}
{\varepsilon_{\chi}^5}\, f_0^\prime(\varepsilon_{\chi})
\int\limits_0^{2 \, \pi} d\phi
\left [B_{\chi x} \cos\phi + B_{\chi y} \sin\phi \right ]
=0 
\text{ and } \sigma_{yx} ^{\mbox{\scriptsize AH}(2), \chi}
= \sigma_{yx} ^{\mbox{\scriptsize AH}(2), \chi} =0 \,.
\end{align}
Only the transverse out-of-plane components are nonzero, viz.
\begin{align}
&  \sigma_{zx} ^{\mbox{\scriptsize AH}(1), \chi} (\mu_{\chi}) =
	\frac{ e^3 \, J \, v_z \, B_{\chi y} } {24\,\pi^2}  
	\int _{0} ^{\infty} d \varepsilon_{\chi}\,
	\frac{f^\prime_{0} (\varepsilon_{\chi} ) } { \varepsilon_{\chi}}
	=  - \frac{ e^3\, J \,v_z \, B_{\chi y} }
	{24\, \pi ^2\, \mu_{\chi} }    
	\left( 1 + \frac{\pi ^{2} \,  }{3 \,  \beta ^{2} \,  \mu_{\chi}^{2} } \right) 
	\text{ and  }
	\nn & \sigma_{zy} ^{\mbox{\scriptsize AH}(1), \chi} (\mu_{\chi}) 
	=   \frac{ e^3\, J \,v_z \, B_{\chi x} }
	{24\, \pi ^2\, \mu_{\chi} }    
	\left( 1 + \frac{\pi ^{2} \,  }{3 \,  \beta ^{2} \,  \mu_{\chi}^{2} } \right)
\text{ [if we have a nonzero $y$-component of } \mathbf E]	. 
\end{align}
We note that $ \sigma_{zx} ^{\mbox{\scriptsize AH}(2), \chi} = \sigma_{zy} ^{\mbox{\scriptsize AH}(2), \chi} = 0 \,.$
Since we are focussing on the in-plane components of the response tensors, we will not discuss further the behaviour of these nonzero out-of-plane components.


\item The Lorentz-force part shows a behaviour analogous to the intrinsic anomalous Hall part described above, with vanishing in-plane components (see Ref.~\cite{nodal_ph} for generic arguments stemming from symmetry considerations).

\end{enumerate}
The expression for $\sigma ^{\chi}_{ij}$, shown above, includes the effects of the BC and the OMM.  
Here, $f_{0} $ is the equilibrium Fermi-Dirac distribution at temperature $T =1/\beta $ and chemical potential $\mu_\chi $. As discussed in Refs.~\cite{onofre, ips-rahul-ph}, while a purely physical magnetic field $\mathbf{B}$ gives a quadratic-dependence of the response on the overall magnetic field, inclusion of a nonzero axial part $\mathbf B_5$ opens up the possibility of generating linear and parabolic behaviour of the response tensors.

Analogous to Eq.~\eqref{eq_elec}, we have the general expressions \cite{girish1}
\begin{align}
{\alpha} ^{\chi}_{ij} &=  e \, \tau  \int \frac{  d^3 \mathbf k } { (2\, \pi )^3  } \, D_{\chi} \,  \left[  v_{\chi i } + e \,  (  {\boldsymbol  v}_{\chi} \cdot \mathbf{\Omega }_{\chi} ) \,  B_{\chi i} \right] \,  \left[  v_{\chi j } + e  \,  (  {\boldsymbol  v}_{\chi} \cdot \mathbf{\Omega }_{\chi} )  \, B_{\chi j} \right] \,  \frac{\mathcal{E}_{\chi}  - \mu }{T} \,  \frac{\partial f_{0} (\mathcal{E}_{\chi}) }{ \partial \mathcal{E}_{\chi}  } , 
\label{eq_thermoel}
\end{align}
and
\begin{align}
\ell^\chi_{ij} &= - \,\tau  
\int \frac{  d^3 \mathbf k } { (2\, \pi )^3  } 
\, D_{\chi} \,  \left[  v_{\chi i } + e \,  (  {\boldsymbol  v}_{\chi} \cdot \mathbf{\Omega }_{\chi} ) \,  B_{\chi i} \right] \,  \left[  v_{\chi j } + e  \,  (  {\boldsymbol  v}_{\chi} \cdot \mathbf{\Omega }_{\chi} )  \, B_{\chi j} \right] \,  \frac{ ( \mathcal{E}_{\chi}  - \mu )^2 }{T} \,  \frac{\partial f_{0} (\mathcal{E}_{\chi}) }{ \partial \mathcal{E}_{\chi}  } , 
\label{eq_thermal}
\end{align}
respectively. Since $\ell^\chi $ determines the first term in the magnetothermal conductivity tensor $\kappa^\chi $, we will often loosely refer to $\ell^\chi $ itself as the magnetothermal coefficient.

\section{Magnetoelectric conductivity}
\label{secsigma}

\begin{figure*}[t]
\includegraphics[width=0.6 \linewidth]{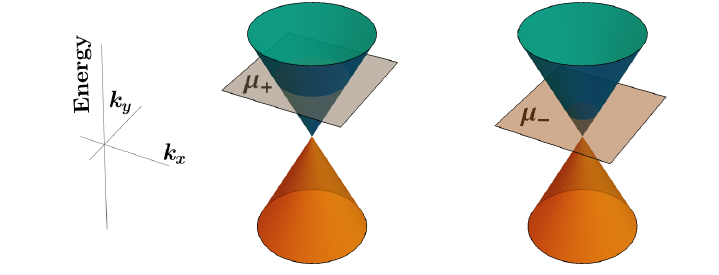}
\caption{\label{fig2node}A pair of conjugate Weyl nodes with the corresponding chemical potentials tuned to two different values, $\mu_+$ and $\mu_-$.}
\end{figure*}

Working in the weak-in-magnetic field limit, which allows expansions in the components of $ \mathbf B_{\chi} $, the terms in Eq.~(\ref{eq_elec}) are disentangled into a sum of three terms, having distinct origins, as follows:
\begin{align}
\sigma ^{\chi}_{ij}  &= \sigma ^{0, \chi}_{ij} + \sigma ^{\Omega , \chi}_{ij}  +  \sigma ^{m, \chi}_{ij} \,.
 \label{totalsigma}
\end{align}
Here,
\begin{align}
\sigma ^{0, \chi}_{ij} = - e^2 \,  \tau 
\int \frac{  d^3 {\mathbf k} } { (2\, \pi )^3  }  
\,  v_{\chi i } ^{(0)} \,  v_{\chi j} ^{(0)} \, 
 f_{0}^\prime (\varepsilon_\chi ) ,  
\label{eq_elec_0B}
\end{align}
is the conductivity surviving in the absence of a magnetic field (i.e., for ${\mathbf B}_{\chi} = {\bf{0}}$),
\begin{align}
\sigma ^{\Omega , \chi}_{ij} = - {e^4 \,  \tau}
\int \frac{  d^3 {\mathbf k} }
 { (2\, \pi )^3  } \,  Q_{\chi i } \, Q_{\chi j } 
\,  f_{0}^\prime (\varepsilon_\chi )  ,
\quad
{\boldsymbol{Q}}_{\chi}  = 
\mathbf{\Omega }_{\chi}  \times
\left  ( {\boldsymbol  v}_{\chi} ^{(0)}   \times \mathbf{B}_{\chi} \right ),
\label{eq_elec_Berry}
\end{align}
is the contribution arising solely from the BC, and  
\begin{align}
\label{eq_elec_OMM}
\sigma ^{m, \chi}_{ij} & =  {2 \, e ^{3} \,  \tau} 
 \int \frac{  d^3 {\mathbf k} }  { (2\, \pi )^3  } 
 \left [    
 {Q}_{\chi i} \,  v^{(m)}_{\chi j}  
  + \frac{ \varepsilon_{\chi}^{(m)}} {e} 
 \left(   \nabla_{{\mathbf k}} \cdot  \boldsymbol{T}_{\chi ij} \right)
  + \frac{ \varepsilon_{\chi}^{(m)}}   {2} \,  
  \left (
  \mathbf{B}_{\chi} \cdot \boldsymbol{V}_{\chi ij } 
  \right)
\frac{\partial  }
  { \partial \varepsilon_\chi  }
  \right ] f_{0}^\prime (\varepsilon_\chi ) \, , \\
 \boldsymbol{T}_{\chi ij} & = e \,  \mathbf{\Omega }_{\chi } \,  B_{\chi i}   \, 
 v_{\chi j}^{(0)} 
 + \frac{1}{2}  \,  \hat{ \mathbf{e} }_{i} \,  v_{\chi j }^{(m)} , 
 \quad  \boldsymbol{V}_{\chi ij} = 
 \mathbf{\Omega }_{\chi} \,  v_{\chi i} ^{(0)} \, 
  v_{\chi j} ^{(0)} - \frac{ {\boldsymbol{m}}_{\chi}} {2\,e}  \,  
 \partial_{k^i} \,  v_{\chi j} ^{(0)} \,,
\end{align}
represents the contribution which goes to zero if OMM is set to zero. The symbol $\hat{ \mathbf{e} }_{i} $ represents
the unit vector along the Cartesian coordinate axis labelled by $i$.

The longitudinal and transverse components of the magnetoelectric conductivity tensor $\sigma^\chi $ (i.e., the LMC and the PHC) are computed from the starting expression shown in Eq.~\eqref{eq_elec}. The details of the intermediate steps are given in Appendices~\ref{app_lmc} and \ref{app_phc}. Before delving into the investigation of the behaviour of these response coefficients in the subsequent subsections, first let us analyze the relation between the current and the axial electromagnetic fields.

For the quasiparticles with chirality $\chi$, we have the electric current contribution
\begin{align}
\label{l1}
J_i^{\chi} (\mu ) =\sigma_{ij}^{\chi} (\mu)\,E_{\chi j}\,,
\end{align}
where $\sigma_{ij}^{\chi}$ is given by Eq.~\eqref{eq_elec}. For two conjugate nodes with chemical potential values $\mu_+$ and $\mu_-$ (as shown schematically in Fig.~\ref{fig2node} for WSMs), we define the total and axial currents as
\begin{align}
\label{l2}
 \boldsymbol{J} (\mu_+, \mu_-) =
\sum_{\chi=\pm 1}\boldsymbol{J}^{\chi}(\mu_\chi)
=\boldsymbol{J}^{+} (\mu_+) + \boldsymbol{J}^{-} (\mu_-)
 \text{ and }
\boldsymbol{J}_5  (\mu_+, \mu_-) =
\sum_{\chi=\pm 1}\chi \, \boldsymbol{J}^{\chi} (\mu_\chi)
=\boldsymbol{J}^{+} (\mu_+)  -\boldsymbol{J}^{-}(\mu_+)  \,,
\end{align}
respectively. These suggest to introduce analogous expressions for the conductivity tensors as
\begin{align}
\label{totalcond}
\sigma_{ij}  (\mu_+, \mu_-) = \sum_{\chi=\pm 1}\sigma_{ij}^{ \chi } (\mu_\chi)
\text{ and }
\sigma_{5ij}=\sum_{\chi=\pm 1}\chi \, \sigma_{ij}^{ \chi }(\mu_\chi) \,.
\end{align}
Using Eq.~\eqref{totalcond}, we find that
\begin{align}
& J_i (\mu_+, \mu_-)
=\sigma_{ij}^{+} (\mu_+)\,E_{+j}
+\sigma_{ij}^{-}  (\mu_-) \, E_{-j}
= \sigma_{ij} (\mu_+, \mu_-)\,E_j + \sigma_{5ij} (\mu_+, \mu_-)\, E_{5j}
\nn & 
\text{and }
J_{5i} (\mu_+, \mu_-) = \sigma_{ij}^{+} (\mu_+)\, E_{+j}-\sigma_{ij}^{-}(\mu_-) \,E_{-j}
=\sigma_{5ij} (\mu_+, \mu_-) \,E_j + \sigma_{ij} (\mu_+, \mu_-) \,E_{5j}\,,
\end{align}
where $\mathbf E_5$ is an axial pseudoelectric field, which can be generated artificially (as explained in the introduction).
For the node with chirality $\chi$, analogous to $\mathbf B_\chi$, the physical $ \mathbf{E} $ and the axial $ \mathbf{E} _5$ add up to give the effective electric field 
$  \mathbf{E} _{\chi}= \mathbf{E} +\chi\,\mathbf{E}_5$, reflecting the dependence on $\chi $. In this paper, we deal with the case where $\mathbf E_5 =\mathbf 0$.

\begin{figure*}[t]
\includegraphics[width=0.8 \linewidth]{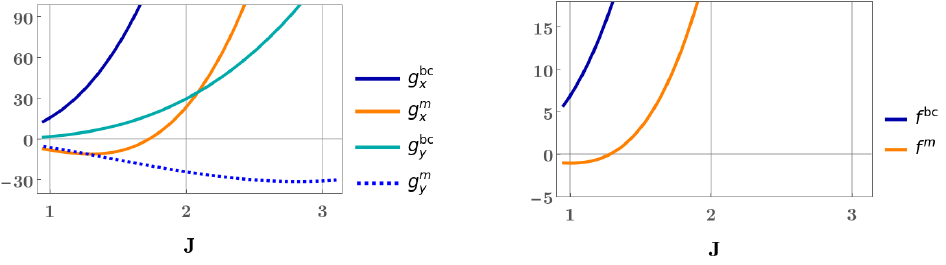}
\caption{\label{figcom}Comparison of the values of the functions defined in Eqs.~\eqref{eqg} and \eqref{eqf} for $J=1, \,2, \,3$.}
\end{figure*}

\begin{figure*}[t]
{\includegraphics[width=0.95 \linewidth]{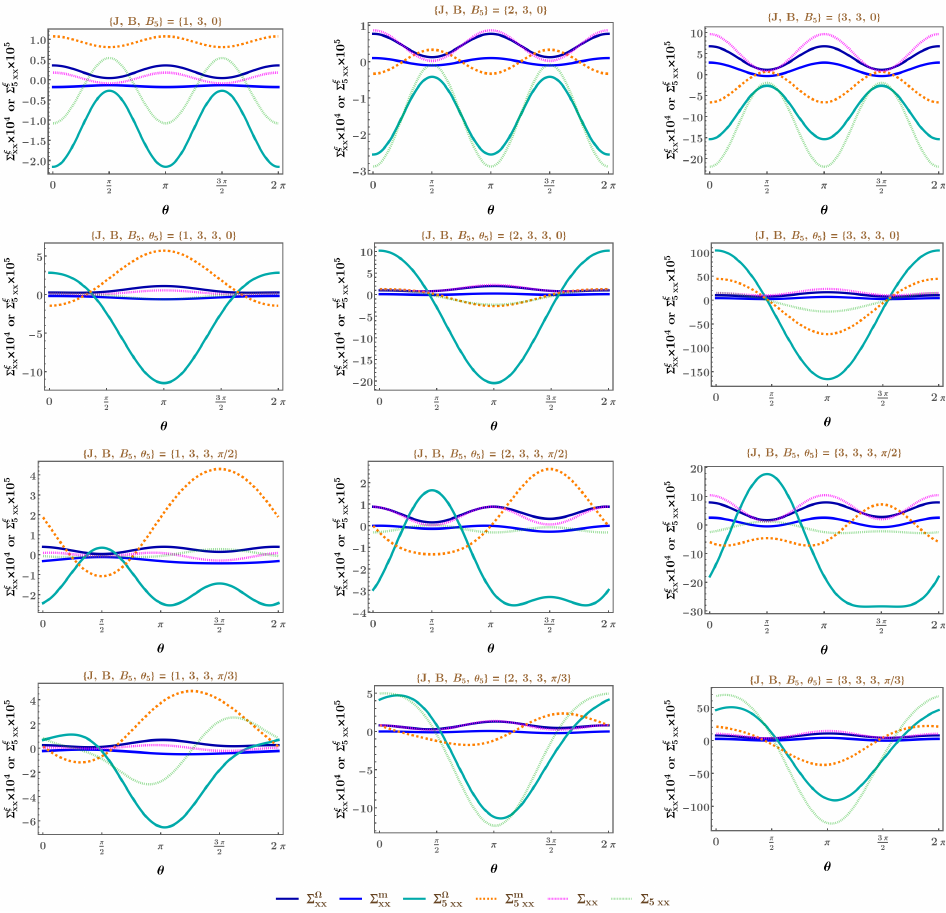}}
\caption{\label{figsxx1}
The total and axial combinations of the LMC (in units of eV) for the two conjugate nodes, defined in Eq.~\eqref{eqSig}, as functions of $ \theta  $, using various values of $B$ (in units of eV$^2$), $B_5$ (in units of eV$^2$), and $\theta_5 $ (as indicated in the plotlabels). We have set $v_z =0.005$, $\tau = 151$ eV$^{-1}$, $\beta = 1160 $ eV$^{-1}$, $\mu_+ = 0.4$ eV, and $\mu_- = 0.2$ eV. As explained below Eq.~\eqref{eqSig}, while $\Sigma^{\Omega}_{xx} $ ($\Sigma^{\Omega}_{5xx} $) represents the part of $\Sigma_{xx} $ ($\Sigma_{5xx} $) originating purely from the BC-contributions (i.e., with no OMM), $\Sigma^{m}_{xx} $ ($\Sigma^{m}_{5xx} $) is the contribution which vanishes if the OMM is not at all considered. We have used the superscript $\xi$ to indicate that, along the vertical axis, we have plotted the BC-only, OMM, and total parts, with the colour-coding shown in the plotlegends.
The values of the maxima and minima of the curves are strongly dependent on the values of $J$.
}
\end{figure*}

\begin{figure*}[t]
{\includegraphics[width=0.95\linewidth]{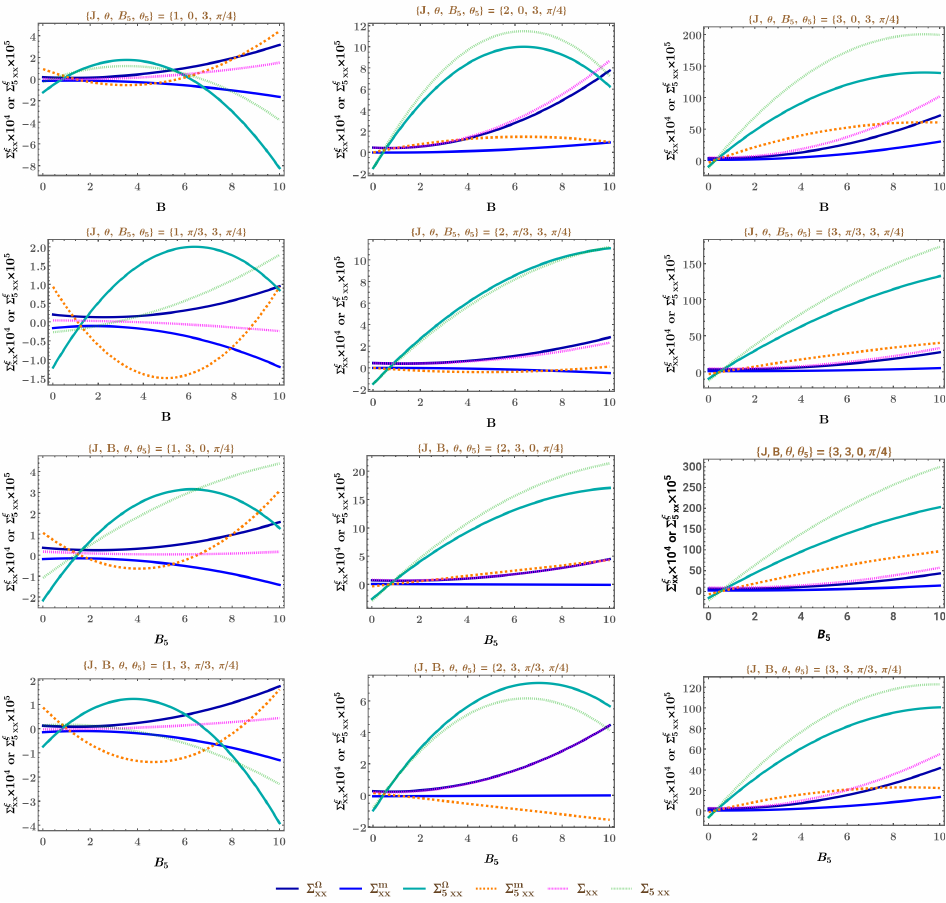}}
\caption{\label{figsxx2}The total and axial combinations of the LMC (in units of eV) for the two conjugate nodes, defined in Eq.~\eqref{eqSig}, as functions of $B$ (in units of eV$^2$) and $B_5$ (in units of eV$^2$), using various values of $\theta$ and $\theta_5 $ (as indicated in the plotlabels). We have set $v_z =0.005$, $\tau = 151$ eV$^{-1}$, $\beta = 1160 $ eV$^{-1}$, $\mu_+ = 0.4$ eV, and $\mu_- = 0.2$ eV. As explained below Eq.~\eqref{eqSig}, while $\Sigma^{\Omega}_{xx} $ ($\Sigma^{\Omega}_{5xx} $) represents the part of $\Sigma_{xx} $ ($\Sigma_{5xx} $) originating purely from the BC-contributions (i.e., with no OMM), $\Sigma^{m}_{xx} $ ($\Sigma^{m}_{5xx} $) is the contribution which vanishes if the OMM is neglected. We have used the superscript $\xi$ to indicate that, along the vertical axis, we have plotted the BC-only, OMM, and total parts, with the colour-coding shown in the plotlegends.
}
\end{figure*}

From the above expressions for the total and axial currents, we now discuss the behaviour of the total and axial LMC and PHC as functions of $\theta$, which is the angle between $\mathbf{E}$ and $\mathbf{B}$, with the former chosen to be directed along the $x$-axis [cf. Fig.~\ref{figsetup}]. For the illustration of the behaviour of the response, we define
\begin{align}
\label{eqSig}
& \Sigma_{ij} (\mathbf{B}_\chi)
 = \sigma_{ij}  (\mu_+, \mu_- )  - \sigma_{ij} ( \mu_+, \mu_- ) 
 \Big \vert_{\mathbf{B}_\chi = \boldsymbol{0} }
\text{ and }
\Sigma_{5ij} (\mathbf{B}_\chi )
 = \sigma_{5ij} (\mu_+, \mu_-) -\sigma_{5ij} (\mu_+, \mu_-) \Big \vert_{\mathbf{B}_\chi = \boldsymbol{0} }
\end{align}
for the total and axial conductivity tensor components, respectively, after subtracting off the $\mathbf B_\chi $-independent parts. We denote the parts connected with $\sigma^{\Omega, \chi}_{ij}$ and $\sigma^{m, \chi}_{ij}$ as $ ( \Sigma_{ij}^\Omega, \, \Sigma_{5ij}^\Omega )$ and $( \Sigma_{ij}^m, \, \Sigma_{5ij}^m )$, respectively. Therefore, if the OMM is not considered, each of $ ( \Sigma_{ij}^m, \, \Sigma_{5ij}^m )$ goes to zero.

\subsection{Longitudinal magnetoconductivity}
\label{secsxx}

\begin{figure*}[t]
{\includegraphics[width=0.95\linewidth]{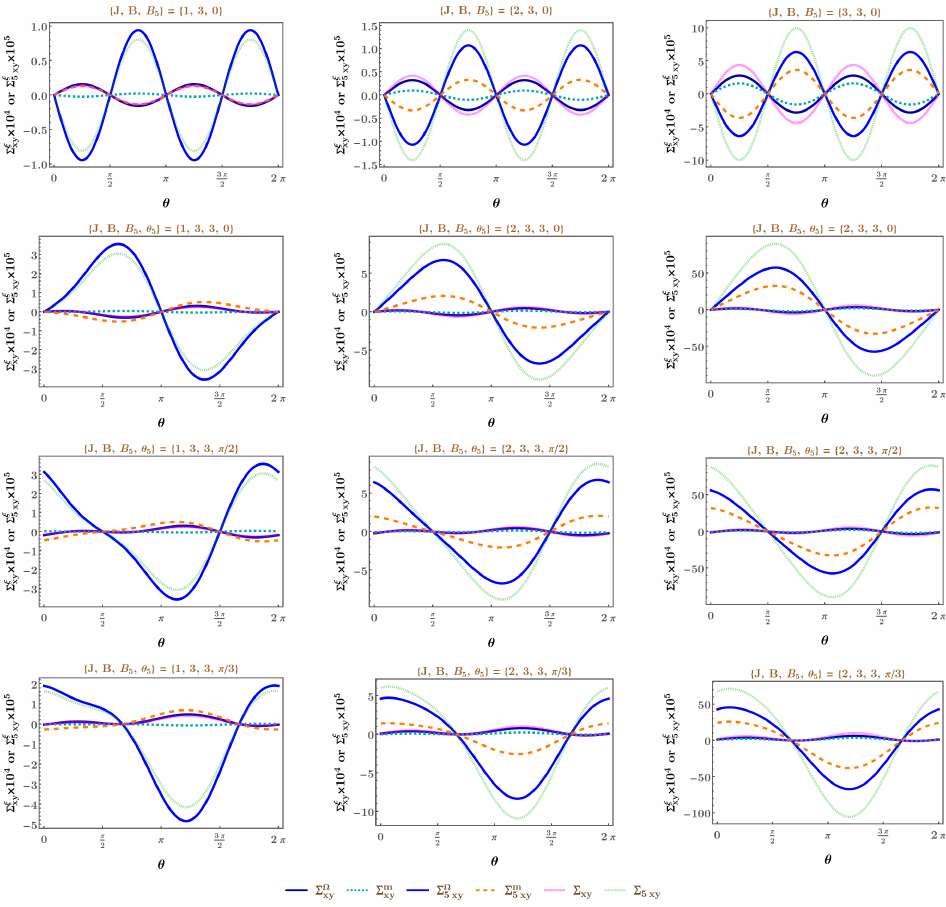}}
\caption{\label{figsxy1}The total and axial combinations of the PHC (in units of eV) for the two conjugate nodes, defined in Eq.~\eqref{eqSig}, as functions of $ \theta  $, using various values of $B$ (in units of eV$^2$), $B_5$ (in units of eV$^2$), and $\theta_5 $ (as indicated in the plotlabels). We have set $v_z =0.005$, $\tau = 151$ eV$^{-1}$, $\beta = 1160 $ eV$^{-1}$, $\mu_+ = 0.4$ eV, and $\mu_- = 0.2$ eV. As explained below Eq.~\eqref{eqSig}, while $\Sigma^{\Omega}_{xy} $ ($\Sigma^{\Omega}_{5xy} $) represents the part of $\Sigma_{xy} $ ($\Sigma_{5xy} $) originating purely from the BC-contributions (i.e., with no OMM), $\Sigma^{m}_{xy} $ ($\Sigma^{m}_{5xy} $) is the contribution which vanishes if the OMM is not at all considered. We have used the superscript $\xi$ to indicate that, along the vertical axis, we have plotted the BC-only, OMM, and (BC + OMM) parts, with the colour-coding shown in the plotlegends.
The values of the maxima and minima of the curves are strongly dependent on the values of $J$.
}
\end{figure*}

\begin{figure*}[t]
{\includegraphics[width = 0.99\linewidth]{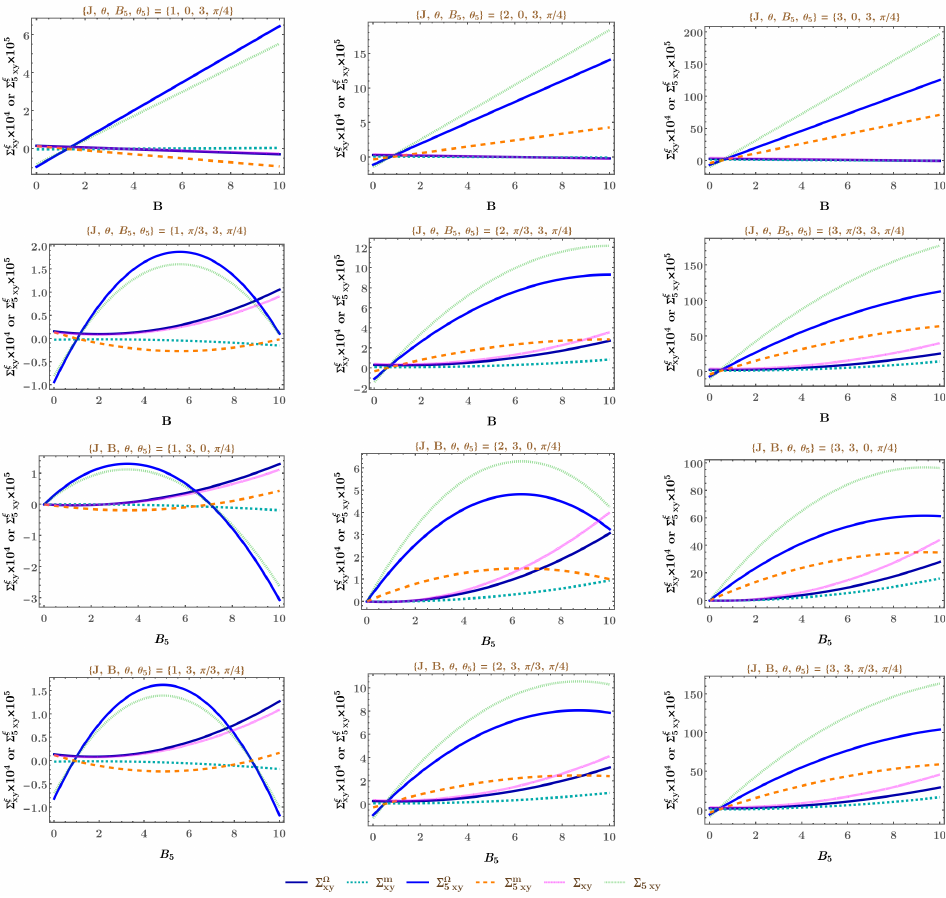}}
\caption{\label{figsxy2}The total and axial combinations of the PHC (in units of eV) for the two conjugate nodes, defined in Eq.~\eqref{eqSig}, as functions of $B$ (in units of eV$^2$) and $B_5$ (in units of eV$^2$), using various values of $\theta$ and $\theta_5 $ (as indicated in the plotlabels). We have set $v_z =0.005$, $\tau = 151$ eV$^{-1}$, $\beta = 1160 $ eV$^{-1}$, $\mu_+ = 0.4$ eV, and $\mu_- = 0.2$ eV. As explained below Eq.~\eqref{eqSig}, while $\Sigma^{\Omega}_{xy} $ ($\Sigma^{\Omega}_{5xy} $) represents the part of $\Sigma_{xy} $ ($\Sigma_{5xy} $) originating purely from the BC-contributions (i.e., with no OMM), $\Sigma^{m}_{xy} $ ($\Sigma^{m}_{5xy} $) is the contribution which vanishes if the OMM is neglected. We have used the superscript $\xi$ to indicate that, along the vertical axis, we have plotted the BC-only, OMM, and (BC + OMM) parts, with the colour-coding shown in the plotlegends.
}
\end{figure*}

Using the explicit expressions derived in Appendix~\ref{app_lmc}, we have
\begin{align}
\sigma^{\chi}_{xx}  (\mu_\chi) &= 
\sigma ^{0, \chi}_{xx}  (\mu_\chi) + \sigma ^{\Omega , \chi}_{xx}  (\mu_\chi) 
+  \sigma ^{m, \chi}_{xx}  (\mu_\chi)\,,
\end{align}
where
\begin{align}
\label{eqsigxxfin}
& \sigma^{0, \chi}_{xx} (\mu_\chi) =
\frac{ e^2 \,  \tau  \,  J   }
{ 6 \,  \pi^2   \,  v_z    }  \,  \Lambda_{2} (\mu_\chi ) \,,\quad
\sigma^{\Omega , \chi}_{xx}   (\mu_\chi)  = 
\frac{e^4 \,  \tau \,  v_{z } \,  \alpha_{J}^{ \frac{2} {J} }   
\, \Lambda_{ - \frac{2} {J} } ( \mu_\chi )
}
{ 128 \,  \pi ^{\frac{3}{2}} } \,  \frac{  \Gamma  \big( 2-\frac{1}{J}\big) }
{  \Gamma \big (  \frac{9}{2} - \frac{1}{J}  \big)}
  \left [ g_x^{bc} (J)\,  B_{\chi x}^2
+ g_y^{bc}  (J)\,  B_{\chi y}^2    \right ]  ,\nn
& \sigma^{m, \chi}_{xx}  (\mu_\chi)
  =   \frac{   e^4 \,  \tau \, v_z \,   \alpha_{J}^{ \frac{2} {J} } 
\Lambda_{ - \frac{2} {J} } (  \mu_\chi )
\, \Gamma \big ( 2 - \frac{1}{J} \big) }
{ 128 \,  \pi^{\frac{3}{2}} 
\, \Gamma \big (  \frac{9}{2} - \frac{1}{J}  \big)  } 
\left [  g_x^m (J) \, B_{\chi x}^2 +   g_y^{m} (J)\,  B_{\chi y}^2  \right ] ,
\end{align}
and
\begin{align}
\label{eqg}
& g_x^{bc}  (J) = J \left ( 32 \,  J^2 -19  \, J + 3  \right ) , \quad
g_y^{bc}  (J) = J \left (3\, J- 1 \right) \left  (2\,J-1 \right)  , \nn &
g_x^m  (J) =  \frac{37 \, J^4 - 100\, J^3 + 74 \, J^{2} -21 \, J +2 } {J} \,,\quad
g_y^{m}  (J) =  \frac{ 3 \,J^4 - 12 \,J^3 - 4 \,J^2  + 9\, J - 2} {J}  \,.
\end{align}
From Fig.~\ref{figcom}, we find that (1) $g_x^{bc}  (J)$ and $g_y^{bc} (J)$ are positive for all $J$-values;
(2) $g_x^m (J)$ is negative for $J=1$ and positive for $J=2, \,3$;
(3) $g_y^m (J)$ is negative for all $J$-values. For $J=1$, the OMM acts in opposition to the BC-only term for the $B_{\chi x}$-part, and reduces the overall response. On the other hand, for the mWSMs, the OMM-part adds up to the BC-only term for the $B_{\chi x}$-part, thus increasing the overall response. For the $B_{\chi y}$-part, the OMM and the BC-only parts always have opposite signs and, thus, tend to reduce the overall response's magnitude. Therefore, for a WSM, the OMM part always reduces the value of $\sigma^{\chi}_{xx} $ by adding a negative contribution.

In order to estimate the effects of the OMM (which was neglected in many earlier works), we plot the behaviour of $ \Sigma_{xx}^\Omega $, $ \Sigma_{xx}^m $, $\Sigma_{5xx}^\Omega  $, and $\Sigma_{5xx}^m   $ in Figs.~\ref{figsxx1} and \ref{figsxx2}. The interplay of the BC-only and the OMM-induced parts are illustrated via some representative parameter values. In agreement with our comparison of the $g$-values, we find that for a WSM, the OMM can even change the sign of the response, depending on the net magnetic field. However, for $J= 2, \, 3$, the $g_x^m$ and the $g_y^m$-parts have opposite signs --- hence, their combined effects may increase or reduce the overall response. From Fig.~\ref{figsxx1}, we find that turning on a nonzero $\mathbf B_5$-part changes the periodicity, with respect to $\theta $, from $\pi $ to $2 \pi $. This is completely expected because the axial pseudomagnetic field causes linear-in-$B \cos \theta $ and/or linear-in-$B\sin \theta $ terms to appear, in addition to the quadratic-in-$B_i$ dependence of the untilted WSMs/mWSMs.

\subsection{Planar Hall conductivity}
\label{secsxy}

Using the explicit expressions derived in Appendix~\ref{app_phc}, we have
\begin{align}
\sigma ^{\chi}_{yx}  (\mu_\chi)  = \sigma ^{\chi}_{xy}  (\mu_\chi)
= \sigma ^{0, \chi}_{xy}  (\mu_\chi) + \sigma ^{\Omega , \chi}_{xy}  (\mu_\chi) 
+  \sigma ^{m, \chi}_{xy}  (\mu_\chi)\,,
\end{align}
where
\begin{align}
\label{eqsigxyfin}
\sigma ^{0, \chi}_{xy} (\mu_\chi) &=   0 \,, \quad
\sigma ^{\Omega , \chi}_{xy} (\mu_\chi)
  =    \frac{e^4 \,  \tau \,  v_z \,    \alpha_{J}^{ \frac{2}{J} } 
 \,   \Lambda_{ - \frac{2} {J} } (\mu_\chi )
 }
{ 64 \,  \pi ^{\frac{3}{2}}  }    \,   \frac{\Gamma \big (  2-\frac{1}{J}  \big)}
{\Gamma \big (  \frac{9}{2} - \frac{1}{J}  \big)}
\,f^{bc} (J) \,  B_{\chi x} \,  B_{\chi y} \, ,\nn
\sigma ^{m, \chi}_{xy} (\mu_\chi)
& =  \frac{e^4  \,  \tau \, v_z
\, \alpha_{J}^{ \frac{2} {J} }  \,  \Lambda_{ - \frac{2} {J} } (\mu_\chi )    }
{64 \,  \pi ^{\frac{3}{2}} } \, 
  \frac{  
  \Gamma \big (  2-\frac{1}{J}  \big)}{  \Gamma \big (  \frac{9}{2} - \frac{1}{J}  \big)}
\, f^m (J) \,  B_{\chi x} \,  B_{\chi y}   \, ,
\end{align}
and
\begin{align}
\label{eqf}
& f^{bc} (J) = J \left ( 13\, J^{2} - 7 \, J +1 \right ) , \quad
f^m (J) = 17 \, J^3-44 \,J^2 + 39\, J -15 + \frac{2}{J}\, .
\end{align}
From Fig.~\ref{figcom}, we find that (1) $ f^{bc}  (J)$ is positive for all $J$-values;
(2) $f^m (J)$ is negative for $J=1$ and positive for $J=2, \,3$. For $J=1$, the OMM thus acts in opposition to the BC-only term, and reduces the magnitude of the overall PHC. In contrast, for the mWSMs, the OMM adds up to the BC-only term, thus increasing the overall response.

In order to estimate the effects of the OMM, we plot the behaviour of $ \Sigma_{xy}^\Omega $, $ \Sigma_{xy}^m $, $\Sigma_{5xy}^\Omega  $, and $\Sigma_{5xy}^m  $ in Figs.~\ref{figsxy1} and \ref{figsxy2}. The interplay of the BC-only and the OMM-induced parts are illustrated via the same parameter values as considered for the LMC. In agreement with our comparison of the $ f $-values, we find that for a WSM, the OMM always reduces the response. On the other hand, for $J= 2, \, 3$, the effect of OMM is to enhance the overall response. Analogous to the LMC, Fig.~\ref{figsxy1} shows that a nonzero $\mathbf B_5$-part changes the periodicity with respect to $\theta $ from $\pi $ to $2 \pi $, which results from the emergence of terms linearly proportional to the components of $\mathbf B $, rather than just the quadratic ones.

\section{Magnetothermoelectric conductivity}
\label{secalpha}

Defining
\begin{align}
F_{ij}^\chi =  D_{\chi}
  \left[  v_{\chi i } + e \,  (  {\boldsymbol  v}_{\chi} \cdot \mathbf{\Omega }_{\chi} ) \,  B_{\chi i} \right] \,  \left[  v_{\chi j } + e  \,  (  {\boldsymbol  v}_{\chi} \cdot \mathbf{\Omega }_{\chi} )  \, B_{\chi j} \right] ,
\end{align}
we expand it as
\begin{align}
\label{eqexpf}
& F_{ij}^\chi  =  F_{ij}^{0,\chi}
 + F_{ij} ^{1,\Omega,\chi} + F_{ij} ^{1, m,\chi} +   F_{ij} ^{2,\Omega,\chi}
 +  F_{ij} ^{2,m,\chi} + F_{ij} ^{2,(\Omega, m),\chi} + \order{ |{ \mathbf B}_{\chi}|^3} \,, 
\end{align}
where
\begin{align}
& F_{ij}^{0, \chi} = v_{\chi i }^{(0)} \, v_{\chi j }^{(0)} \,, \quad
F_{ij}^{1,\Omega, \chi} =   e  \left [  {\boldsymbol  v}_{\chi} ^{(0)} \cdot \mathbf{\Omega }_{\chi}  \right ] 
 \left [ v_{\chi i }^{(0)} \, B_{\chi j} +  B_{\chi i} \, v_{\chi j }^{(0)} \right ]
 - e   \left( {\mathbf B}_{\chi} \cdot \mathbf{\Omega }_{\chi}  \right)  
  v_{\chi i } ^{(0)}  \,   v_{\chi j } ^{(0)}  \,,\quad
F_{ij}^{1,m, \chi} = \left [ v_{\chi i } ^{(0)} \,v_{\chi j }^{(m)}  
 + v_{\chi i }^{(m)} \, v_{\chi j } ^{(0)} \right ] ,
\nn &  F_{ij} ^{2,\Omega,\chi} =   e^2 \,  Q_{\chi i } \, Q_{\chi j } \,,\quad
 F_{ij} ^{2,m,\chi} =  v_{\chi i } ^{(m)} \,v_{\chi j }^{(m)}\,, \nn
& F_{ij} ^{2,(\Omega, m),\chi} = 
e \left  [  {\boldsymbol  v}_{\chi} ^{(0)} \cdot \mathbf{\Omega }_{\chi} \right ]
\left [ v_{\chi i }^{(m)} \, B_{\chi j} +  B_{\chi i}  \,  v_{\chi j } ^{(m)} \right ] 
  + e  \left  [   {\boldsymbol  v}_{\chi} ^{(m)} \cdot \mathbf{\Omega }_{\chi} \right  ]
  \left  [ v_{\chi i } ^{(0)} \,B_{\chi j}  +   B_{\chi i} \,v_{\chi j} ^{(0)} \right ]
 - e \left( {\mathbf B}_{\chi} \cdot \mathbf{\Omega }_{\chi}  \right) 
  \left  [ v_{\chi i } ^{(0)} \, v_{\chi j} ^{(m)} 
 + v_{\chi i }^{(m)} \, v_{\chi j } ^{(0)} \right ] .
\end{align}
Here, $F_{ij}^{0, \chi}$ is $\mathbf B_\chi $-independent, $ F_{ij} ^{1,\Omega, \chi}$ is linear in the components of the BC,
$ F_{ij} ^{1,m, \chi}$ is linear in the OMM-contributions, $ F_{ij} ^{2,\Omega, \chi}$ is quadratic in the components of the BC,
$ F_{ij} ^{1,m, \chi}$ is quadratic in the OMM-contributions, and $ F_{ij} ^{2,(\Omega, m),\chi} $ is a mixed term which contains products of the BC-components and the OMM-contributions.

Using the above expressions for the function defined as
$ G_{ij}^\chi = F_{ij}^\chi \,  \frac{\mathcal{E}_{\chi}  - \mu }{T} $, the weak-field expansion gives us
\begin{align}
G_{ij}^\chi & =
F_{ij}^\chi \,\frac{\varepsilon_\chi  - \mu} {T} 
 + \left[ F_{ij}^{0, \chi}
+ F_{ij}^{1,\Omega, \chi} + F_{ij}^{1,m, \chi}  \right]  
 \frac{  \varepsilon_\chi ^{ (m) } }{T} 
 + \order{ |{ \mathbf B}_{\chi}|^3 } \nn
& = G_{ij}^{0,\chi}
 + G_{ij} ^{1,\Omega,\chi} + G_{ij} ^{1, m,\chi} +  G_{ij} ^{2,\Omega,\chi}
 +  G_{ij} ^{2,m,\chi} + G_{ij} ^{2,(\Omega, m),\chi} + \order{ |{ \mathbf B}_{\chi}|^3 } \,.
\end{align}
Here,
\begin{align}
& G_{ij}^{0,\chi} = \frac{  F_{ij}^{0, \chi} \left( \varepsilon_\chi  - \mu \right) } {T}
 \,,\quad
 G_{ij} ^{1,\Omega ,\chi} = \frac{  F_{ij} ^{1,\Omega, \chi}  
 \left( \varepsilon_\chi  - \mu \right) } {T} \,,\quad
 G_{ij} ^{1,m,\chi} =  \frac{ F_{ij}^{0, \chi}\, \varepsilon_\chi ^{ (m) } }{T} 
 + \frac{  F_{ij} ^{1,m, \chi}  \left( \varepsilon_\chi  - \mu \right) } {T} \,,
 \quad G_{ij} ^{2,\Omega ,\chi} = \frac{  F_{ij} ^{2,\Omega, \chi}  
 \left( \varepsilon_\chi  - \mu \right) } {T} \,,\nn
& G_{ij} ^{2,m,\chi} =  \frac{ F_{ij}^{1, m, \chi}\, \varepsilon_\chi ^{ (m) } }{T} 
 + \frac{  F_{ij} ^{ 2,m, \chi}  \left( \varepsilon_\chi  - \mu \right) } {T} \,, 
\quad G_{ij} ^{2,(\Omega, m),\chi} = 
\frac{ F_{ij}^{1, \Omega, \chi}\, \varepsilon_\chi ^{ (m) } }{T} 
+
\frac{  F_{ij} ^{ 2,(\Omega, m), \chi}  \left( \varepsilon_\chi  - \mu \right) } {T}\,,
\end{align}
with the parts named in a similar spirit as done for the case of $F_{ij}^\chi$.

Defining a third function $ H_{ij}^\chi = G_{ij} \, f'_{0} ( \mathcal{E}_{\chi}   )  $, and using Eq.~\eqref{Exp_f}, its expansion turns out to be
\begin{align}
H_{ij}^\chi & =
G_{ij}^\chi \, f'_{0} ( \varepsilon_\chi )
 + \left[ G_{ij}^{0, \chi}
+ G_{ij}^{1,\Omega, \chi} + G_{ij}^{1,m, \chi}  \right]  
 \, \varepsilon_\chi ^{ (m) } \, f^{\prime \prime}_{0} ( \varepsilon_\chi )
+ \frac{ G_{ij}^{0, \chi} \,\left( \varepsilon_\chi ^{ (m) } \right)^2 
\, f^{\prime \prime \prime }_{0} ( \varepsilon_\chi )} 
{2} 
 + \order{ |{ \mathbf B}_{\chi}|^3}  \nn
& = G_{ij}^{0,\chi}
 + G_{ij}^{1,\Omega,\chi} + G_{ij}^{1, m,\chi} +  G_{ij}^{2,\Omega,\chi}
 +  G_{ij}^{2,m,\chi} + G_{ij}^{2,(\Omega, m),\chi} + \order{ |{ \mathbf B}_{\chi}|^3} \,.
\end{align}
Here,
\begin{align}
& H_{ij}^{0,\chi} = G_{ij}^{0, \chi} \, f'_{0} ( \varepsilon_\chi )\,, \quad
H_{ij}^{1,\Omega,\chi} = G_{ij}^{1,\Omega,\chi} \, f'_{0} ( \varepsilon_\chi )\,,\quad
H_{ij}^{1,m,\chi} = 
G_{ij}^{0, \chi}  \, \varepsilon_\chi ^{ (m) } \, f^{\prime \prime}_{0} ( \varepsilon_\chi ) 
+ G_{ij}^{1,m,\chi} \, f'_{0} ( \varepsilon_\chi )\,,\nn
& H_{ij}^{2 ,\Omega ,\chi} = G_{ij}^{2 ,\Omega ,\chi} \, f'_{0} ( \varepsilon_\chi )\,,
\quad H_{ij}^{2 ,m ,\chi} =
\frac{G_{ij}^{0, \chi} \left( \varepsilon_\chi ^{ (m) } \right)^2 
f^{\prime \prime \prime}_{0} ( \varepsilon_\chi ) 
} {2} +
G_{ij}^{ 1, m, \chi}  \, \varepsilon_\chi ^{ (m) } \, f^{\prime \prime}_{0} ( \varepsilon_\chi ) 
+ G_{ij}^{2 , m ,\chi} \, f'_{0} ( \varepsilon_\chi )\,, \nn &
H_{ij}^{2,(\Omega, m),\chi} = 
 G_{ij}^{1,\Omega,\chi} \, \varepsilon_\chi ^{ (m) } \, f^{\prime \prime}_{0} ( \varepsilon_\chi ) 
+ G_{ij}^{2,(\Omega, m),\chi} \, f'_{0} ( \varepsilon_\chi )\,.
\end{align}
The nomenclature of these parts follow the same scheme as that for $ F_{ij}^\chi $ and $ G_{ij}^\chi$.

\begin{figure*}[t]
{\includegraphics[width = 0.99\linewidth]{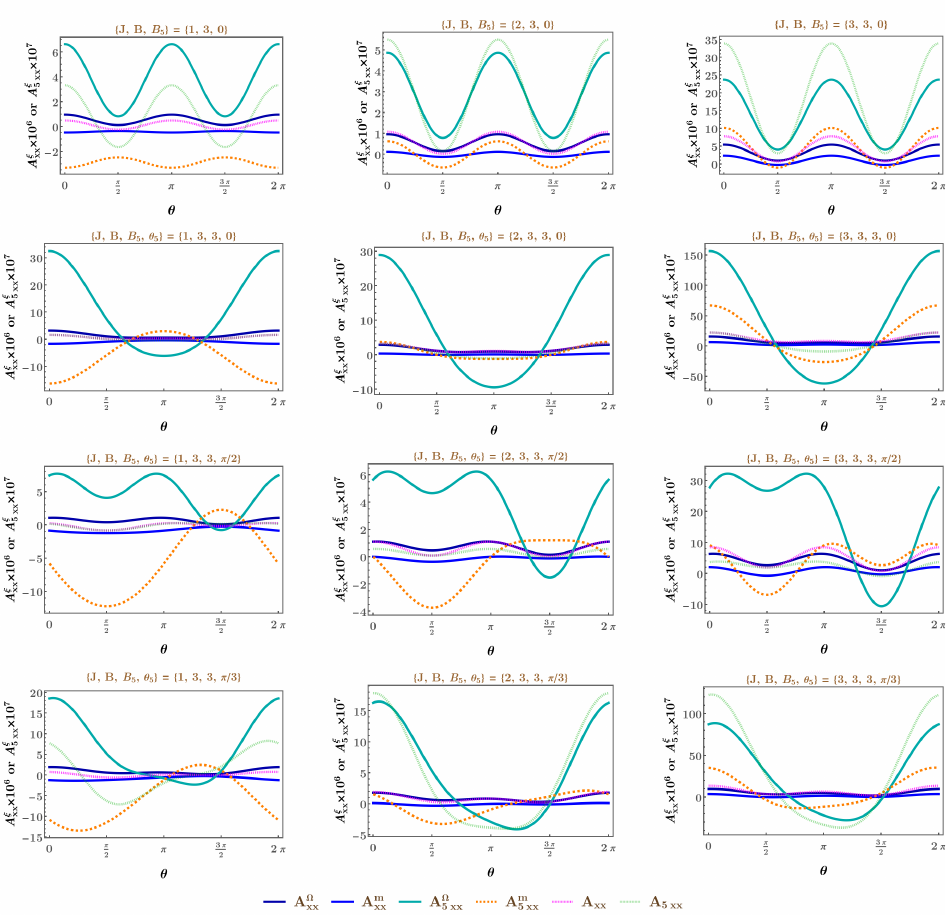}}
\caption{\label{figaxx1}The total and axial combinations of the LTEC (in units of eV) for the two conjugate nodes, defined in Eq.~\eqref{eqA}, as functions of $ \theta  $, using various values of $B$ (in units of eV$^2$), $B_5$ (in units of eV$^2$), and $\theta_5 $ (as indicated in the plotlabels). We have set $v_z =0.005$, $\tau = 151$ eV$^{-1}$, $\beta = 1160 $ eV$^{-1}$, $\mu_+ = 0.2 $ eV, and $\mu_- = 0.35 $ eV. As explained below Eq.~\eqref{eqA}, while $ A^{\Omega}_{xx} $ ($ A^{\Omega}_{5xx} $) represents the part of $ A_{xx} $ ($A_{5xx} $) originating purely from the BC-contributions (i.e., with no OMM), $ A^{m}_{xx} $ ($ A^{m}_{5xx} $) is the contribution which vanishes if OMM is not at all considered. We have used the superscript $\xi$ to indicate that, along the vertical axis, we have plotted the BC-only, OMM, and (BC + OMM) parts, with the colour-coding shown in the plotlegends.
The values of the maxima and minima of the curves are strongly dependent on the values of $J$.
}
\end{figure*}

Starting from Eq.~\eqref{eq_thermoel}, applying the weak-in-magnetic-field limit using the expansions outlined above, the expression for the magnetoelectric conductivity tensor $\alpha^\chi$ is dissociated into
\begin{align}
& \alpha^{\chi}_{ij}  = \alpha^{0, \chi}_{ij} + 
+ \alpha^{\Omega , \chi}_{ij}  +  \alpha^{m, \chi}_{ij} \,,
\text{ where } 
\alpha^{\Omega , \chi}_{ij}  = \alpha^{1,\Omega , \chi}_{ij} + \alpha^{2,\Omega , \chi}_{ij}
\,, \quad
\alpha^{m, \chi}  =  \alpha^{1, m , \chi}_{ij} + \alpha^{2, m , \chi}_{ij} 
+ \alpha^{2, (\Omega, m), \chi}_{ij} \,.
\label{eqalphatot}
\end{align}
The terms $\alpha^{0, \chi}_{ij}$, $\alpha^{1,\Omega , \chi}_{ij} $, $\alpha^{2,\Omega , \chi}_{ij} $, $\alpha^{1,m , \chi}_{ij} $, $\alpha^{2,m , \chi}_{ij} $, and $\alpha^{2, (\Omega, m), \chi}_{ij} $ consist of the integrands $ H^{0, \chi}_{ij}$, $ H^{1,\Omega , \chi}_{ij} $, $ H^{2,\Omega , \chi}_{ij} $, $ H^{1,m , \chi}_{ij} $, $ H^{2,m , \chi}_{ij} $, and $ H^{2, (\Omega, m), \chi}_{ij} $, respectively. Analogous to the case of $\sigma^\chi_{ij}$, the term $  \alpha_{ij}^{m, \chi}  $ goes to zero if the OMM is set to zero.

The longitudinal and transverse components of the magnetothermoelectric conductivity tensor $\alpha^\chi $ (i.e., the LTEC and the TTEC) are computed from the expressions shown above. The details of the intermediate steps have been relegated to Appendices~\ref{appalphaxx} and \ref{appalphayx}.

\begin{figure*}[t]
{\includegraphics[width = 0.99\linewidth]{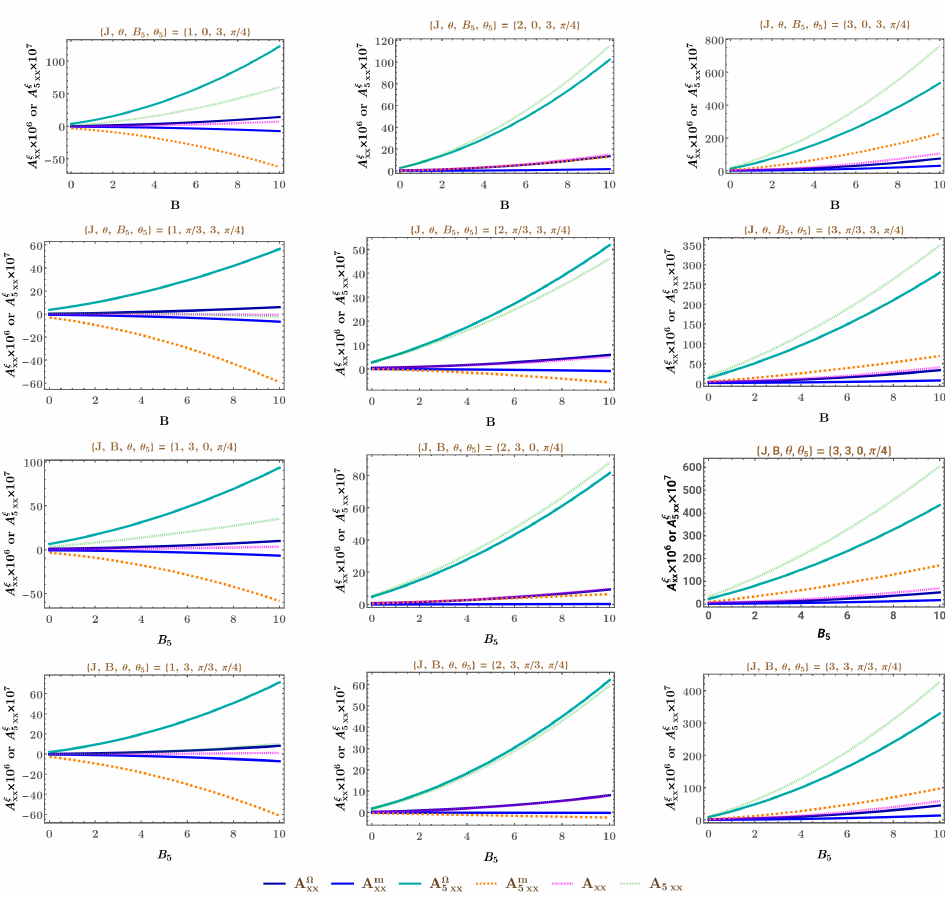}}
\caption{\label{figaxx2}The total and axial combinations of the LTEC (in units of eV) for the two conjugate nodes, defined in Eq.~\eqref{eqA}, as functions of $B$ (in units of eV$^2$) and $B_5$ (in units of eV$^2$), using various values of $\theta$ and $\theta_5 $ (as indicated in the plotlabels). We have set $v_z =0.005$, $\tau = 151$ eV$^{-1}$, $\beta = 1160 $ eV$^{-1}$, $\mu_+ = 0.2$ eV, and $\mu_- = 0.35$ eV. As explained below Eq.~\eqref{eqA}, while $ A^{\Omega}_{xx} $ ($A^{\Omega}_{5xx} $) represents the part of $ A_{xx} $ ($ A_{5xx} $) originating purely from the BC-contributions (i.e., with no OMM), $A^{m}_{xx} $ ($ A^{m}_{5xx} $) is the contribution which vanishes if the OMM is neglected. We have used the superscript $\xi$ to indicate that, along the vertical axis, we have plotted the BC-only, OMM, and (BC + OMM) parts, with the colour-coding shown in the plotlegends.
}
\end{figure*}

Akin to the magnetoconductivity tensors, we define the total and axial magnetothermoelectric tensors as 
\begin{align}
\alpha_{ij}=\sum_{\chi}\alpha_{ij}^{ \chi }
\text{ and }
\alpha_{5ij}=\sum_{\chi}\chi\, \alpha_{ij}^{ \chi } \,,
\end{align}
respectively.
Subtracting off the $\mathbf B_\chi $-independent parts, we define
\begin{align}
\label{eqA}
& A_{ij} (\mathbf{B}_\chi)
 = \alpha_{ij}  (\mu_+, \mu_- )  - \alpha_{ij} ( \mu_+, \mu_- ) 
 \Big \vert_{\mathbf{B}_\chi = \boldsymbol{0} }
\text{ and }
A_{5ij} (\mathbf{B}_\chi )
 = \alpha_{5ij} (\mu_+, \mu_-) - \alpha_{5ij} (\mu_+, \mu_-) 
 \Big \vert_{\mathbf{B}_\chi = \boldsymbol{0} } .
\end{align}
We denote the parts connected with $\alpha^{\Omega, \chi}_{ij}$ and $\alpha^{m, \chi}_{ij}$ as $ ( A_{ij}^\Omega, \, A_{5ij}^\Omega )$
and $( A_{ij}^m, \, A_{5ij}^m )$, respectively. Therefore, if the OMM is not considered, each of
$ ( A_{ij}^m, \, A_{5ij}^m )$ goes to zero. 

\begin{figure*}[t]
{\includegraphics[width = 0.99\linewidth]{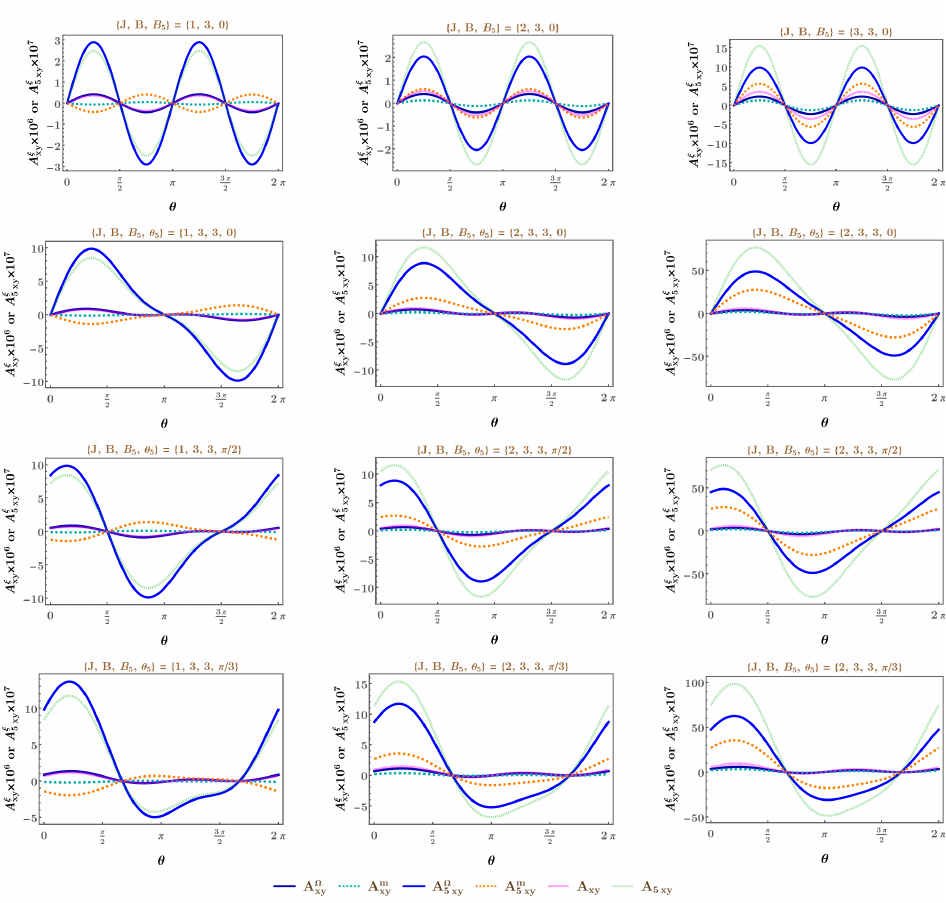}}
\caption{\label{figaxy1}The total and axial combinations of the TTEC (in units of eV) for the two conjugate nodes, defined in Eq.~\eqref{eqA}, as functions of $ \theta  $, using various values of $B$ (in units of eV$^2$), $B_5$ (in units of eV$^2$), and $\theta_5 $ (as indicated in the plotlabels). We have set $v_z =0.005$, $\tau = 151$ eV$^{-1}$, $\beta = 1160 $ eV$^{-1}$, $\mu_+ = 0.2 $ eV, and $\mu_- = 0.35 $ eV. As explained below Eq.~\eqref{eqA}, while $ A^{\Omega}_{xy} $ ($ A^{\Omega}_{5xy} $) represents the part of $ A_{xy} $ ($A_{5xy} $) originating purely from the BC-contributions (i.e., with no OMM), $ A^{m}_{xy} $ ($ A^{m}_{5xy} $) is the contribution which vanishes if OMM is not at all considered. We have used the superscript $\xi$ to indicate that, along the vertical axis, we have plotted the BC-only, OMM, and (BC + OMM) parts, with the colour-coding shown in the plotlegends.
The values of the maxima and minima of the curves are strongly dependent on the values of $J$.
}
\end{figure*}

\subsection{Longitudinal magnetothermoelectric coefficient}
\label{secaxx}

\begin{figure*}[t]
{\includegraphics[width = 0.99\linewidth]{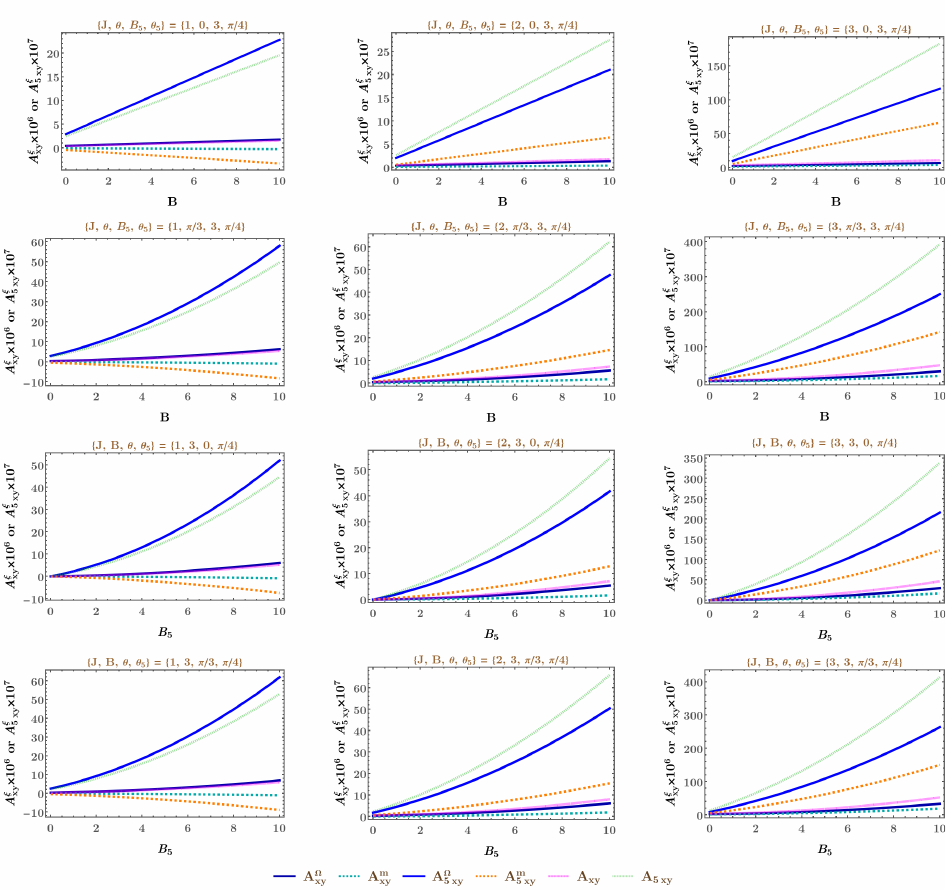}}
\caption{\label{figaxy2}The total and axial combinations of the TTEC (in units of eV) for the two conjugate nodes, defined in Eq.~\eqref{eqA}, as functions of $B$ (in units of eV$^2$) and $B_5$ (in units of eV$^2$), using various values of $\theta$ and $\theta_5 $ (as indicated in the plotlabels). We have set $v_z =0.005$, $\tau = 151$ eV$^{-1}$, $\beta = 1160 $ eV$^{-1}$, $\mu_+ = 0.2$ eV, and $\mu_- = 0.35$ eV. As explained below Eq.~\eqref{eqA}, while $ A^{\Omega}_{xy} $ ($A^{\Omega}_{5xy} $) represents the part of $ A_{xy} $ ($ A_{5xy} $) originating purely from the BC-contributions (i.e., with no OMM), $A^{m}_{xy} $ ($ A^{m}_{5xy} $) is the contribution which vanishes if the OMM is neglected. We have used the superscript $\xi$ to indicate that, along the vertical axis, we have plotted the BC-only, OMM, and (BC + OMM) parts, with the colour-coding shown in the plotlegends.
}
\end{figure*}

Using the explicit expressions derived in Appendix~\ref{appalphaxx}, we have
\begin{align}
\alpha ^{\chi}_{xx}  (\mu_\chi)  
= \alpha^{0, \chi}_{xx}  (\mu_\chi) + \alpha^{\Omega , \chi}_{xx}  (\mu_\chi) 
+  \alpha^{m, \chi}_{xx}  (\mu_\chi)\,,
\end{align}
where
\begin{align}
\label{eqalphaxxfin}
\alpha^{0, \chi}_{xx} (\mu_\chi) &= 
 - \frac{ e \,  \tau  \,  J \, \mu_\chi  }
 { 9   \,  v_z \, \beta  } \,, \quad
\alpha^{\Omega , \chi}_{xx} (\mu_\chi)
  =  \frac{ \sqrt{\pi } \, e^3 \,  \tau \,   v_{z } \,  \alpha_{J}^{ \frac{2} {J} } \, 
 \Gamma  \big( 2-\frac{1}{J}\big) }
 { 192 \,   \beta \, \mu_\chi^{1 + \frac{2} { J} }
\, \Gamma \big (  \frac{9}{2} - \frac{1}{J}  \big)} \,    
  \frac{ g_x^{bc} \, B_{\chi x}^2 
  + g_y^{bc}  \,  B_{\chi y}^2} {J} \, ,\nn
\alpha^{m, \chi}_{xx} (\mu_\chi)
& =   \frac{ \sqrt{\pi} \, e^3 \,  \tau \, v_z \,   \alpha_{J}^{ \frac{2} {J} } 
\, \Gamma \big ( 2 - \frac{1}{J} \big) }
{ 192  \,   \mu_\chi^{1+\frac{2} {J}} \,   \beta
\,  \Gamma \big (  \frac{9}{2} - \frac{1}{J}  \big)  }  
  \,      \frac{ g_x^m \, B_{\chi x}^2  + g_y^m  \,  B_{\chi y}^2} {J} \, . 
\end{align}

\subsection{Transverse magnetothermoelectric coefficient}
\label{secaxy}

Using the explicit expressions derived in Appendix~\ref{appalphayx}, we have
\begin{align}
\alpha ^{\chi}_{yx}  (\mu_\chi)  = \alpha ^{\chi}_{xy}  (\mu_\chi)
= \alpha^{0, \chi}_{xy}  (\mu_\chi) + \alpha^{\Omega , \chi}_{xy}  (\mu_\chi) 
+  \alpha^{m, \chi}_{xy}  (\mu_\chi)\,,
\end{align}
where
\begin{align}
\label{eqalphaxyfin}
& \alpha^{0, \chi}_{xy} (\mu_\chi) =   0 \,, \quad
\alpha^{\Omega , \chi}_{xy} (\mu_\chi)
  =   \frac{ \sqrt{\pi } \, 
  e^3 \,  \tau \,  v_z    \,  \alpha_{J}^{ \frac{2}{J} }
\, \Gamma \big (  2-\frac{1}{J}  \big)   }
  { 96    \,   \beta  \, \mu_\chi^{ 1+\frac{2} {J}}  
\,\Gamma \big (  \frac{9}{2} - \frac{1}{J}  \big)   }  
\, \frac{ f^{bc} (J) } {J} 
  \,    B_{\chi x} \,  B_{\chi y} \, ,\nn
& \alpha^{m, \chi}_{xy} (\mu_\chi)
 =    \frac{ \sqrt{\pi } \, e^3  \,  \tau \, v_z \, \alpha_{J}^{ \frac{2} {J} }  
 \, \Gamma \big (  2-\frac{1}{J}  \big)}
{96  \,    \beta  \, \mu_\chi^{ 1+\frac{2} {J}} 
\, \Gamma \big (  \frac{9}{2} - \frac{1}{J}  \big)} 
\, \frac{ f^m (J) } {J} \,  B_{\chi x} \,  B_{\chi y} \,. 
\end{align}

\subsection{Mott relation}

From the explicit expressions of the in-plane longitudinal and transverse components of $\sigma^\chi $ and $\alpha^\chi $, which we have derived [cf. Eqs.~\eqref{eqsigxxfin}, \eqref{eqsigxyfin}, \eqref{eqalphaxxfin}, and \eqref{eqalphaxyfin}], we can immediately infer that the relation
\begin{align}
\partial_{\mu_\chi}  \sigma_{ij}^{\chi} (\mu_\chi)
 = - \frac  {3\, e\, \beta }  {\pi^2} \,  \alpha_{ij}^{\chi} (\mu_\chi)
	+ \order{\beta^{-2}}
\label{eqmott}
\end{align}
is satisfied. This is equivalent to satisfying the Mott relation $\alpha_{ij}^\chi  (\mu_\chi) = - \frac{ \pi^2  }{3 \,  e \,   \beta } \,\partial_{\mu_\chi} \sigma_{ij}^\chi  (\mu_\chi) $, which holds in the limit $\beta \rightarrow \infty $ \cite{mermin}, and can be derived mathematically through the Sommerfeld expansion. Hence, we find that the Mott relation continues to hold even in the presence of OMM, agreeing with the results in Ref.~\cite{prl_niu}, where generic settings have been considered.
	
Due to the Mott relation, the nature of $ A_{ij}^\Omega  $ and $ A_{ij}^m  $ can be readily inferred from that of $ \Sigma_{ij}^\Omega  $ and $ \Sigma_{ij}^m  $, which we have already discussed in the preceding section.
Nevertheless, we provide here some representative plots of the (1) longitudinal planar components $ A_{xx}^\Omega  $, $ A_{xx}^m $, $ A_{5xx}^\Omega $, and $ A_{5xx}^m  $ in Figs.~\ref{figaxx1} and \ref{figaxx2}; (2) transverse planar components
$ A_{xy}^\Omega  $, $ A_{xy}^m  $, $ A_{5xy}^\Omega  $, and $ A_{5xy}^m$ in Figs.~\ref{figaxy1} and \ref{figaxy2}.
We choose a different set of values for $\mu_+$ and $\mu_-$, compared to those chosen for Figs.~\ref{figsxx1}--\ref{figsxy2}, so as to cover a somewhat different parameter range for illustrative purposes. In fact, we have chosen here $\mu_+ - \mu_- < 0$, compared to the chosen value of $\mu_+ - \mu_- > 0$ for the earlier section, which leads to a sign-flip of the cyan, orange, and light-green curves representing the axial combinations of the response tensors.

\section{Magnetothermal coefficient}
\label{secell}

Using the function $F_{ij}$, whose Taylor-expanded form has been shown in Eq.~\eqref{eqexpf},
we now define the function
$ \tilde G_{ij}^\chi = F_{ij}^\chi \left ( \mathcal{E}_{\chi}  - \mu \right)^2 / \,T
= F_{ij}^\chi 
\left [  \varepsilon_\chi + \varepsilon_\chi ^{ (m) }  - \mu \right ]^2 /\,T
$. Its weak-field expansion is given by
\begin{align}
\tilde G_{ij}^\chi & =
F_{ij}^\chi  \,\frac{\left( \varepsilon_\chi  - \mu \right)^2 } {T}
+ 2
\left[ F_{ij}^{0, \chi}
+ F_{ij}^{1,\Omega, \chi} + F_{ij}^{1,m, \chi}  \right] 
\, \frac{ \left( \varepsilon_\chi  - \mu \right)    \varepsilon_\chi ^{ (m)} } {T}   
 +  F_{ij}^{0, \chi}   \,\frac{ \left( \varepsilon_\chi ^{ (m)} \right)^2 
 } {T} 
 + \order{ |{ \mathbf B}_{\chi}|^3 } \nn
& = \tilde G_{ij}^{0,\chi}
 + \tilde G_{ij} ^{1,\Omega,\chi} + G_{ij} ^{1, m,\chi} + \tilde G_{ij} ^{2,\Omega,\chi}
 + \tilde G_{ij} ^{2,m,\chi} + \tilde G_{ij} ^{2,(\Omega, m),\chi} + \order{ |{ \mathbf B}_{\chi}|^3} \,.
\end{align}
Here,
\begin{align}
& \tilde G_{ij}^{0,\chi} = \frac{  F_{ij}^{0, \chi} \left( \varepsilon_\chi  - \mu \right)^2 } {T}
 \,,\quad
\tilde  G_{ij} ^{1,\Omega ,\chi} = \frac{  F_{ij} ^{1,\Omega, \chi}  
 \left( \varepsilon_\chi  - \mu \right)^2 } {T} \,,\quad
 \tilde G_{ij} ^{1,m,\chi} =  \frac{2 \, F_{ij}^{0, \chi}
 \left( \varepsilon_\chi  - \mu \right) \varepsilon_\chi ^{ (m) } }{T} 
 + \frac{  F_{ij} ^{1,m, \chi}  \left( \varepsilon_\chi  - \mu \right)^2 } {T} \,,
 \nn & \tilde G_{ij} ^{2,\Omega ,\chi} = \frac{  F_{ij} ^{2,\Omega, \chi}  
 \left( \varepsilon_\chi  - \mu \right)^2 } {T} \,,
\quad \tilde G_{ij} ^{2,m,\chi} =  \frac{ 2\, F_{ij}^{1, m, \chi}
\left( \varepsilon_\chi  - \mu \right)  \varepsilon_\chi ^{ (m) } }{T} 
 + \frac{  F_{ij} ^{ 2,m, \chi}  \left( \varepsilon_\chi  - \mu \right)^2 } {T} 
 + \frac{  F_{ij}^{0, \chi}  \left( \varepsilon_\chi ^{ (m)} \right)^2  } {T}\,, 
\nn & 
\tilde G_{ij} ^{2,(\Omega, m),\chi} = 
\frac{ 2\, F_{ij}^{1, \Omega, \chi} \left( \varepsilon_\chi  - \mu \right)  \varepsilon_\chi ^{ (m) } }
{T} 
+
\frac{  F_{ij} ^{ 2,(\Omega, m), \chi}  \left( \varepsilon_\chi  - \mu \right)^2 } {T}\,.
\end{align}

We need to define one last function $ \tilde H_{ij}^\chi = \tilde  G_{ij} \, f'_{0} ( \mathcal{E}_{\chi}   )  $. Using Eq.~\eqref{Exp_f}, its expansion turns out to be
\begin{align}
\tilde  H_{ij}^\chi & =
\tilde  G_{ij}^\chi \, f'_{0} ( \varepsilon_\chi )
 + \left[ \tilde  G_{ij}^{0, \chi}
+ \tilde  G_{ij}^{1,\Omega, \chi} + \tilde  G_{ij}^{1,m, \chi}  \right]  
 \, \varepsilon_\chi ^{ (m) } \, f^{\prime \prime}_{0} ( \varepsilon_\chi )
+ \frac{ \tilde  G_{ij}^{0, \chi} \,\left( \varepsilon_\chi ^{ (m) } \right)^2 
\, f^{\prime \prime \prime }_{0} ( \varepsilon_\chi )} 
{2} 
 + \order{ |{ \mathbf B}_{\chi}|^3}  \nn
& = \tilde G_{ij}^{0,\chi}
 +\tilde  G_{ij}^{1,\Omega,\chi} + \tilde  G_{ij}^{1, m,\chi} + \tilde  G_{ij}^{2,\Omega,\chi}
 +  \tilde G_{ij}^{2,m,\chi} + \tilde  G_{ij}^{2,(\Omega, m),\chi} + \order{ |{ \mathbf B}_{\chi}|^3} \,,
\end{align}
where
\begin{align}
& \tilde  H_{ij}^{0,\chi} = \tilde  G_{ij}^{0, \chi} \, f'_{0} ( \varepsilon_\chi )\,, \quad
\tilde  H_{ij}^{1,\Omega,\chi} = \tilde  G_{ij}^{1,\Omega,\chi} \, f'_{0} ( \varepsilon_\chi )\,,\quad
\tilde  H_{ij}^{1,m,\chi} = 
\tilde  G_{ij}^{0, \chi}  \, \varepsilon_\chi ^{ (m) } \, f^{\prime \prime}_{0} ( \varepsilon_\chi ) 
+ \tilde  G_{ij}^{1,m,\chi} \, f'_{0} ( \varepsilon_\chi )\,,\nn
& \tilde H_{ij}^{2 ,\Omega ,\chi} = \tilde  G_{ij}^{2 ,\Omega ,\chi} \, f'_{0} ( \varepsilon_\chi )\,,
\quad \tilde H_{ij}^{2 ,m ,\chi} =
\frac{ \tilde  G_{ij}^{0, \chi} \left( \varepsilon_\chi ^{ (m) } \right)^2 
f^{\prime \prime \prime}_{0} ( \varepsilon_\chi ) 
} {2} +
\tilde G_{ij}^{ 1, m, \chi}  \, \varepsilon_\chi ^{ (m) } \, f^{\prime \prime}_{0} ( \varepsilon_\chi ) 
+ \tilde G_{ij}^{2 , m ,\chi} \, f'_{0} ( \varepsilon_\chi )\,, \nn &
\tilde H_{ij}^{2,(\Omega, m),\chi} = 
 \tilde G_{ij}^{1,\Omega,\chi} \, \varepsilon_\chi ^{ (m) } \, f^{\prime \prime}_{0} ( \varepsilon_\chi ) 
+ \tilde  G_{ij}^{2,(\Omega, m),\chi} \, f'_{0} ( \varepsilon_\chi )\,.
\end{align}

Using the expressions shown above, the integrand in Eq.~\eqref{eq_thermal} is expanded in the weak-in-magnetic-field limit, leading to
\begin{align}
& \ell^{\chi}_{ij}  = \ell^{0, \chi}_{ij} + 
+ \ell^{\Omega , \chi}_{ij}  +  \ell^{m, \chi}_{ij} , \text{ where }
\ell^{\Omega , \chi}_{ij}  = \ell^{1,\Omega , \chi}_{ij} + \ell^{2,\Omega , \chi}_{ij}
\,, \quad
\ell^{m, \chi}_{ij}  =  \ell^{1, m , \chi}_{ij} + \ell^{2, m , \chi}_{ij} 
+ \ell^{2, (\Omega, m), \chi}_{ij} \,.
\label{eqelltot}
\end{align}
Analogous to $ \alpha^\chi_{ij}$, the terms $ \ell^{0, \chi}_{ij}$, $ \ell^{1,\Omega , \chi}_{ij} $, $ \ell^{2,\Omega , \chi}_{ij} $, $ \ell^{1,m , \chi}_{ij} $, $ \ell^{2,m , \chi}_{ij} $, and $ \ell^{2, (\Omega, m), \chi}_{ij} $ have been labelled such that they are defined by the integrands $ \tilde  H^{0, \chi}_{ij}$, $ \tilde  H^{1,\Omega , \chi}_{ij} $, $ \tilde  H^{2,\Omega , \chi}_{ij} $, $ \tilde  H^{1,m , \chi}_{ij} $, $ \tilde  H^{2,m , \chi}_{ij} $, and $ \tilde  H^{2, (\Omega, m), \chi}_{ij} $, respectively. Analogous to the cases of $\sigma^\chi_{ij}$ and $ \alpha^\chi_{ij}$, the term $  \ell_{ij}^{m, \chi}  $ goes to zero if the OMM is set to zero.

The expressions are cumbersome and, hence, it is useful to break them up into smaller bits. In particular, we define
\begin{align}
\label{l2}
\ell_{ij}^{2, m, \chi} = I^{ \ell, 2}_{1, i j} + I^{ \ell, 2}_{ 2, ij} + I^{ \ell, 2}_{ 3, ij} \,,
\end{align}
where
\begin{align}
\label{l3}
I^{ \ell, 2}_{ 1, ij} & = -\,
\tau \int\frac{d^3 \mathbf{ k }}{\left( 2 \, \pi \right)^3} \, \Bigg[ v_{\chi i}^{\left( 0 \right)} \, v_{\chi j}^{\left( 0 \right)} \, \frac{ \left( \varepsilon_{\chi}^{\left( m \right)} \right)^2}{T} + 2 \left( v_{\chi i}^{(0)} \, v_{\chi j}^{(m)} + v_{\chi i}^{(m)} \, v_{\chi j}^{(0)} \right) \, \frac{\left( \varepsilon_{\chi} - \mu \right) \, \varepsilon_{\chi}^{(m)}}{T} + v_{\chi i}^{( m )} \, v_{\chi j}^{( m )} \, \frac{ \left( \varepsilon_{\chi} - \mu \right)^2 }{T} \Bigg] \, f_0^{ \prime } (\varepsilon_{\chi} ) \,,\nn
I^{ \ell, 2}_{ 2, ij} & = -\,\tau 
\int\frac{d^3 \mathbf{ k }}{\left( 2 \, \pi \right)^3} \, \Bigg[ \left( v_{\chi i}^{( 0 )} \, v_{\chi j}^{( m )} + v_{\chi i}^{( m )} \, v_{\chi j}^{ (0) } \right) \, \frac{ \left( \varepsilon_{\chi} - \mu \right)^2 \varepsilon_{\chi}^{ (m) } }{T} + 2 \, v_{\chi i}^{ (0) } \, v_{\chi j}^{ (0) } \, \frac{ \left( \varepsilon_{\chi} - \mu \right) \left( \varepsilon_{\chi}^{ (m) } \right)^2}{T} \Bigg] \, f_0^{ \prime \prime} (\varepsilon_{\chi} )\,,
\nn
I^{ \ell, 2}_{ 3, ij} &= -\,
\tau \int\frac{d^3 \mathbf{ k }}{\left( 2 \, \pi \right)^3} \, 
\frac{v_{\chi i}^{ (0) } \, v_{\chi j}^{ (0) } } {2} \,  
\frac{ \left( \varepsilon_{\chi} - \mu \right)^2 \left( \varepsilon_{\chi}^{ (m) } \right)^2 }{T} 
\, f_0^{ \prime \prime \prime } (\varepsilon_{\chi} ) \,. 
\end{align}

In a similar spirit, we define
\begin{align}
\label{l4}
\ell_{ij}^{2, ( m , \Omega ), \chi} = I^{ \ell, 3}_{ 1, ij}  
+  I^{ \ell, 3}_{ 2, ij} +  I^{ \ell, 3}_{ 3, ij} \,,
\end{align}
with
\begin{align}
\label{l5}
I^{ \ell, 3}_{ 1, ij}  &= - \,2 \, e \, \tau
 \int \frac{ d^3 \mathbf { k }}{( 2 \,\pi )^3} \,  
\left [
  \left ( \boldsymbol{ v }_{\chi}^{ (0) } \cdot  \mathbf{\Omega }_{\chi} \right) 
  \left( v_{\chi i}^{( 0 )} \, B_{\chi j} + B_{\chi i} \, v_{\chi j}^{ ( 0 ) }  \right)
  - \left( \mathbf{ B }_{\chi} \cdot \boldsymbol{ \Omega}_{\chi} \right) 
  \, v_{\chi i}^{( 0 )} \, v_{\chi j}^{ (0) } \right ]
  \frac{ \left( \varepsilon_{\chi} - \mu \right) \, \varepsilon_{\chi}^{( m )}}{T} 
 \, f_0^{ \prime } (\varepsilon_{\chi} ) \,,\nn
I^{ \ell, 3}_{ 2, ij}  &= - \,e \, \tau
 \int \frac{ d^3 \mathbf { k }}{( 2 \,\pi )^3} \, 
\Big [ \left ( \boldsymbol{ v }_{\chi}^{ (0) } \cdot  \mathbf{\Omega }_{\chi} \right ) 
 \left( v_{\chi i}^{( m )} \, B_{\chi j} + B_{\chi i} \, v_{\chi j}^{ ( m ) }  \right) 
 + \left ( \boldsymbol{ v }_{\chi}^{ ( m ) } \cdot  \mathbf{\Omega }_{\chi} \right ) 
 \left( v_{\chi i}^{( 0 )} \, B_{\chi j} + B_{\chi i} \, v_{\chi j}^{ (0) }  \right) 
\nn  & 
\hspace{ 3 cm}
-  \left( \mathbf{ B }_{\chi} \cdot \boldsymbol{ \Omega}_{\chi} \right) 
 \left( v_{ \chi i }^{ (0) } \, v_{ \chi j }^{ (m) }
  + v_{ \chi i }^{ (m) } \, v_{ \chi j }^{ (0) } \right) \Big ]
\, \frac{ \left( \varepsilon_{\chi} - \mu \right)^2 }{T}  \, f_0^{ \prime } (\varepsilon_{\chi} ) \,,
\nn
I^{ \ell, 3}_{ 3, ij}  &= - \, e \, \tau 
\int \frac{ d^3 \mathbf { k }}{( 2 \,\pi )^3} \, \bigg[ \left( \boldsymbol{ v }_{\chi}^{ (0) } \cdot  \mathbf{\Omega }_{\chi} \right) \, \left( v_{\chi i}^{ (0) } \, B_{\chi j} + B_{\chi i} \, v_{\chi j}^{ (0) } \right) 
- \left(  \mathbf{ B  }_{ \chi } \cdot  \mathbf{\Omega }_{ \chi } \right) \, v_{ \chi i }^{ (0) } \, v_{ \chi j }^{ (0) } \bigg] \,
 \frac{ \left( \varepsilon_{\chi} - \mu \right)^2 
 \varepsilon_{ \chi }^{ (m) } }{ T } \, f_0^{ \prime \prime} (\varepsilon_{\chi} ) \,.
\end{align}

The longitudinal and transverse components of the tensor $\ell^\chi $ are computed from the expressions shown above, in the same way as we have done for $\sigma^\chi$ and $\alpha^\chi$.  The details of the intermediate steps have been relegated to Appendices~\ref{appellxx} and \ref{appellyx}.

\subsection{Longitudinal magnetothermal coefficient}
\label{secellxx}

Using the explicit expressions derived in Appendix~\ref{appellxx}, we have
\begin{align}
\ell^{\chi}_{xx}  (\mu_\chi)  
= \ell^{0, \chi}_{xx}  (\mu_\chi) + \ell^{\Omega , \chi}_{xx}  (\mu_\chi) 
+  \ell^{m, \chi}_{xx}  (\mu_\chi)\,,
\end{align}
where
\begin{align}
\label{eqellxxfin}
\ell^{0, \chi}_{xx} (\mu_\chi) & = \frac{J \, \tau\,\mu_\chi^2\, T} { 18 \, v_z }  \,, \quad
\ell^{\Omega, \chi}_{xx} (\mu_\chi)  =  
\frac{	\sqrt \pi \, e^2\, \tau \, v_z \, \alpha_{J}^{ \frac{2} {J} } \, T}
{3 \times 128 \,    \mu_\chi^{\frac{2} {J}} } 
\,\frac{ \Gamma \big (2 - \frac{1}{J} \big) }{ \Gamma \big( \frac{9}{2} - \frac{1}{J} \big) }
 \left [  g_x^{bc} (J) \, B_{\chi x}^2 +   g_y^{bc} (J)\,  B_{\chi y}^2  \right ] , \nn
\ell^{ m, \chi}_{xx}  (\mu_\chi) &= 
\frac{  \sqrt \pi \,
e^2 \,  \tau \, v_z \,   \alpha_{J}^{ \frac{2} {J} }
\,T  }
{ 3 \times 128 \,      \mu_\chi^{ \frac{2} {J} }
 } 
\,\frac{ \Gamma \big (2 - \frac{1}{J} \big) }{ \Gamma \big( \frac{9}{2} - \frac{1}{J} \big) } 
  \left [  g_x^m (J) \, B_{\chi x}^2 +   g_y^{m} (J)\,  B_{\chi y}^2  \right ]  .
\end{align}

\subsection{Transverse magnetothermal coefficient}
\label{secellxy}

Using the explicit expressions derived in Appendix~\ref{appellyx}, we have
\begin{align}
\label{eqellxyfin}
\ell^{\chi}_{yx}  (\mu_\chi)  = \ell^{\chi}_{xy}  (\mu_\chi)
= \ell^{0, \chi}_{xy}  (\mu_\chi) + \ell^{\Omega , \chi}_{xy}  (\mu_\chi) 
+  \ell^{m, \chi}_{xy}  (\mu_\chi)\,,
\end{align}
where
\begin{align}
\ell^{0, \chi}_{xy} (\mu_\chi) &=   0 \,, \quad
\ell^{\Omega , \chi}_{xy} (\mu_\chi)
  = \frac{	\sqrt \pi \, e^2\, \tau \, v_z \, \alpha_{J}^{ \frac{2} {J} } \, T}
{3 \times 64 \,    \mu_\chi^{\frac{2} {J}} } 
\,\frac{ \Gamma \big (2 - \frac{1}{J} \big) }
{ \Gamma \big( \frac{9}{2} - \frac{1}{J} \big) }
\, f^{bc} (J) \,B_{\chi x} \,  B_{\chi y}   \,,\nn
\ell^{m, \chi}_{xy} (\mu_\chi)
& =     \frac{	\sqrt \pi \, e^2\, \tau \, v_z \, \alpha_{J}^{ \frac{2} {J} } \, T}
{3 \times 64 \,    \mu_\chi^{\frac{2} {J}} } 
\,\frac{ \Gamma \big (2 - \frac{1}{J} \big) }{ \Gamma \big( \frac{9}{2} - \frac{1}{J} \big) }
\, f^{m} (J) \,B_{\chi x} \,  B_{\chi y}   \,.
\end{align}

\subsection{Wiedemann-Franz law}

From the explicit expressions of the in-plane longitudinal and transverse components of $\sigma^\chi $ and $\ell^\chi $, which we have derived [cf. Eqs.~\eqref{eqsigxxfin}, \eqref{eqsigxyfin}, \eqref{eqellxxfin}, and \eqref{eqellxyfin}], we immediately find that the relation
\begin{align}
\sigma^{\chi} _{ij} =
 \frac{ 3 \,e^2 } {\pi^2\, T}   \, \ell^{ \chi} _{ij}   + \order{\beta ^{-2}}
\label{eqwf}
\end{align}
is satisfied. This is equivalent to satisfying the Wiedemann-Franz law, which holds in the limit $\beta \rightarrow \infty $ \cite{mermin}. Hence, we have demonstrated that the Wiedemann-Franz law continues to hold even in the presence of OMM. 
This relation tells us that, knowing the nature of the magnetoelectric conductivity, we can infer the behaviour of the magnetothermal coefficient. Therefore, it is not necessary to provide any separate plots for this response.

\section{Summary and future perspectives}
\label{secsum}

In this paper, we have considered planar Hall (or planar thermal Hall) configurations such that a 3d Weyl or multi-Weyl semimetal is subjected to a conjunction of an electric field $\mathbf E $ (and/or temperature gradient $\nabla_{\mathbf r } T$) and an effective magnetic field $\mathbf B_\chi $, oriented at a generic angle with respect to each other. The $z$-axis is chosen to be along the direction along which the mWSM shows a linear-in-momentum dispersion, and is perpendicular to the plane of $\mathbf E $ (or $\nabla_{\mathbf r } T$) and $\mathbf B_\chi $. The effective magnetic field consists of two parts --- (a) an actual/physical magnetic field $\mathbf B $, and (b) an emergent magnetic field $\mathbf B_5 $ which arises if the sample is subjected to elastic deformations (strain tensor field). Since $\mathbf B_5 $ exhibits a chiral nature, because it couples to conjugate nodal points of opposite chiralities with opposite signs, $\mathbf B_\chi $ is given by $\mathbf B + \chi \, \mathbf B_5 $. The relative orientations of these two constituents of ${\mathbf B}_\chi$, with respect to the direction of the electric field (or temperature gradient), give rise to a rich variety of possibilities in the characteristics of the electric, thermal, and thermoelectric response tensors.  We have derived explicit expressions for these response coefficients, which have helped us to identify unambiguously the interplay of the BC- and OMM-contributions. In addition, we have illustrated the overall behaviour of the response in some realistic parameter regimes. We have found that the total (i.e., sum) and the axial (i.e., difference) combinations of the response, from the two conjugate nodes, depend strongly on the specific value of $J$. In particular, for the planar transverse components of the response tensors, while the OMM part acts exclusively in opposition with the BC-only part for the Weyl semimetals, the former syncs with the latter for $J>1$, thereby enhancing the overall response. The strain-induced $\mathbf B_5 $ provides a way to have linear-in-$ B $ terms in the response coefficients for untilted nodes, in addition to the quadratic-in-$ B$ dependence.

The $B^2$ dependence and the $\pi$-periodic behaviour (with respect to $\theta$) of the magnetoelectric conductivity, in the absence of $B_5$, have been observed in numerous experiments, which involve materials like ZrTe$_{5}$ \cite{li_2016}, TaAs \cite{cheng-long}, NbP and NbAs \cite{li_nmr17}, and Co$_3$Sn$_2$S$_2$ \cite{shama}, known to host Weyl nodes. Furthermore, the magnetothermal coefficient has also been measured in materials such as NdAlSi \cite{marcin}, which again shows the expected $B^2$-dependence. It is possible to modify these experimental set-ups to devise the mechanism for applying strain gradients to the samples, thus leading to the realization of $B_5$~\cite{diaz_2022} in addition.

For the case of a negative chemical potential, we need to focus on the valence band at the corresponding node. The energy, band velocity, and Berry curvature will have opposite signs compared to the case of the positive chemical potential. For the magnetic-field-independent ``Drude'' parts, we have found that, independent of the $J$-values, $\sigma^{0, \chi}_{xx} (\mu_\chi)$  and $ \ell^{0, \chi}_{xx} (\mu_\chi)$ are proportional to $\mu_\chi^2$, while $ \alpha^{0, \chi}_{xx} (\mu_\chi) \propto \mu_\chi $. For the magnetic-field-dependent parts, the leading-order $\mu_\chi$-dependence goes as (i) $\mu_\chi ^{ -\frac{2}{J} }$ for $\sigma^{\chi}_{ij}$,
(ii) $\mu_\chi^{ -1- \frac{2}{J} }$ for $\alpha^{\chi}_{ij}$, and (iii) $\mu^{  \frac{2}{J} }$ for $\ell^{\chi}_{ij}$. Hence, for $J=3$, the non-Drude parts depend on a fractional power of $\mu_\chi$, which physically makes no sense for $\mu_\chi < 0$. Now, we have to remember that, in our formalism, we have not included $\mu_\chi$ in the starting Hamiltonian $\mathcal{H}_\chi$, but have included it in the Fermi distribution function. This has allowed us to derive analytical expressions by applying the Sommerfeld expansion, and the forms of the final expressions (summarized above) are the artifacts of our specific procedure. Therefore, we conclude that it is possible to derive the results for $ \mu_\chi < 0 $ by implementing the same procedure, but considering the transport contributed by a valence band (with the sign changes quoted above), for $J=1, 2$. However, it will not work for $J=3$ due to the presence of fractional powers of $\mu_\chi$. The results for $J=1,2$ are thus expected to have a dependence on the various parameters similar to the $\mu_\chi >0$ case that we have considered here, modulo possible minus signs for the various contributing parts.
As for the $J=3$ case, we can include the negative chemical potential at the Hamiltonian level, such that it directly enters into the energy-eigenvalue expressions, and then numerically evaluate the behaviour of the various response tensors.

In Refs.~\cite{das19_linear, das19_linear2}, the authors have included a momentum-independent internode scattering time $\tau_v$, in addition to the intranode scattering time $\tau$. The inclusion of the internode processes results in stabilizing the chiral anomaly, as it inherently leads to different chemical potentials at the two conjugate WSM/mWSM nodes. By plugging in $\tau_v$ as a phenomenological constant (because of ignoring its momentum dependence), Ref.~\cite{das19_linear} has considered the resulting semiclassical Boltzmann equations for the two nodes in an untilted WSM. The authors have shown that a finite internode-scattering-induced amplitude is tied to a difference in the chemical potential between the two nodes, given by $\Delta \mu \propto \left( \mathbf{E} \cdot \mathbf B \right) \tau_v$ (in the absence of a pseudomagnetic field). They have inferred that this results in the magnetoelectric conductivity tensor components acquiring extra contributions $ \propto \tau_v$ and normal physical conditions dictate that $\tau_v\gg \tau $. The same behaviour is found for the case of the thermoelectric conductivity tensor. A more complete treatment can be found in \cite{timm_omm}, where the authors do not assume a momentum-independent $\tau_v$ and go beyond the relaxation-time approximation.
It remains to be seen how the above calculations pan out for the mWSMs and multifold semimetals \cite{ips-rsw}, and what are the forms of the resulting final expressions.

In the future, it will be worthwhile to perform the same calculations by including tilted nodes \cite{das19_linear, das19_linear2, ips-rahul-tilt, timm_omm}, because tilting is applicable for generic scenarios. A tilt can induce terms which are linearly-dependent on $B$, even in the absence of a strain-induced $\mathbf B_5$-part. Furthermore, a more realistic calculational set-up should include internode scatterings \cite{das19_linear2} and going beyond the relaxation-time approximation \cite{timm_omm}. 
Although we have considered the weak-magnetic-field scenario in this paper, under the influence of a strong quantizing magnetic field, we have to incorporate the effects of the discrete Landau levels while computing the linear response \cite{ips-kush, fu22_thermoelectric, staalhammar20_magneto, yadav23_magneto}.
Other auxiliary directions include the study of linear and nonlinear response in the presence of disorder and/or strong correlations \cite{ips-seb, ips_cpge, ips-biref, ips-klaus, rahul-sid, ips-rahul-qbt, ips-qbt-sc, ips-hermann-review}. One could also explore the effects of a time-periodic drive \cite{ips-sandip, ips-sandip-sajid, ips-serena}, for instance, by shining circularly polarized light.

\section*{Acknowledgments}
LM and AMR acknowledge support from CONACyT (México) under project
number CF- 428214, and DGAPA-UNAM under project number AG100224.
IM's research has received funding from the European Union's Horizon 2020 research and innovation
programme under the Marie Skłodowska-Curie grant agreement number 754340.


\appendix

\section{Some useful integrals and identitites} 
\label{app_int}

In order to obtain the final expressions of the response tensors, we have to deal with integrals of the form:
\begin{align}
I = \int \frac{  d^3 {\bf{k}} }{(2 \, \pi )^3 } \, {\mathcal F }({\bf{k}} , \varepsilon_\chi ) \, 
 f^\prime_{0} (\varepsilon_\chi)  \,,
\end{align}
where $\varepsilon_\chi = \epsilon_{\mathbf k}$, according to the conventions set in Sec.~\ref{secmodel}.
Taking advantage of the azimuthal symmetry about the $z$-axis, we take advantage of the cylindrical coordinates defined by
\begin{align}
k_{x} = k_{\perp} \cos \phi \,, \quad
k_{y} = k_{\perp} \sin \phi\,, \text{ and } k_{z} = k_{z}\,, 
\end{align}
where $k_{\perp} \in [0, \infty )$ and $\phi \in [0, 2 \pi )$.
Hence, we rewrite the integral as
\begin{align}
I = \frac{1}{(2\, \pi )^3 }  \int_{- \infty} ^{\infty} dk_{z}  
\int_{0} ^{\infty} dk_{\perp} \, k_{\perp} 
\, {\mathcal G } (k_{\perp} , k_{z} , \varepsilon_\chi )\,
f_{0}^\prime (\varepsilon_\chi) \,,
\text{ where }
{\mathcal G } (k_{\perp} , k_{z} , \varepsilon_\chi ) = 
\int_{0} ^{2 \pi }  d \phi \, 
{\mathcal F } (k_{\perp} , \phi , k_{z} , \varepsilon_\chi ) \, .
\end{align}

In the next step, we change variables from $(k_{\perp} , k_{z})$ to $(\varepsilon_\chi , \varphi )$ by the coordinate transformation
\begin{align}
k_{\perp} = \left( \frac{\varepsilon_\chi}{\alpha_{J} } \sin \varphi \right)  ^{1/J} , \qquad  k_{z} = \frac{\varepsilon_\chi}{v_z } \cos \varphi ,  
\end{align}
where $\varepsilon_\chi \in [0, \infty )$ and $\varphi \in [0, \pi )$. The Jacobian of the transformation is $\mathcal{J} (\varepsilon_\chi , \varphi ) =   \frac{1}{J v_z \sin \varphi } \left( \frac{\varepsilon_\chi \sin \varphi}{\alpha_{J} } \right) ^{1/J} $, leading to
\begin{align}
I = \frac{1}{(2 \,\pi )^3 }  \int_{0} ^{\infty} d \varepsilon_\chi \,   
{\mathcal K } (\varepsilon_\chi  )   \, f^\prime_{0} (\varepsilon_\chi) \,,
\end{align}
where
\begin{align}
{\mathcal K } (\varepsilon_\chi  ) &=  \int_{0} ^{\pi} d \varphi \, 
 \mathcal{J} (\varepsilon_\chi , \varphi ) \,  {\mathcal H } (\varepsilon_\chi , \varphi ) \,,\nn
{\mathcal H } (\varepsilon_\chi , \varphi ) & = k_{\perp} \, 
{\mathcal G } (k_{\perp} , k_{z} , \varepsilon_\chi ) 
= \left( \frac{\varepsilon_\chi}{\alpha_{J} } \sin \varphi \right)^{\frac{1}{J}} \,
{\mathcal G } \Bigg( \left( \frac{\varepsilon_\chi}{\alpha_{J} } 
\sin \varphi \right)^{\frac{1}{J}}, \,  \frac{\varepsilon_\chi}{v_z } 
\cos \varphi , \varepsilon_\chi \Bigg )\, ,
\end{align}
whose final evaluation can be implemented via the Sommerfeld expansion \cite{mermin} under the condition $\beta \mu \gg 1$. This is because the integral will turn out to consist of terms of the form
\begin{align}
\Lambda_{n} (\mu ) \equiv 
- \int_{0} ^{ \infty} d \varepsilon_\chi    \,    \varepsilon_\chi ^{n}  \,   f_{0}^\prime (\varepsilon_\chi )  \text{ for } n \geq 0\,,
\end{align}
which, upon using the Sommerfeld expansion, yields
\begin{align}
\Lambda_{n} (\mu )=  \mu ^{n} \,  \left[ 1 + \frac{\pi^2 \,  n \,  (n-1)}
{6 \left(  \beta \,  \mu \right)^{2} } 
+ \order{\left( {\beta \,\mu}\right)^{-3}} \right] .
\end{align}

We now list some identities used in the paper, along with their derivations:
\begin{enumerate}

\item 
\begin{align}
\frac{\partial ^{l} }{\partial \mu ^{l} } \Lambda_{n} (\mu ) 
 = \frac{n!}{(n-l)!}  \, \Lambda_{n-l} (\mu ) \,.  
 \label{Identity1}
\end{align}
\\Proof:
We start with the relation
\begin{align}
\frac{\partial}{\partial \mu} \Lambda_{n} (\mu ) &=   n \, \mu ^{n-1} \,  \left[ 1 + \frac{\pi^2 \,  n \,  (n-1)}{6 \,  \beta^2 \,  \mu^2 }   \right] + \mu ^{n} \,  \left[ -  \frac{2 \pi^2 \,  n \,  (n-1)}{6 \,  \beta^2 \,  \mu ^{3} }   \right]  \notag \\ &=   n \, \mu ^{n-1} \,  \left[ 1 + \frac{\pi^2  \,  (n-1) \,  (n - 2 ) }{6 \,  \beta^2 \,  \mu^2 }   \right] \notag \\ &=   n \,   \Lambda_{n-1} (\mu ) . 
\end{align}
Using this result recursively, we arrive at
\begin{align}
\frac{\partial^2 }{\partial \mu^2 } \Lambda_{n} (\mu ) 
&=  n \,  \frac{\partial}{\partial \mu} \Lambda_{n-1} (\mu ) = n \, (n-1) \, \Lambda_{n-2} (\mu ) ,  \notag \\ 
\frac{\partial ^{3} }{\partial \mu ^{3} } \Lambda_{n} (\mu ) 
&=  n \, (n-1) \,  \frac{\partial}{\partial \mu} \Lambda_{n-2} (\mu ) 
= n \, (n-1) \,  (n-2 ) \,  \Lambda_{n-3} (\mu ) \,,
\notag \\ & \phantom{=} \vdots \notag \\ 
\frac{\partial ^{l} }{\partial \mu ^{l} } \Lambda_{n} (\mu ) 
&=  n \, (n-1) \,  \cdots \, (n-l+2) \, \frac{\partial}{\partial \mu} \Lambda_{n-l+1} (\mu )
 = n \, (n-1) \,  \cdots \, (n-l+1) \, \Lambda_{n-l} (\mu )\,.
\end{align}

\item
\begin{align}
\int_{0} ^{\infty} d \varepsilon_\chi \,  \varepsilon_\chi  ^{n}  \,  ( \varepsilon_\chi - \mu ) \,  \frac{\partial  f_{0} ( \varepsilon_\chi ) }{ \partial \varepsilon_\chi  }  = -  \frac{  n \,  \pi^2  }{3 \,  \beta^2 } \, \mu ^{n-1} \,. \label{Identity2}
\end{align}
\\Proof: 
\begin{align}
\int_{0} ^{\infty} d \varepsilon_\chi \,  \varepsilon_\chi  ^{n}  \,  ( \varepsilon_\chi - \mu ) \frac{\partial  f_{0} ( \varepsilon_\chi ) }{ \partial \varepsilon_\chi  }  &= \int_{0} ^{\infty} d \varepsilon_\chi \,    ( \varepsilon_\chi  ^{n+1} - \mu \varepsilon_\chi  ^{n} ) \frac{\partial  f_{0} ( \varepsilon_\chi ) }{ \partial \varepsilon_\chi  } \notag \\  &= - \left[  \Lambda_{n+1} (\mu) - \mu \, \Lambda_{n} (\mu) \right] \notag \\ &= - \mu ^{n+1} \,  \left[ 1 + \frac{\pi^2 \,  n \,  (n+1)}{6 \,  \beta^2 \,  \mu^2 }   \right] + \mu ^{n+1} \,  \left[ 1 + \frac{\pi^2 \,  n \,  (n-1)}{6 \,  \beta^2 \,  \mu^2 }   \right].
\end{align}

\item
\begin{align}
\int_{0} ^{ \infty} d \varepsilon_\chi    \,    \varepsilon_\chi ^{n}  \,(-1) ^{l+1}  \,
\frac{\partial ^{l+1} 
\, f_{0} ( \varepsilon_\chi ) } { \partial \varepsilon_\chi ^{l+1} }  
= \frac{n!}{(n-l)!}  \, \Lambda_{n-l} (\mu )\,. 
\label{Identity3}
\end{align}
\\Proof:
Using Eq.~\eqref{Identity1} by setting $l=1$, we get
\begin{align}
 \int_{0} ^{ \infty} d \varepsilon_\chi    \,    \varepsilon_\chi ^{n}  \, 
  \frac{\partial^2 f_{0} ( \varepsilon_\chi ) }{ \partial \varepsilon_\chi^2 }  
=
\frac{\partial}{\partial \mu} \int_{0} ^{ \infty} d \varepsilon_\chi    \,    \varepsilon_\chi ^{n}  \,  (  -1)
\,   f_{0}^\prime (\varepsilon_\chi )    
 \equiv \frac{\partial  }{\partial \mu } \Lambda_{n} (\mu )  
   =   n \,   \Lambda_{n-1} (\mu ) .
\end{align}
Taking a derivative with respect to $\mu$ once more, we get
\begin{align}
\int_{0} ^{ \infty} d \varepsilon_\chi    \,    \varepsilon_\chi ^{n}  \,  ( -1)
\,  \frac{\partial ^{3} f_{0} ( \varepsilon_\chi ) }{ \partial \varepsilon_\chi ^{3} }  
=   n \,  \frac{\partial}{\partial \mu }  \Lambda_{n-1} (\mu ) = n \, (n-1 ) \, \Lambda_{n-2} .
\end{align}
Performing this procedure recursively yields Eq.~\eqref{Identity3}.

\item 
\begin{align}
\int_{0} ^{ \infty} d \varepsilon_\chi    \,   
\varepsilon_\chi ^{n}  \,  (\varepsilon_\chi - \mu ) 
\,  (-1) ^{l+1}  
\, \frac{\partial ^{l+1} f_{0} ( \varepsilon_\chi ) }
{ \partial \varepsilon_\chi ^{l+1} } 
 = \frac{(n+1)!}{(n+1-l)!}  \, \Lambda_{n+1-l} (\mu )- \mu \, \frac{n!}{(n-l)!}  \, \Lambda_{n-l} (\mu ) \,. 
\label{Identity4}
\end{align}
\\ Proof:
The identity follows directly from Eq.~\eqref{Identity3}.

\end{enumerate}

\section{Computation of the longitudinal components of the magnetoelectric conductivity tensor}
\label{app_lmc}

In this section, we show the derivation of the explicit expression of the longitudinal component of the magnetoelectric conductivity tensor $\sigma^\chi $ (i.e., the LMC), shown in Sec.~\ref{secsxx}, starting from the integrals appearing in Eqs.~\eqref{eq_elec} and \eqref{totalsigma}.

\subsection{Field-independent contribution}

The $\mathbf B_\chi$-independent contribution, shown in Eq.~(\ref{eq_elec_0B}), is given by
\begin{align}
\sigma ^{0, \chi}_{xx} &= - e^2 \,  \tau  \int \frac{  d^3 {\mathbf k} }{(2 \,\pi )^{3} }  
\left( v_{\chi x } ^{(0)} \right)^2 \, f_{0}^\prime (\varepsilon_\chi ) 
= - e^2 \,  \tau  \,  J^2  \, 
 \alpha ^{4}_{J}  \int \frac{  d^3 {\mathbf k} }{(2 \,\pi )^3 }  
   \left( \frac{ k_{x} \,  k_{\perp} ^{2J-2}  }
{   \varepsilon_\chi  }   \right)^2 \,  f_{0}^\prime (\varepsilon_\chi )\, . 
\end{align}
To evaluate this integral, we use the coordinate transformations described in Appendix \ref{app_int}. Switching to the cylindrical coordinates $(k_{\perp} , \,k_{z}, \,\phi)$ and integrating over the azimuthal angle $\phi$, we get
\begin{align}
\sigma ^{0, \chi}_{xx} &= - \frac{ e^2 \,  \tau  \,  J^2 \,  \alpha ^{4}_{J} } { (2\, \pi )^3  } 
\int_{0} ^{2 \pi}  d \phi  \cos^2 \phi 
\int_{- \infty} ^{\infty} dk_{z}  \int_{0} ^{\infty}  dk_{\perp} \,    \frac{   k_{\perp} ^{4\, J-1}  }{ \varepsilon_\chi^2 } \,  f_{0}^\prime (\varepsilon_\chi ) \notag \\ &
= - \frac{ e^2 \,  \tau  \,  J^2 \,  \alpha ^{4}_{J} }{8 \,  \pi^2  }  
 \int_{- \infty} ^{\infty} dk_{z}  \int_{0} ^{\infty}  
 dk_{\perp} \,    \frac{   k_{\perp} ^{4\, J-1}  }{ \varepsilon_\chi^2 } \,  f_{0}^\prime (\varepsilon_\chi ) . \label{Sigma_XX_Berry}
\end{align}
We now change variables from $(k_{\perp} , k_{z})$ to $(\varepsilon_\chi , \varphi )$, and arrive at the expression
\begin{align}
\label{eqsigmaxx0}
\sigma ^{0, \chi}_{xx} &=  - \frac{ e^2\,  \tau \,  J   }
{ 8 \,  \pi^2  \,  v_z    }     \int_{0} ^{\pi} \, 
 d\varphi  \, \sin^3 \varphi    
 \int_{0} ^{ \infty} d \varepsilon_\chi    \,    \varepsilon_\chi^2   
   \,  f_{0}^\prime (\varepsilon_\chi )  
= \frac{ e^2 \,  \tau  \,  J   }
{ 6 \,  \pi^2   \,  v_z    }  \,  \Lambda_{2} (\mu ) \,.
\end{align}
This result reproduces the field-independent part of Eq.~(17) of Ref. \cite{ips-rahul-ph} and, on setting $J=1$, also agrees with Eq. (17) of Ref.~\cite{onofre}.

\subsection{Contribution solely from BC (no OMM)}

The contribution to $\sigma_{xx}^\chi $ solely from the BC, captured by Eq.~\eqref{eq_elec_Berry}, is given by
\begin{align}
\label{eqomegabc}
\sigma ^{\Omega , \chi}_{xx} &= - 
e^4\,  \tau 
\int \frac{  d^3 {\mathbf k} } { (2\, \pi )^3  } \,  Q_{\chi x }^2 \,  f_{0}^\prime (\varepsilon_\chi ) = 
- {e^4 \,  \tau}  \int \frac{  d^3 {\mathbf k} } { (2\, \pi )^3  } 
  \left[ \mathbf{\Omega }_{\chi}  \times ( {\boldsymbol  v}_{\chi} ^{(0)}   \times \mathbf{B}_{\chi} ) \right]_{x}^2 \,  f_{0}^\prime (\varepsilon_\chi ) \nn
 &= - \frac{ \alpha_J^4 \,  
 J ^4 \,  v_z^2  \, e^4 \,  \tau
 }
 {128 \,  \pi^2 } 
 \int_{- \infty} ^{\infty}  dk_{z}   \int_{0}^{\infty}  dk_{\perp} \,  \frac{k_{\perp} ^{4 J -3}}
 {  \varepsilon_\chi ^{8} } \left[ \alpha_J^4 \,  k_{\perp} ^{4 J}  \,  \left( 3 \,  B_{ \chi x}^2 + B_{ \chi y}^2 \right) + 8 \,  B_{ \chi x}^2 \,  v_z^2 \,  k_{z}^2 \,  \varepsilon_\chi^2 \right]   f_{0}^\prime (\varepsilon_\chi )  . 
\nn &= -
\frac{ J ^{3} \,  v_z \,  \alpha_{J}^{ \frac{2}{J} }  }{128 \,  \pi^2 } \, e^4 \,  \tau
\int_{0} ^{\pi} \,  d \varphi \,    \sin ^{3-\frac{2} {J}}  \varphi \left[  \sin ^{4} \varphi 
\left( 3 \,  B_{ \chi x}^2 + B_{ \chi y}^2 \right) + 8 \,  B_{ \chi x}^2 \,   \cos^2 \varphi   \right] 
 \int_{0} ^{\infty} d \varepsilon_\chi \,  \varepsilon_\chi ^{ -\frac{2} {J}}  \,   f_{0}^\prime (\varepsilon_\chi ) 
\nn & = 
\frac{e^4 \,  \tau \,  J \,  v_{z } \,  \alpha_{J}^{ \frac{2} {J} }   }
{ 128 \,  \pi ^{\frac{3}{2}} } \,  \frac{  \Gamma  \big( 2-\frac{1}{J}\big) }
{  \Gamma \big (  \frac{9}{2} - \frac{1}{J}  \big)}
  \left[ \left ( 32 \,  J^2 -19  \, J + 3  \right )  
B_{\chi x}^2 + (3\, J- 1)\,  (2\,J-1) \,  B_{\chi y}^2     \right] 
 \Lambda_{ - \frac{2} {J} } (\mu ) \,.
\end{align}
This result is consistent with Eq. (17) of Ref. \cite{ips-rahul-ph} and, on setting $J=1$, also agrees with Eq. (19) of Ref.~\cite{onofre}.

\subsection{Contribution with the integrand proportional to nonzero powers of OMM}

From Eq.~\eqref{eq_elec_OMM}, the OMM-dependent part is given by
\begin{align}
\sigma ^{m, \chi}_{xx} =  
\sigma ^{m, \chi}_{1xx} + \sigma ^{m, \chi}_{2xx} + \sigma ^{m, \chi}_{3xx}\,, 
\end{align}
where
\begin{align}
\sigma ^{m, \chi}_{1xx} &= {2  \,  e ^{3} \,  \tau}
\int \frac{  d^3 {\mathbf k} }  { (2\, \pi )^3  } \,     {Q}_{\chi x} \, 
 v_{\chi x}^{(m)}  \,  f_{0}^\prime (\varepsilon_\chi ) = -
{2 \,  e^4 \,  \tau}
 \int \frac{  d^3 {\mathbf k} } { (2\, \pi )^3  } 
 \left[ \mathbf{\Omega }_{\chi}  \times ( {\boldsymbol  v}_{\chi} ^{(0)}  
  \times \mathbf{B}_{\chi} ) \right ]_{x} 
  \left[ {\bf{B}}  \cdot \partial_{x} 
  \left(  \varepsilon_\chi \,  \mathbf \Omega_{\chi}  \right) \right] \,  
  f_{0}^\prime (\varepsilon_\chi ) \,,
\end{align}
\begin{align}
\sigma ^{m, \chi}_{2xx} =  2 \,  e^2 \,  \tau
\int \frac{  d^3 {\mathbf k} } { (2\, \pi )^3  } \,   
 \varepsilon_{\chi}^{(m)}    
 \left(  \nabla_{{\mathbf k}}  \cdot  \boldsymbol{T}_{\chi xx} \right) 
    f_{0}^\prime (\varepsilon_\chi )  \, , 
\end{align}
and
\begin{align}
\sigma ^{m, \chi}_{3xx} =  {  e ^{3} \,  \tau}  
\int \frac{  d^3 {\mathbf k} } { (2\, \pi )^3  } \,   \varepsilon_{\chi}^{(m)}  \left( 
\mathbf{B}_{\chi} \cdot \boldsymbol{V}_{\chi xx } \right )   
 f_{0}^{ \prime \prime}  (\varepsilon_\chi ) 
 =   \frac{  e^4 \,  \tau}{ 8 \,  \pi ^{3}   }  
 \int d^{3} {\mathbf k} \,   \varepsilon_\chi 
   \left( \mathbf{B}_{\chi} \cdot \mathbf{\Omega }_{\chi} \right)^2 
 \left[  \left( \partial_{k_x} \varepsilon_\chi  \right)^2 
  - \frac{ \varepsilon_\chi} {2 }   
    \left( \partial_{k_{x}}^2  \,   \varepsilon_\chi  \right)  \right]   
f_{0}^{\prime \prime} (\varepsilon_\chi ) \,.
\end{align}

The first term evaluates to
\begin{align}
\sigma ^{m, \chi}_{1xx} &= \frac{  e^4 \,  \tau \,  \alpha_J^4  \,  J ^{3} \,  v_z^2 }{ 32\,  \pi^2  }  
\int_{- \infty} ^{\infty}  dk_{z}  \int_{0}^{\infty}  dk_{\perp} \,   
\frac{ k_{\perp} ^{4 J - 3} }{  \varepsilon_\chi  ^{8} } \,  
\,  f_{0}^\prime (\varepsilon_\chi ) 
\nn & \hspace{ 5.5 cm } \times
\left [ 
\alpha_J^4 \,  k_{\perp} ^{4 J} 
\left( B_{ \chi x}^2 + B_{ \chi y}^2 \right) + \alpha_{J}^2 \,  k_{\perp} ^{2J} \, 
 v_z^2 \,  k_{z}^2   \left \lbrace  (J+1) \,  B_{ \chi x}^2  - (J-1) \, 
  B_{ \chi y}^2 \right \rbrace 
 + 4 \,  B_{ \chi x}^2 \,  J \,  v_z ^{4} \,  k_{z} ^{4} 
 \right ]  \nn
&= \frac{  e^4 \,  \tau \,  \alpha_{J} ^{2J} \,  J^2  \,  v_z  }{ 32 \,  \pi^2   } 
  \int_{0}^{\pi} \,  d\varphi  \left( \sin  \varphi \right)^{3-\frac{2} {J}}    
 \left [  \sin^4 \varphi 
  \left( B_{ \chi x}^2 + B_{ \chi y}^2 \right) +  \sin^2 \varphi  \cos^2 \varphi 
   \left \lbrace  (J+1) \,  B_{ \chi x}^2  - (J-1) 
 \,  B_{ \chi y}^2 \right \rbrace 
 + 4 \,  B_{ \chi x}^2  \, J \,    \cos ^{4} \varphi  \right ] \notag \\ 
 & \hspace{ 0.5 cm} 
 \times  \int_{0} ^{\infty}  d \varepsilon_\chi  \, 
  \varepsilon_\chi ^{ - \frac{2} {J}} \,   f_{0}^\prime (\varepsilon_\chi )  \nn
& = - \frac{  e^4 \,  \tau \,  v_z \,   \alpha_{J}^{ \frac{2} {J} }  }
{ 64 \,  \pi^{\frac{3}{2}}   } \, 
  \frac{  \Gamma  \left( 2-\frac{1}{J}\right) }
{  \Gamma \big (  \frac{9}{2} - \frac{1}{J}  \big)} \,  \left[  (8 \,  J^3+13 \,  J^2-11 \,  J+2) \,   B_{\chi x}^2  + (1-2 \,  J) \,  \left(J^2 -7 \,  J+2\right)\,   
B_{\chi y}^2 \right]  \,  \Lambda_{ - \frac{2} {J} } (\mu ) \,. 
\end{align}

The second term evaluates to
\begin{align}
\sigma ^{m, \chi}_{2xx} &= 
\frac{  e^4 \,  \tau \,  \alpha_J^4 \,  J ^{3} \,  v_z^2 }
{32 \,  \pi^2 } 
 \int_{- \infty} ^{\infty}  dk_{z}   \int_{0}^{\infty}  dk_{\perp} \,   
 \frac{k_{\perp}^{4 J-3}}{\varepsilon_\chi ^{8}} 
\, f_{0}^\prime (\varepsilon_\chi )
 \nn & \hspace{ 3 cm } \times
\left [ 
-  \alpha_{J}^2 \,  k_{\perp} ^{2J}  \, 
  v_z^2 \,  k_{z}^2   
    \left \lbrace (5\, J+1)\,  B_{ \chi x}^2  + (J+1) \,  B_{ \chi  y}^2 \right \rbrace  
    +(J-1) \,  v_z ^{4} \,  k_{z} ^{4} 
  \left(3 \, B_{ \chi  x}^2 + B_{ \chi  y}^2  \right) - 2 \, (J-1) \,  B_{ \chi  x}^2 
  \,   \alpha_J^4 \,  k_{\perp} ^{4\, J} 
   \right ] \nn
 &= \frac{  e^4 \,  \tau \,  \alpha_{J}^{ \frac{2} {J} } \,  J^2 \,  v_z }
 {32 \,  \pi^2 } 
  \int_{0} ^{\pi} \,  d \varphi  \left( \sin  \varphi \right)^{3-\frac{2} {J}}  
 \Big [ -    \sin^2 \varphi  \cos^2 \varphi  
 \left \lbrace  (5 \, J+1) \,  B_{ \chi  x}^2  + (J+1) \,  B_{ \chi  y}^2 \right \rbrace 
\nn & \hspace{ 6 cm}
 + (J-1)   \cos ^{4} \varphi  \,  \left(3 \,  B_{ \chi  x}^2 + B_{ \chi y}^2  \right) 
  - 2\,  (J-1) \,  B_{ \chi x}^2    \sin ^{4} \varphi   
  \Big ] 
\int_{0}^{\infty}  d \varepsilon_\chi  \,
   \varepsilon_\chi ^{-\frac{2} {J}}  \,  f_{0}^\prime (\varepsilon_\chi ) \nn
&  =
 \frac{  e^4 \,  \tau \,  v_z \,  \alpha_{J} ^{ \frac{2}  {J}  }
}
 {128 \,  \pi^{\frac{3}{2}} }  
  \frac{   \Gamma \big ( 2 - \frac{1}{J} \big)  }
  {    \Gamma \big (  \frac{9}{2} - \frac{1}{J}  \big)}  \left[ \left (59\, J ^{3} - 85\,
   J^2 + 46\, J - 8 \right )   B_{ \chi x}^2
    + J \left( J^2 + 5 \, J - 2 \right)  \,  B_{ \chi y}^2  
   \right] \,   \Lambda_{ - \frac{2} {J} } (\mu )  \,.
\end{align}

The third term evaluates to 
\begin{align}
\sigma ^{m, \chi}_{3xx}
& =  \frac{  e^4 \,  \tau  \,  \alpha_{J} ^{6} \,   J ^{3} \,  v_z^2 }
{256 \,  \pi^2  } 
 \int_{- \infty} ^{\infty}  dk_{z}  \int_{0}^{\infty}  dk_{\perp} \,   \frac{k_{\perp} ^{6J-3} }
 {  \varepsilon_\chi  ^{7}} \,   f_{0}^{ \prime \prime}  (\varepsilon_\chi )
 \Big [ 
 2\,  v_z^2 \,  k_{z}^2   
 \left \lbrace (3\, J-1) \,  B_{\chi x}^2 
+   (J+1) \,  B_{\chi y}^2 \right \rbrace 
 \nn & \hspace{ 8 cm}
- \alpha_{J}^2 \,  k_{\perp} ^{2J} 
\left \lbrace (3\, J+2) \,  B_{\chi x}^2 + (J-2) \, 
 B_{\chi y}^2 \right \rbrace \Big ]
\nn & = 
  \frac{  e^4 \,  \tau \,  \alpha_{J} ^{ \frac{2} {J} } \,   J^2 \,  v_z }
  {256 \,  \pi^2 } 
    \int_{0}^{\pi} \,  d \varphi   \left( \sin  \varphi \right)^{5-\frac{2} {J}}  
  \left [ 2  \cos^2 \varphi 
  \left \lbrace (3\, J-1) \,  B_{\chi x}^2
  +   (J+1) \,  B_{\chi y}^2 \right \rbrace
     -   \sin^2 \varphi   \left \lbrace
   (3\, J+2) \,  B_{\chi x}^2 + (J-2) \,  B_{\chi y}^2 
\right \rbrace \right]
   \notag \\ 
  & \qquad  \times  
  \int_{0}^{\infty}  d \varepsilon_\chi 
  \,  \varepsilon_\chi ^{1-\frac{2} {J}} \, f_{0}^{ \prime \prime}  (\varepsilon_\chi )
\nn & = 
-  \frac{  e^4 \,  \tau \,  \alpha_{J} ^{ \frac{2} {J} }  \,  J  \,  v_z }
{128  \, \pi^2} 
 \frac{  \sqrt{\pi} \,  \Gamma \big ( 3 - \frac{1}{J} \big)  }
 {    \Gamma \big (  \frac{9}{2} - \frac{1}{J}  \big)}  \,   \left[   (J+1) (3 \,J-1) \,  B_{\chi x}^2 + (J^{2}-4\, J+1) \,  B_{\chi y}^2 \right] 
  \int_{0}^{\infty}  d \varepsilon_\chi \,  
  \varepsilon_\chi ^{1-\frac{2} {J}} \, f_{0}^{ \prime \prime}  (\varepsilon_\chi )  
\nn & =  -  \frac{  e^4 \,  \tau \,  \alpha_{J} ^{ \frac{2} {J} }  \,  J  \,  v_z }
{128 \,  \pi ^{\frac{3}{2}} }  \,
 \frac{  
 \left( 1 - \frac{2}{J} \right) \left( 2 - \frac{1}{J} \right) 
  \Gamma \big( 2 - \frac{1}{J} \big )  }
  {    \Gamma \big( \frac{9}{2} - \frac{1}{J} \big)}   
    \left[  \left  (J+1 \right) \left (3 \, J-1 \right )   B_{\chi x}^2 + 
    \left (J^2-4\,J+1 \right ) \,  B_{\chi y}^2 \right] \,  \Lambda_{ - \frac{2} {J} } (\mu )  \,.  
\end{align}

Adding up the three terms, we get
\begin{align}
\label{eqomegaOMM}
& \sigma ^{m, \chi}_{xx}
\nn & =   \frac{   e^4 \,  \tau \, v_z \,   \alpha_{J}^{ \frac{2} {J} } 
\,\Lambda_{ - \frac{2} {J} } (\mu )
\, \Gamma \big ( 2 - \frac{1}{J} \big) }
{ 128 \,  \pi ^{\frac{3}{2}}  \,  J
\, \Gamma \big (  \frac{9}{2} - \frac{1}{J}  \big)  } 
\left[ \left( 37 \, J^4 - 100\, J^3 + 74 \, J^{2} -21 \, J +2 \right)  
 B_{\chi x}^2 +  \left (   3 \,J^4 - 12 \,J^3 - 4 \,J^2  + 9\, J - 2  \right )    B_{\chi y}^2  \right] .
\end{align}
As a consistency check, we set $J = 1$, which gives us
\begin{align}
\sigma ^{m, \chi}_{xx} \Big \vert_{J=1} = -  \frac{  e^4 \,  \tau \,  v_z \,  
 \alpha_{J}^2  \,    \Lambda_{-2} (\mu ) }
 { 120 \,  \pi^2 }  
 \left (4 \,  B_{\chi x}^2 + 3 \,  B_{\chi y}^2  \right ),
\end{align}
agreeing with Eq. (20) of Ref.~\cite{onofre}.

\section{Computation of the transverse components of the magnetoelectric conductivity tensor}
 \label{app_phc}

In this section, we show the derivation of the explicit expressions for the transverse components $\sigma^\chi_{yx} = \sigma^\chi_{xy}$ of the magnetoelectric conductivity tensor $\sigma^\chi $ (i.e., the PHC) shown in Sec.~\ref{secsxy}, starting from the integrals appearing in Eqs.~\eqref{eq_elec} and \eqref{totalsigma}.

\subsection{Field-independent contribution}

The field-independent contribution of Eq.~(\ref{eq_elec_0B}) vanishes due to the azimuthal symmetry about the $z$-axis, as shown below:
\begin{align}
\sigma ^{0, \chi}_{xy} &= - e^2 \,  \tau  \int \frac{  d^3 {\mathbf k} } { (2\, \pi )^3  }  \,  v_{\chi x } ^{(0)} 
\,  v_{\chi y} ^{(0)} \,  
 f_{0}^\prime (\varepsilon_\chi )
 = - e^2 \,  \tau \,  J^2 \,  \alpha ^{4}_{J} 
  \int \frac{  d^3 {\mathbf k} } { (2\, \pi )^3  }  
 \left( \frac{ k_{\perp} ^{2J-2}  } {  \varepsilon_\chi }   \right)^2 \,  k_{x} \,   k_{y}  \,  
 f_{0}^\prime (\varepsilon_\chi )  = 0 \,.
\end{align}

\subsection{Contribution solely from BC (no OMM)}

The contribution to $\sigma_{xy}^\chi $ solely from the BC, captured by Eq.~\eqref{eq_elec_Berry}, is given by
\begin{align}
\label{eqsigmaxyBC}
\sigma ^{\Omega , \chi}_{xy} &= - e^4 \,  \tau
  \int \frac{  d^3 {\mathbf k} } { (2\, \pi )^3  } \,  Q_{\chi x } \,  Q_{\chi y } \, 
   f_{0}^\prime (\varepsilon_\chi ) 
 = -  e^4 \,  \tau
   \int \frac{  d^3 {\mathbf k} } { (2\, \pi )^3  } 
     \left[ \mathbf{\Omega }_{\chi}  \times ( {\boldsymbol  v}_{\chi} ^{(0)}   \times \mathbf{B}_{\chi} ) \right]_{x} 
   \left[ \mathbf{\Omega }_{\chi}  \times ( {\boldsymbol  v}_{\chi} ^{(0)}   \times \mathbf{B}_{\chi} ) \right]_{y} 
  \,  f_{0}^\prime (\varepsilon_\chi )  \nn
 &= - \frac{e^4 \,  \tau \,  \alpha_J^4  \, J ^4 \, v_z^2   }
{ 64 \,  \pi^2   } 
\,  B_{\chi x} \,  B_{\chi y}  
 \int_{- \infty} ^{\infty}  dk_{z}  
 \int_{0}^{\infty}  dk_{\perp} \,    
 \frac{ k_{\perp}^{4 J-3}  } { \varepsilon_\chi ^{8} } 
 \left( \varepsilon_\chi^2 +   v_z^2 \,  k_{z}^2 \,  \right)^2  \,  f_{0}^\prime (\varepsilon_\chi )   
\nn
&=  - \frac{e^4 \,  \tau  \,  \alpha_{J}^{ \frac{2} {J} } \,  J ^{3} \,  v_z   }
{ 64 \,  \pi^2    } \,  B_{\chi x} \,  B_{\chi y} 
 \int_{0} ^{\pi} \,  d \varphi  \left( \sin  \varphi \right)^{3-\frac{2} {J}} 
  \left( 1 +  \cos^2 \varphi \right)^2 
 \, \int_{0} ^{  \infty} \,  d \varepsilon_\chi \, \varepsilon_\chi ^{ -\frac{2} {J}}  \,  f_{0}^\prime (\varepsilon_\chi ) \nn
&  =    \frac{e^4 \,  \tau \,  v_z \,  J \,  \alpha_{J}^{ \frac{2}{J} } 
 \,   \Lambda_{ - \frac{2} {J} } (\mu )
 }
{ 64 \,  \pi ^{\frac{3}{2}}  }     
\,  \frac{\Gamma \big (  2-\frac{1}{J}  \big)}{\Gamma \big (  \frac{9}{2} - \frac{1}{J}  \big)}
\,  B_{\chi x} \,  B_{\chi y} 
\left ( 13\, J^{2} - 7 \, J +1 \right )  .
\end{align}
This result is consistent with Eq. (20) of Ref. \cite{ips-rahul-ph} and, on setting $J=1$, also agrees with Eq. (19) of Ref.~\cite{onofre}.

\subsection{Contribution with the integrand proportional to nonzero powers of OMM}

From Eq.~\eqref{eq_elec_OMM}, the OMM-dependent part is given by
\begin{align}
\sigma ^{m, \chi}_{xy} =  
\sigma ^{m, \chi}_{1xy} + \sigma ^{m, \chi}_{2xy} + \sigma ^{m, \chi}_{3xy}\,, 
\end{align}
where
\begin{align}
\sigma ^{m, \chi}_{1xy} &=  2 \,  e ^{3} \,  \tau
  \int \frac{  d^3 {\mathbf k} } { (2\, \pi )^3  } \,     {Q}_{\chi x} \,  v_{\chi y}^{(m)} 
   \,  f_{0}^\prime (\varepsilon_\chi ) 
   = - {2 \,  e^4 \,  \tau} 
    \int \frac{  d^3 {\mathbf k} } { (2\, \pi )^3  } 
    \left[ \mathbf{\Omega }_{\chi}  \times ( {\boldsymbol  v}_{\chi} ^{(0)}   \times \mathbf{B}_{\chi} ) \right]_{x}
     \left[ {\bf{B}}  \cdot \partial_{y} \left(  \varepsilon_\chi\,   \mathbf \Omega_{\chi}  \right) \right] 
     f_{0}^\prime (\varepsilon_\chi ) \,,
\end{align}
\begin{align}
\sigma ^{m, \chi}_{2xy} =  2 \, e ^{3} \,\tau
 \int \frac{  d^3 {\mathbf k} } { (2\, \pi )^3  } \,     
 \frac{\varepsilon_{\chi}^{(m)}}{e}   
 \left( \nabla_{{\mathbf k}} \cdot  \boldsymbol{T}_{\chi xy} \right)
  f_{0}^\prime (\varepsilon_\chi ) \,  , 
\end{align}
and
\begin{align}
\sigma ^{m, \chi}_{3xy} =   e ^{3}\,  \tau
  \int \frac{  d^3 {\mathbf k} } { (2\, \pi )^3  } \,  
  \varepsilon_{\chi}^{(m)}  \left(
   \mathbf{B}_{\chi} \cdot \boldsymbol{V}_{\chi xy } \right)
 f_{0}^{\prime \prime} (\varepsilon_\chi )  \, .
\end{align}

The first term evaluates to
\begin{align}
\sigma ^{m, \chi}_{1xy} &=  \frac{e^4 \,  \tau \,   \alpha_J^4  \,  J ^4  \, v_z ^4 }
{16 \,  \pi^2 } 
\,  B_{\chi x}  \,  B_{\chi y}  
\int_{- \infty} ^{\infty}  dk_{z}  \int_{0}^{\infty}  
dk_{\perp} \,   \frac{ k_{z}^2 \,  k_{\perp}^{4 J-3}  } { \varepsilon_\chi^8 } 
\left( \varepsilon_\chi^2 +   v_z^2 \,  k_{z}^2 \right) 
f_{0}^\prime (\varepsilon_\chi ) \nn
 &=  \frac{e^4 \,  \tau \,  \alpha_{J}^{ \frac{2} {J} }  \,  J ^{3} \,  v_z  }
 {16 \,  \pi ^2} 
 \,  B_{\chi x} \,  B_{\chi y}  \int_{0}^{\pi} d \varphi   
 \cos^2 \varphi  \left( \sin  \varphi  \right)^{3-\frac{2}{J} } 
  \left( 1 + \cos^2 \varphi \right)  \int_{0} ^{\infty}  d  \varepsilon_\chi \,  \varepsilon_\chi ^{ -\frac{2} {J}}  
  \,  f_{0}^\prime (\varepsilon_\chi ) \nn
&  = -   \frac{e^4 \,  \tau \,  J^2 \left  (5 \, J-1 \right ) 
\alpha_{J}^{ \frac{2} {J} }  \,  v_z   \,  \Lambda_{ - \frac{2} {J} } (\mu )
}
{32 \,  \pi^{\frac{3}{2}}   } \,  
\frac{ \Gamma \big (  2-\frac{1}{J}  \big)}{  \Gamma \big (  \frac{9}{2} - \frac{1}{J}  \big)}
\,B_{\chi x} \,  B_{\chi y} \, .  
\end{align}

The second term evaluates to
\begin{align}
\sigma ^{m, \chi}_{2xy} & =
 -  \frac{  e^4 \,  \tau \,  \alpha_J^4  \,  J ^{3} \,  v_z^2  }{16 \,  \pi^2   } \,  
 B_{\chi x}  \,  B_{\chi y}  \int_{- \infty} ^{\infty}  dk_{z}  \int_{0}^{\infty}  dk_{\perp} \,  \frac{ k_{\perp}^{4 J-3} }{\varepsilon_\chi ^{8}}  \,     \left[  (J-1) \,  \alpha_J^4 \,  k_{\perp} ^{4 J} + 2 \,  J \,  v_z^2 \,  k_{z}^2 \,  \alpha_{J}^2 \,  k_{\perp} ^{2J} -(J-1) \,  v_z^{4} \,  k_{z}^{4} \right]  \,   f_{0}^\prime (\varepsilon_\chi )   
\nn & =  -  \frac{  e^4 \,  \tau \, J ^2 \,   \alpha_{J} ^{ \frac{2} {J} }  \,   v_z 
}
{16 \,  \pi^2   } \,  B_{\chi x} \,  B_{\chi y} 
 \int_{0} ^{\pi}   d \varphi \left( \sin  \varphi \right)^{3-\frac{2} {J}}     
  \left[  (J-1)   \sin^4 \varphi  
  + 2 \,  J   \cos^2 \varphi   
   \sin^2 \varphi 
    - (J-1)  \cos ^{4} \varphi   \right]
   \int_{0}^{\infty}  d \varepsilon_\chi \, 
 \varepsilon_\chi ^{-\frac{2}{J}}  \,
  f_{0}^\prime (\varepsilon_\chi ) \nn
& =  -  \frac{  e^4 \,  \tau \,  \alpha_{J} ^{ \frac{2} {J} }  \,  v_z 
\,  \Lambda_{ - \frac{2} {J} } (\mu )
}
{64  \,  \pi ^{\frac{3}{2}}  }
\,  \frac{ \Gamma \big (  2-\frac{1}{J}  \big)}
{ \Gamma \big (  \frac{9}{2} - \frac{1}{J}  \big)} 
 \,  B_{\chi x} \,  B_{\chi y} \left  ( 45 \, J^2 -29\, J ^{3} -24\, J +4 \right )  .  
\end{align}

The third term evaluates to
\begin{align}
\sigma ^{m, \chi}_{3xy} &= - \frac{  e^4 \,  \tau \,  \alpha_{J} ^{6} \,  J ^{3}  \,  v_z^2 }
{ 128 \,  \pi^2   } \,  B_{\chi x}  \, B_{\chi y}  
\int_{- \infty} ^{\infty}  dk_{z}  
\int_{0}^{\infty}  dk_{\perp} \,   
\frac{k_{\perp} ^{6J-3}}{ \varepsilon_\chi ^{7} }  
\left[ (J+2) \,  \alpha_{J}^2 \,  
k_{\perp} ^{2J}  - 2 \,  (J-1) \,  v_z^2 \,  k_{z}^2 \right]  
f_{0}^{ \prime \prime}  (\varepsilon_\chi )  \nn
& = - \frac{  e^4 \,  \tau \,  \alpha_{J}^{ \frac{2} {J} } \,  J^2 \,  v_z }
{ 128 \,  \pi^2    } \,  B_{\chi x} \,  B_{\chi y}  
 \int_{0}^{\infty}  d \varphi \left( \sin  \varphi \right)^{5-\frac{2} {J}} 
    \left[ (J+2)  \sin^2 \varphi  - 2 \,  (J-1)  \cos^2 \varphi \right]  
     \int_{0} ^{\pi} d \varepsilon_\chi  \, \varepsilon_\chi  ^{1-\frac{2} {J}} 
  \,  f_{0}^{ \prime \prime}  (\varepsilon_\chi )
\nn
& =  -  \frac{  e^4 \,  \tau \,  J\,\alpha_{J}^{ \frac{2} {J} } \,  v_z
\,  \Lambda_{ - \frac{2} {J} } (\mu )  }
{ 64 \,  \pi ^{\frac{3}{2}}   }  \,   
 \frac{ 
 \left(  J^2 + 3\, J -1 \right ) 
 \left(2-\frac{1}{J}\right) \,  \left( 1 - \frac{2}{J} \right) 
  \Gamma \big (  2-\frac{1}{J}  \big)}{  \Gamma \big (  \frac{9}{2} - \frac{1}{J}  \big)}  
  \,  B_{\chi x} \,  B_{\chi y}\,.
\end{align}

Finally, summing up the three contributions we obtain
\begin{align}
\label{eqomegaOMM_xy}
\sigma ^{m, \chi}_{xy} =  \frac{e^4  \,  \tau \,  \alpha_{J}^{ \frac{2} {J} }  \,  v_z
\,  \Lambda_{ - \frac{2} {J} } (\mu )     }
{64 \,  \pi ^{\frac{3}{2}}
} \, 
  \frac{  
  \left( 17 \, J^3-44 \,J^2 + 39\, J + \frac{2}{J}-15 \right) 
  \Gamma \big (  2-\frac{1}{J}  \big)}{  \Gamma \big (  \frac{9}{2} - \frac{1}{J}  \big)}
  \,  B_{\chi x} \,  B_{\chi y}   \, ,
\end{align}
which is consistent with Eq.~(20) of Ref.~\cite{onofre}.

\section{Computation of the longitudinal components of the magnetothermoelectric coefficient}
\label{appalphaxx}

In this section, we outline the derivation of the explicit expression of the longitudinal component of the magnetothermoelectric conductivity tensor $\alpha^\chi $ (i.e., the LTEC), shown in Sec.~\ref{secalpha}, starting from the integrals appearing in Eq.~\eqref{eq_thermoel}.

\subsection{Field-independent contribution}

The $\mathbf B_\chi$-independent contribution, shown in Eq.~\eqref{eqalphatot}, is given by
\begin{align}
 {\alpha} ^{0, \chi}_{xx} & = e  \,  \tau  \int \frac{  d^3 {\mathbf k} }
 {(2\, \pi ) ^{3} }  
 \left ( v_{\chi x } ^{(0)}  \right )^{2} 
  \frac{\varepsilon_\chi  - \mu}{T}   f_{0} ^{\prime} ( \varepsilon_\chi )
=   e  \,  \tau  \,  J^2  \,  \alpha^{4}_{J} 
\int \frac{  d^3 {\mathbf k} } { (2\, \pi )^3  }  
      \left( \frac{ k_{x} \,  k_{\perp} ^{2J-2}  }
    {  \varepsilon_\chi }   \right)^{2} 
      \left( \frac{ \varepsilon_\chi - \mu }{T} \right) \,  
   f_{0} ^{\prime} ( \varepsilon_\chi ) \nn
&=   \frac{ e  \,  \tau  \,  J^2 \,  \alpha ^{4}_{J} } {(2\, \pi )^3   }  
\int_{0}^{2 \pi} d \phi  \cos^2 \phi 
 \int_{- \infty} ^{\infty} dk_{z}  
\int_{0} ^{\infty}  dk_{\perp} \,    \frac{   k_{\perp} ^{4J-1}  }
{ \varepsilon_\chi^2 } \,  f_{0}^\prime (\varepsilon_\chi ) \nn  &
=   \frac{ e   \,  \tau  \,  J^2 \,  \alpha ^{4}_{J} }{8 \,  \pi^2    }  
 \int_{- \infty} ^{\infty} dk_{z}  \int_{0} ^{\infty}  dk_{\perp} \,   
  \frac{   k_{\perp} ^{4 J-1}  }{ \varepsilon_\chi^2 } \,  \frac{ \varepsilon_\chi - \mu }{T} 
   \,  f_{0}^\prime (\varepsilon_\chi )
\nn & = - \frac{ e \,  \tau  \,  J   }
{ 6 \,  \pi^2   \,  v_z    }  \,  \frac{\Lambda_{3} (\mu ) - \mu \, \Lambda_{2} (\mu )}{T} 
 = - \frac{ e \,  \tau  \,  J \, \mu  }{ 9   \,  v_z \, \beta  }   \,.
\end{align}
This result reproduces the field-independent part of Eq.~(23) of Ref.~\cite{ips-rahul-ph}.

\subsection{Contribution solely from BC (no OMM)}

The contribution to $\alpha_{xx}^\chi$ solely from the BC, captured by the term $\alpha^{\Omega , \chi}_{xx}  \equiv {\alpha} ^{1,\Omega , \chi}_{xx}
+ {\alpha} ^{2,\Omega , \chi}_{xx}  $ of Eq.~\eqref{eqalphatot}, is given by the sum of
\begin{align}
{\alpha} ^{1,\Omega , \chi}_{xx} =   e^2 \,\tau 
\int \frac{  d^3 {\mathbf k} }{(2\, \pi ) ^{3} }  
  \left[ 2  \left  (  {\boldsymbol  v}_{\chi} ^{(0)} \cdot \mathbf{\Omega }_{\chi}  \right )  
  v_{\chi x } ^{(0)} \, B_{\chi x} 
    -   \left( {\mathbf B}_{\chi} \cdot \mathbf{\Omega }_{\chi}  \right)   
    \left (  v_{\chi x } ^{(0)} \right  )^2  \right]
   \frac{\varepsilon_\chi  - \mu}{T}  \,  f_{0} ^{\prime} ( \varepsilon_\chi ) = 0 ,
\end{align}
and
\begin{align}
{\alpha} ^{2,\Omega , \chi}_{xx} &=   e ^{3}\,  \tau
  \int \frac{  d^3 {\mathbf k} } { (2\, \pi )^3  } \,  Q_{\chi x }^2 \, 
  \frac{\varepsilon_\chi  - \mu}{T}  \,   f_{0} ^{\prime} ( \varepsilon_\chi ) \nn
&=    \frac{ \alpha_J^4 \,  J ^4 \,  v_z^2}
{128 \,  \pi^2 } \, e^3 \,  \tau
 \int_{- \infty} ^{\infty}  dk_{z}   \int_{0}^{\infty}  dk_{\perp} \,  \frac{k_{\perp} ^{4 J -3}}{  \varepsilon_\chi ^{8} } \left[ \alpha_J^4 \,  k_{\perp} ^{4 J}  \,  \left( 3 \,  B_{\chi x}^2 + B_{\chi y}^2 \right) 
 + 8 \,  B_{\chi x}^2 \,  v_z^2 \,  k_{z}^2 \,  \varepsilon_\chi^2 \right]  \frac{\varepsilon_\chi  - \mu}{T}  \,  f_{0} ^{\prime} ( \varepsilon_\chi ) 
 \nn
 &=  \frac{ J^3 \,  v_z \,  \alpha_{J}^{ \frac{2}{J} }  }
{128 \,  \pi^2 } \,{e^3 \,  \tau}
   \int_{0} ^{\pi} \,  d \varphi 
  \left(  \sin   \varphi \right)^{3-\frac{2}{J} } 
  \left[  \sin^4 \varphi \,  \left( 3 \,  B_{\chi x}^2 + B_{\chi y}^2 \right) 
  + 8 \,  B_{\chi x}^2 \,   \cos^2 \varphi   \right] 
   \int_{0} ^{\infty} d \varepsilon_\chi \,  \varepsilon_\chi^{ - \frac{2}{J}}  \,
 \frac{ \varepsilon_\chi  - \mu }{T}
  \,  f_{0} ^{\prime} ( \varepsilon_\chi ) 
\nn & =
 - \frac{e^3 \,  \tau \,  J \,  v_{z } \,  \alpha_{J}^{ \frac{2} {J} }   }
 { 128 \,  \pi^2  } \, 
  \frac{ \sqrt{\pi } \,  \Gamma  \left( 2-\frac{1}{J}\right) }
  {  \Gamma \big (  \frac{9}{2} - \frac{1}{J}  \big)} 
  \left[ \left ( 32 \,  J^2 -19  \, J + 3  \right )  B_{\chi x}^2 + (3\,J- 1)\,  (2\,J-1)
   \,  B_{\chi y}^2     \right]  
  \frac{\Lambda_{1-\frac{2} {J}} (\mu ) - \mu \, \Lambda_{ - \frac{2} {J} } (\mu ) }{T} \nn
& =  \frac{ \sqrt{\pi } \, e^3 \,  \tau \,   v_{z } \,  \alpha_{J}^{ \frac{2} {J} } \, 
 \Gamma  \big( 2-\frac{1}{J}\big) }
 { 192 \,   \beta \, \mu ^{1 + \frac{2} { J} }
\, \Gamma \big (  \frac{9}{2} - \frac{1}{J}  \big)} \,    
  \left[ \left ( 32 \,  J^2 -19  \, J + 3  \right )  B_{\chi x}^2 
  + (3 \,J- 1)\,  (2\,J-1) \,  B_{\chi y}^2     \right]  .
\end{align}
This result reproduces Eq.~(25) of Ref.~\cite{ips-rahul-ph}.

\subsection{Contribution with the integrand proportional to nonzero powers of OMM}

The contribution to $\alpha_{xx}^\chi $, which vanishes if OMM is omitted, is captured by the term $\alpha^{m , \chi}_{xx}
\equiv   \alpha^{1, m , \chi}_{xx} + \alpha^{2, m , \chi}_{xx} 
+ \alpha^{2, (\Omega, m), \chi}_{xx} $ of Eq.~\eqref{eqalphatot}. We find that
\begin{align}
 {\alpha} ^{1,m, \chi}_{xx} = e  \,  \tau 
  \int \frac{  d^3 {\mathbf k} }{(2\, \pi ) ^{3} } 
    \left [  v_{\chi x } ^{(0)}  
   \left ( 2  \, v_{\chi x } ^{(m)}  
   \, \frac{\varepsilon_\chi  - \mu}{T}
    +   v_{\chi x } ^{(0)} \,
    \frac{  \varepsilon_\chi ^{ (m) } }{T}    \right )
   f_{0} ^{\prime} ( \varepsilon_\chi )
     + \left (v_{\chi x } ^{(0)}  \right )^2 \frac{\varepsilon_\chi  - \mu}{T}
     \varepsilon_\chi ^{ (m) } 
    \, f_{0} ^{\prime \prime} (\varepsilon_\chi)   \right ]  = 0\, .  
\end{align}
The remaining part is obtained from a sum of the following integrals, which we evaluate explicitly:
\begin{align}
I^\alpha_{1, xx} & \equiv  
  \frac{e  \,  \tau}{2}   \int \frac{d^3 {\mathbf k} }{(2 \pi )^3 }  
 \left  (  v_{\chi x } ^{(0)} \right )^2 \frac{\varepsilon_\chi  - \mu}{T} \,
 \left(  \varepsilon_\chi ^{(m )} \right )^ 2
 \, f_{0} ^{\prime \prime \prime} ( \varepsilon_\chi )    
\nn & = 
  \frac{e^3 \, \tau \,  v_z \, J^3\, \alpha_{J}^{ \frac{2} {J} } }
{ 256 \, \pi^{\frac{3}{2}} \, T } \,
\frac{   \Gamma \big(4-\frac{1}{J} \big)   }
{  \Gamma \big (  \frac{9}{2} - \frac{1}{J}  \big)} 
\left (3 \,B_{ \chi x}^2  +  B_{ \chi y} \right ) 
\int_{0} ^{\infty} d \varepsilon_\chi \,    \varepsilon_\chi  ^{2-\frac{2} {J}} (\varepsilon_\chi  - \mu ) \,  f_{0} ^{\prime \prime \prime} ( \varepsilon_\chi )  \nn
&  = 
-  \frac{e^3 \, \tau \,  v_z \, \alpha_{J}^{ \frac{2} {J} } \,  \mu ^{1-\frac{2} {J}} }
{ 64 \, \pi^{\frac{3}{2}} \, T  } \,
 \frac{   \Gamma \big (4-\frac{1}{J} \big)   }
{  \Gamma \big (  \frac{9}{2} - \frac{1}{J}  \big)}  
\left (3 \, B_{ \chi x}^2  +  B_{ \chi y}^2 \right ) 
  \left( J-1 \right)     
   \left [  J^2 - \frac{2 \left( J-2 \right)  \pi^2    }
{3 \,  \beta^2 \,  \mu^2 }  \right ] ,
\end{align}
\begin{align}
I^\alpha_{ 2, xx} & \equiv 
  \frac{e^3 \, \tau \,  v_z \, J \, \alpha_{J}^{ \frac{2} {J} }   }
  { 128 \, \pi^{\frac{3}{2}}\, T } \,
  \frac{\Gamma \big (3-\frac{1}{J} \big )   }
  {  \Gamma \big (  \frac{9}{2} - \frac{1}{J}  \big)}    
   \int_{0} ^{\infty} d \varepsilon_\chi 
  \left\lbrace \left[   \left(13\, J^2+ 9\, J-4\right) \mu -4 \left(J^2+3\, J-1\right)
   \varepsilon_\chi   \right] B_{ \chi x}^2 \right. \notag \\ & \hspace{4cm} \left. + \left[   \left(-5 \, J^2+15 \, J-4\right) \mu +4 \left(2 J^2-4 J+1\right) \varepsilon_\chi   \right] B_{ \chi y}^2  \right\rbrace \,  \varepsilon_\chi ^{1-\frac{2} {J}} f_{0} ^{\prime \prime} (\varepsilon_\chi)  
\nn &=  \frac{e^3 \, \tau \,  v_z \, \alpha_{J}^{ \frac{2} {J} }   }
{ 384 \, \pi^{\frac{3}{2}} \,  \beta } \,
\frac{\Gamma \big (3-\frac{1}{J}\big)   }
{  \Gamma \big (  \frac{9}{2} - \frac{1}{J}  \big)}  \,  \mu ^{-(1+\frac{2} {J})} 
 \frac{1}{ J }     \left [
 \left \lbrace
  3 \,\beta ^2 J^2 \left(5 \, J^2-33\, J + 10\right) \mu ^2
 + 3 \, \pi ^2 \left(7 \, J^3+3\, J^2-40 J+12\right) \right \rbrace
  B_{ \chi x}^2 \right. \notag \\ & \hspace{3cm} \left. 
 + \left \lbrace 3 \beta ^2 J^2 \left(11 J^2-23\, J+6\right) \mu ^2
 + \pi ^2 \left(-21 J^3+95 \, J^2-120 J+28\right) \right\rbrace B_{ \chi y}^2  \right ]  ,
\end{align} 
\begin{align}
I^\alpha_{ 3, xx} & \equiv 
   e  \,  \tau  \int \frac{d^3 {\mathbf k} }{(2 \,\pi )^3 }   \left [  
\left \lbrace
 \left ( v_{\chi x } ^{(m)} \right )^2 + { 2 \, e} \left  (  {\boldsymbol  v}_{\chi} ^{(0)} 
      \cdot \mathbf{\Omega }_{\chi} \right )    v_{\chi x} ^{(m)}\, B_{\chi x} 
     +  2 \, e  \left(   {\boldsymbol  v}_{\chi} ^{(m)} \cdot \mathbf{\Omega }_{\chi}  \right )  
      v_{\chi x } ^{(0)} B_{\chi x}    
-  2 \, e  \left( {\mathbf B}_{\chi} \cdot \mathbf{\Omega }_{\chi}  \right)  v_{\chi x} ^{(0)} \, v_{\chi x } ^{(m)}  
 \right \rbrace    \frac{\varepsilon_\chi  - \mu}{T} \right. \notag \\ & \hspace{3cm} \left.  
+ \left \lbrace   2 \, e  \,  (  {\boldsymbol  v}_{\chi} ^{(0)} \cdot \mathbf{\Omega }_{\chi}   )  v_{\chi x } ^{(0)} \,B_{\chi x}    - e \,  \left( {\mathbf B}_{\chi} \cdot \mathbf{\Omega }_{\chi}  \right) 
\left  (   v_{\chi x } ^{(0)} \right )^2 
 + {2   v_{\chi x } ^{(0)} v_{\chi x } ^{(m)}    }  \right \rbrace   
 \frac{  \varepsilon_\chi ^{ (m) } }{T} \right ]  
f_{0} ^{\prime} ( \varepsilon_\chi ) 
\nn
 &=    \frac{e^3 \, \tau \,  v_z \, \alpha_{J}^{ \frac{2} {J} }   
\, \Gamma \big (  2-\frac{1}{J}  \big) }
{ 128 \, \pi^{\frac{3}{2}}  \,  T 
\, J\, \Gamma \big (  \frac{9}{2} - \frac{1}{J}  \big)} 
\int_{0} ^{\infty} d \varepsilon_\chi  \left [
 \left \lbrace \left(7 \, J^4-16 \,J^3-17 \, J^2+20\, J-4\right) \mu 
 + \left(-33\, J^4+ 11 \,J^3 + 34\, J^2-24 \,J+4\right) 
 \varepsilon_\chi  \right \rbrace B_{ \chi x}^2 \right.  \notag \\ & \hspace{3. cm} 
  \left.  + \left \lbrace \left(-7 \, J^4+44 \,J^3-61 \,J^2+28 \,J-4\right) \mu 
 +\left(17 \, J^4-79 \,J^3 + 84 \,J^2-32\, J+4\right) \varepsilon_\chi \right \rbrace 
 B_{ \chi y}^2 \right ]
  \varepsilon_\chi ^{-\frac{2} {J}}  f_{0} ^{\prime} ( \varepsilon_\chi ) \nn 
  &= -   \frac{e^3 \, \tau \,  v_z \, \alpha_{J}^{ \frac{2} {J} }   }
  { 384 \, J^2 \, \pi^{\frac{3}{2}} \,  \beta 
 \,\mu^{1+\frac{2} {J}}  } \,
\frac{  \Gamma \big (  2-\frac{1}{J}  \big)   }
    {  \Gamma \big (  \frac{9}{2} - \frac{1}{J}  \big)}    
   \left [ \left\lbrace  \pi ^2 \left(40 \,J^4-79 \,J^3-61 \,J^2 +78 \,J-16\right)
    -3 \,\beta ^2 J^2 \mu ^2 \left(26\, J^3+5 \, J^2-17 \, J+4\right) \right \rbrace 
  B_{ \chi x}^2  \right.  
     \notag \\ & \hspace{1.5cm}   \left.   
   + \left[ 3 \, \beta ^2 J^2 \mu ^2 \left(10 \,J^3-35 \, J^2+23\, J-4\right) 
   +\pi ^2 \left(-24 \,J^4 + 143\, J^3-215 \, J^2+106 J-16\right)\right]  B_{ \chi y}^2 \right ]  .
\end{align}

Summing up the three nonzero contributions, we finally obtain
\begin{align}
\alpha_{xx} ^{m, \chi} =   \frac{ \sqrt{\pi} \, e^3 \,  \tau \, v_z \,   \alpha_{J}^{ \frac{2} {J} } 
\, \Gamma \big ( 2 - \frac{1}{J} \big) }
{ 192  \,  J^2 \, \mu ^{1+\frac{2} {J}} \,   \beta
\,  \Gamma \big (  \frac{9}{2} - \frac{1}{J}  \big)  }  
  \left[ \left( 37 \,J^4-100\, J^3+74 \,J^{2} -21\, J +2 \right) \,  B_{\chi x}^2
 +  (-2 + 9 \, J - 4 \,J^2 - 12 \,J^3 + 3 \,J^4)  \,  B_{\chi y}^2  \right] . 
\end{align}

\section{Computation of the transverse components of the magnetothermoelectric coefficient}
\label{appalphayx}

In this section, we show the derivation of the explicit expressions for the transverse components $\alpha^\chi_{yx} = \alpha^\chi_{xy}$ of the magnetothermoelectric conductivity tensor $\alpha^\chi $ (i.e., the TTEC) shown in Sec.~\ref{secalpha}, starting from the integrals appearing in Eq.~\eqref{eq_thermoel}.

\subsection{Field-independent contribution}

The $\mathbf B_\chi$-independent contribution of Eq.~\eqref{eq_thermoel} vanishes, as shown below:
\begin{align}
{\alpha} ^{0, \chi}_{xy} &=   e   \,  \tau  \int \frac{d^3 {\mathbf k} }
{(2 \,\pi )^3 }  \,  v_{\chi x } ^{(0)} \,  v_{\chi y} ^{(0)}  \, \frac{\varepsilon_\chi  - \mu }{T}  \,   f_{0}^\prime (\varepsilon_\chi )  
= e  \,  \tau \,  J^2 \,  \alpha ^{4}_{J}  \int \frac{d^3 {\mathbf k} }
{(2 \, \pi )^3 }  
 \left( \frac{ k_{\perp} ^{2J-2}  } {  \varepsilon_\chi }   \right)^2   k_{x} \,   k_{y} 
\, \frac{ \epsilon_{\mathbf k}  - \mu } {T} 
  \,   f_{0}^\prime (\varepsilon_\chi )  = 0 \,.
\end{align}
This is expected due to the azimuthal symmetry about the $z$-axis.

\subsection{Contribution solely from BC (no OMM)}

The contribution to $\alpha_{xy}^\chi$ solely from the BC is captured by 
\begin{align}
{\alpha} ^{\Omega , \chi}_{xy} &=  e^3 \,  \tau 
\int \frac{d^3 {\mathbf k} }{(2 \pi )^3 } \,  Q_{\chi x } \,  Q_{\chi y } \,
 \frac{ \varepsilon_\chi  - \mu }{T}  \,  
 f_{0}^\prime (\varepsilon_\chi ) \nn
& =  \frac{e^3 \,  \tau \,  \alpha_J^4  \, J ^4 \, v_z^2   } { 64 \,  \pi^2   }
 \,  B_{\chi x} \,  B_{\chi y}\,   \int_{- \infty} ^{\infty}  dk_{z}  \int_{0}^{\infty}  dk_{\perp} \,    \frac{ k_{\perp}^{4 J-3}  }{ \varepsilon_\chi ^{8} } \left( \varepsilon_\chi^2 +   v_z^2 \,  k_{z}^2 \,  \right)^2   \,  \frac{ \varepsilon_\chi  - \mu }{T}  \,
  f_{0}^\prime (\varepsilon_\chi )  \nn
&=  \frac{e^3 \,  \tau  \,  \alpha_{J}^{ \frac{2} {J} } \,  J^3 \,  v_z   }
{ 64 \,  \pi^2    } \,  B_{\chi x} \,  B_{\chi y} \,   \int_{0} ^{\pi} \,  d \varphi \,   
\left( \sin  \varphi \right) ^{3-\frac{2} {J}}  \left( 1 +  \cos^2 \varphi \right)^2 \, \int_{0} ^{  \infty} 
  d \varepsilon_\chi \, \varepsilon_\chi ^{ -\frac{2} {J}}  \,  \frac{ \varepsilon_\chi  - \mu }{T}  \,
  f_{0}^\prime (\varepsilon_\chi )    
\nn & = 
  \frac{ \sqrt{\pi } \, 
  e^3 \,  \tau \,  v_z    \,  \alpha_{J}^{ \frac{2}{J} } }{ 96    \,   \beta  \, \mu ^{ 1+\frac{2} {J}}   }    
  \,    B_{\chi x} \,  B_{\chi y} \left  ( 13\, J^{2} - 7 \, J +1 \right) 
  \frac{\ \Gamma \big (  2-\frac{1}{J}  \big)}{\Gamma \big (  \frac{9}{2} - \frac{1}{J}  \big)}   \,,
\end{align}
which agrees with Eq.~(27) of Ref.~\cite{ips-rahul-ph}.

\subsection{Contribution with the integrand proportional to nonzero powers of OMM}

The contribution to $\alpha_{xy}^\chi $, which vanishes if OMM is omitted, is captured by the sum of the following three integrals, which we evaluate explicitly:
\begin{align}
I^\alpha_{1, xy} & \equiv   e  \,  \tau    
 \int \frac{d^3 {\mathbf k} }{(2 \,\pi )^3 }  \,  
 \frac{ v_{\chi x } ^{(0)} \,v_{\chi y } ^{(0)}}{2}    
 \, \frac{\varepsilon_\chi  - \mu}{T}   \left( \varepsilon_\chi ^{(m)}  \right)^2 
 f_{0} ^{\prime \prime \prime} ( \varepsilon_\chi ) 
\nn &  = \frac{e^3 \,  \tau \,  v_z \, J^3\, 
  \alpha_{J}^{ \frac{2}{J} } \, \mu ^{-1-\frac{2} {J}} }
{ 128 \, \pi^{\frac{3}{2}} \,T }  \,
  \frac{  \Gamma \left(4-\frac{1}{J}\right) }
  { \Gamma \big (  \frac{9}{2} - \frac{1}{J}  \big)}  \,B_{\chi x} \, B_{\chi y}  
  \int_{0} ^{  \infty} \,  d \varepsilon_\chi \,  \varepsilon_\chi ^{2-\frac{2} {J}} 
  \left (\varepsilon_\chi -\mu \right )  
   f_{0} ^{\prime \prime \prime} ( \varepsilon_\chi )  \nn
& = - \frac{  e^3 \,  \tau \,  v_z \,    (J-1) 
\, \alpha_{J}^{ \frac{2}{J} } \, \mu ^{ 1-\frac{2} {J}} }
{ 32 \,  \pi^{\frac{3}{2}}\,T  }  \,
  \frac{ \Gamma \left(4-\frac{1}{J}\right) }
{ \Gamma \big (  \frac{9}{2} - \frac{1}{J}  \big)}  B_{\chi x} \, B_{\chi y}        
 \left[ 1 -  \frac{2 \, (J-2) \,  \pi^2     }{3 \,  J^2  \, \beta^2 \,  \mu^2 }  \right] , 
\end{align}

\begin{align}
I^\alpha_{2, xy} & \equiv    
 e  \,  \tau  \int \frac{d^3 {\mathbf k} }{(2 \pi )^3 }  
   \left [  
   \left \lbrace  e  \left  (  {\boldsymbol  v}_{\chi} ^{(0)} \cdot \mathbf{\Omega }_{\chi} \right  ) 
   \left ( v_{\chi x}^{(0)} \, B_{\chi y} +  B_{\chi x} \, v_{\chi y }^{(0)}  \right) 
    - e  \left( {\mathbf B}_{\chi} \cdot \mathbf{\Omega }_{\chi}  \right)    
     v_{\chi x } ^{(0)}   \,  v_{\chi y } ^{(0)}   \right \rbrace  \frac{\varepsilon_\chi  - \mu}{T} 
    \right. \notag \\ & 
   \hspace{ 2.5 cm} \left.   +   
   \left  ( v_{\chi x } ^{(0)} \, v_{\chi y } + v_{\chi x } \, v_{\chi y } ^{(0)} \right )  
     \frac{\varepsilon_\chi  - \mu}{T} +  v_{\chi x } ^{(0)} \, v_{\chi y } ^{(0)}  \,
    \frac{  \varepsilon_\chi ^{ (m) } }{T}     \right ]
   \varepsilon_\chi ^{ (m) } \, f_{0} ^{\prime \prime} (\varepsilon_\chi) 
\nn
 &=  \frac{e^3 \,  \tau \,  v_z \,   J^3\, \alpha_{J}^{ \frac{2}{J} }  }
 { 64 \,  \pi^{\frac{3}{2}} \, T  }  \,
 \frac{ \Gamma \left(4-\frac{1}{J}\right) }
 { \Gamma \big (  \frac{9}{2} - \frac{1}{J}  \big)} \,B_{\chi x} \, B_{\chi y} 
   \int_{0} ^{  \infty} \,  d \varepsilon_\chi \,   
  \varepsilon_\chi ^{1-\frac{2} {J}} \left (3 \,\mu -2 \, \varepsilon_\chi  \right)  
  f_{0} ^{\prime \prime} (\varepsilon_\chi)  \nn
& =
  \frac{e^3 \,  \tau \,  v_z \,    
\alpha_{J}^{ \frac{2}{J} } }
{ 192 \,  \pi^{\frac{3}{2}} \, \beta \,  \mu ^{1+\frac{2} {J}}  }  \,
\,\frac{ \Gamma \big(4-\frac{1}{J}\big ) }
{ \Gamma \big (  \frac{9}{2} - \frac{1}{J}  \big)}
\, B_{\chi x} \, B_{\chi y}   
   \left[ \pi ^2 \left(7 \, J^2-12 J-4\right)-3\, \beta ^2 \,\mu ^2 \,J^2 \,(J+2)  \right]  , 
\end{align}
\begin{align}
I^\alpha_{3, xy} & \equiv 
 e  \,  \tau  \int \frac{d^3 {\mathbf k} }{(2 \,\pi )^3 }  
 \left\lbrace \left[  v_{\chi x } ^{(m)} v_{\chi y } ^{(m)}
  + e  \left  (  {\boldsymbol  v}_{\chi} ^{(0)} \cdot \mathbf{\Omega }_{\chi} \right )  
  \left ( v_{\chi x } ^{(m)} \, B_{\chi y} 
  +  B_{\chi x}    v_{\chi y } ^{(m)} \right )  
   + e  \,  (   {\boldsymbol  v}_{\chi} ^{(m)} \cdot \mathbf{\Omega }_{\chi}  ) ( v_{\chi x} ^{(0)} B_{\chi y}  
+   B_{\chi x} \, v_{\chi y } ^{(0)}  ) \right. \right. \notag \\ & \phantom{=} \left.  
 - e \left( {\mathbf B}_{\chi} \cdot \mathbf{\Omega }_{\chi}  \right) 
 \left (v_{\chi x } ^{(0)} \, v_{\chi y } ^{(m)} + v_{\chi x } \, v_{\chi y } ^{(0)} \right ) \right] 
 \frac{\varepsilon_\chi  - \mu}{T} + \left[  e  \left  (  {\boldsymbol  v}_{\chi} ^{(0)} \cdot \mathbf{\Omega }_{\chi}  \right )
 \left ( v_{\chi x } ^{(0)} B_{\chi y} +  B_{\chi x}  v_{\chi y } ^{(0)}  \right ) 
  - e   \left( {\mathbf B}_{\chi} \cdot \mathbf{\Omega }_{\chi}  \right)     v_{\chi x} ^{(0)}  \,   v_{\chi y } ^{(0)}   \right. \notag \\ & \phantom{=} \left.  \left.  +   ( v_{\chi x } ^{(0)} \, v_{\chi y } + v_{\chi x } \, v_{\chi y } ^{(0)} )   \right]   \frac{  \varepsilon_\chi ^{ (m) } }{T}   \right\rbrace f_{0} ^{\prime} ( \varepsilon_\chi )  
\nn  &
=  \frac{e^3 \,  \tau \,  v_z \,    \alpha_{J}^{ \frac{2}{J} }  }
{ 64 \,  \pi^2 \, T  } 
\, \frac{ \sqrt{\pi} \, 
\Gamma \big (  2-\frac{1}{J}  \big) }{ \Gamma \big (  \frac{9}{2} - \frac{1}{J}  \big)} 
\, B_{\chi x}  \, B_{\chi y} 
  \int_{0} ^{  \infty}   d \varepsilon_\chi 
\left[ \left (7\, J-2) ( J^{2} - 4\, J +2 \right) \mu 
-\left (5 \, J-4 \right ) 
\left (5 \, J^2 - 5 \, J +1 \right)
 \varepsilon_\chi \right] \varepsilon_\chi ^{-\frac{2} {J}} \, f_{0} ^{\prime} (\varepsilon_\chi)  
\nn &  = 
- \frac{e^3 \,  \tau \,  v_z \,    \alpha_{J}^{ \frac{2}{J} } }
{ 192 \, J \,  \pi^{\frac{3}{2}} \,  \beta\, \mu ^{1+\frac{2} {J}} }  
\,\frac{  
\Gamma \big (  2-\frac{1}{J}  \big) }
{ \Gamma \big (  \frac{9}{2} - \frac{1}{J}  \big)} \,
B_{\chi x}  \, B_{\chi y}      
 \left[ \pi ^2 \left(32\, J^3-111 \,J^2+77 \,J-14\right)-9 \,\beta ^2\, J^2
  \left(6 J^2-5 \, J+1\right) \mu ^2 \right] . 
\end{align}

Adding the three contributions, we obtain the final result as
\begin{align}
{\alpha}^{m, \chi}_{xy} =  \frac{ \sqrt{\pi } \, e^3  \,  \tau \,  \alpha_{J}^{ \frac{2} {J} }  \,  v_z  
 \, \Gamma \big (  2-\frac{1}{J}  \big)}
{96  \,  J \,   \beta  \, \mu^{ 1+\frac{2} {J}} 
\, \Gamma \big (  \frac{9}{2} - \frac{1}{J}  \big)} 
\,  B_{\chi x} \,  B_{\chi y} 
 \left( 17 \,J^3-44 \,J^2 + 39 \, J+\frac{2}{J}-15 \right)  .
\end{align}

\section{Computation of the longitudinal components of the magnetothermal coefficient}
\label{appellxx}

In this section, we show the derivation of the explicit expression of the longitudinal component of the magnetothermal coefficient tensor $ \ell^\chi $, shown in Sec.~\ref{secell}, starting from the integrals appearing in Eq.~\eqref{eq_thermal}.

\subsection{Field-independent contribution}

The $\mathbf B_\chi$-independent contribution, shown in Eq.~\eqref{eqelltot}, is given by
\begin{align}
\ell^{0, \chi}_{xx} & = - \,\tau  
\int \frac{d^3 {\mathbf k} } {(2 \,\pi )^3 } 
 \left( v_{\chi x } ^{(0)} \right)^2 \,  \frac{ ( \varepsilon_\chi - \mu   )^2   } {T} 
 \,  f_{0} ^{\prime} ( \varepsilon_\chi )  \nn
 & =   \frac{J \, \tau}{6 \, \pi^2 \,  v_z \, T} 
 \left[ \Lambda_{4} (\mu) - 2 \,\mu \, \Lambda_{3} (\mu)  + \mu^2 \, \Lambda_{2} (\mu) \right]
=  \frac{J \, \tau\,\mu^2\, T} { 18 \, v_z } \,. 
\end{align}
Comparing with Eq.~\eqref{eqsigmaxx0}, we readily find that the relation
\begin{align}
\sigma^{ 0, \chi} _{ij} =
 \frac{ 3 \,e^2 } {\pi^2\, T}   \, \ell^{ 0,\chi} _{ij}   + \order{\beta ^{-2}}
\end{align}
is satisfied.

\subsection{Contribution solely from BC (no OMM)}

The contribution to $\ell_{xx}^\chi$ solely from the BC, captured by the term $\ell^{\Omega , \chi}_{xx}  \equiv 
\ell^{1,\Omega , \chi}_{xx}
+ \ell^{2,\Omega , \chi}_{xx}  $ of Eq.~\eqref{eqelltot}, is given by the sum of
\begin{align}
\label{long1}
\ell^{1,\Omega , \chi}_{xx} & =
-\, \tau
\int\frac{d^{3}{\mathbf k}}{(2 \, \pi) ^{3}}
\left[2\, e\left ({\boldsymbol  v}_{\chi} ^{(0)} \cdot \mathbf{\Omega } _{\chi} \right )
v _{\chi x } ^{(0)} \, B _{\chi x}-e\,  \left( {\mathbf B} _{\chi} 
\cdot \mathbf{\Omega } _{\chi}  \right)v _{\chi x} ^{(0)} \,v _{\chi x} ^{(0)} \right]
\frac{\left( \varepsilon_{\chi}-\mu\right)^{2}}{T}
\, f_{0} ^{\prime } (\varepsilon_\chi) \nn
& =
\frac{\chi\, e\,\tau  \, J^3 \,v_z \, \alpha_J^4 \,B_{\chi x}}
{32\, \pi^3 \, T}
\int\limits_{-\infty}^{\infty}
dk_z
\int\limits_0^{\infty} dk_{\perp} \,
k_{\perp}^{4J-2}\,
\frac{\left(\varepsilon_{\chi}-\mu\right)^2}{\varepsilon_{\chi}^5}\,
 f_0^\prime (\varepsilon_{\chi} )
\int\limits_0^{2 \, \pi} d\phi 
\left [4\,v_z^2\,k_z^2 \cos\phi
+\alpha_J^2 \, k_{\perp}^{2J}
\left \lbrace 3\cos\phi-\cos\phi\cos(2\phi)\right \rbrace \right ]\nn
&
\qquad -\frac{\chi\, e\,\tau\, J^3
\,v_z\, \alpha_J^6 \,B_{\chi y} }
{16\,\pi^3 \,T}
\int\limits_{-\infty}^{\infty} dk_z
\int\limits_0^{\infty} dk_{\perp} \,
k_{\perp}^{6J-2}\,
\frac{\left(\varepsilon_{\chi}-\mu\right)^2}{\varepsilon_{\chi}^5}
\,f_{0} ^{\prime } (\varepsilon_\chi)
\int\limits_0^{2\,\pi} d\phi
\cos^2\phi \, \sin\phi
\nn & = 0\,,
\end{align}
and
\begin{align}
\ell^{ 2,\Omega , \chi}_{xx} & =  - \,e ^2 \, \tau   
\int \frac{d^3 {\mathbf k} }{(2 \,\pi )^3 }  \, Q^2_{\chi x }    \, 
 \frac{ ( \varepsilon_\chi - \mu   )^2   }{T} 
 \, f_{0}^{\prime} ( \varepsilon_\chi ) 
\nn & =  - \frac{	e^2 \,\tau \, v_z \, J \,\alpha_{J}^{ \frac{2} {J} } }
{128 \, \pi^{\frac{3}{2}} \,  T } 
\,\frac{ \Gamma \big (2 - \frac{1}{J} \big) }{ \Gamma \big( \frac{9}{2} - \frac{1}{J} \big) }  
  \left[ \left ( 32 \,  J^2 -19  \, J + 3  \right )  B_{\chi x}^2 + \left (3 \,J- 1 \right )
  \left  (2 \,J-1 \right )   B_{\chi y}^2     \right]   
   \int_{0} ^{  \infty}   d \varepsilon_\chi \,  \varepsilon_\chi ^{-\frac{2} {J}}  
   \left (\varepsilon_\chi -\mu \right)^2   f_{0} ^{\prime } ( \varepsilon_\chi ) 
\nn &=  
\frac{	\sqrt \pi \, e^2\, \tau \, v_z \, J \,\alpha_{J}^{ \frac{2} {J} } \, T}
{3 \times 128 \,    \mu^{\frac{2} {J}} } 
\,\frac{ \Gamma \big (2 - \frac{1}{J} \big) }{ \Gamma \big( \frac{9}{2} - \frac{1}{J} \big) }
  \left[ \left ( 32 \,  J^2 -19  \, J + 3  \right )  B_{\chi x}^2 +
  \left (3 \,J- 1 \right )
  \left  (2 \,J-1 \right )  B_{\chi y}^2     \right]   . 
\end{align}
Comparing with Eq.~\eqref{eqomegabc}, we find that
\begin{align}
\sigma ^{ \Omega , \chi}_{xx} = \frac{ 3\, e^2 } {\pi^2\, T}  \,  \ell^{ \Omega , \chi}_{xx}  
 + \mathcal{O} (\beta ^{-2}) \,.  
\end{align}
 
\subsection{Contribution with the integrand proportional to nonzero powers of OMM}

The contribution to $\ell_{xx}^\chi$, which vanishes if the OMM is omitted, is captured by the term $
\ell^{m, \chi}_{xx}  =  \ell^{1, m , \chi}_{xx} + \ell^{2, m , \chi}_{xx} 
+ \ell^{2, (\Omega, m), \chi}_{xx} $ of Eq.~\eqref{eqelltot}.

\subsubsection{Part one}

We find that the part contributed by a linear-in-OMM term in the integrand is given by
\begin{align}
 \ell^{1, m , \chi}_{xx} 
=  I^{\ell, 1}_{1, xx} + I^{\ell, 1}_{2, xx} + I^{\ell, 1}_{3, xx} \,,
\end{align}
where
\begin{align}
& I^{\ell, 1}_{ 1, xx} \equiv -\, 2 \,\tau
\int\frac{  d^3{\mathbf k} }{(2 \,\pi ) ^{3} }
\left( v _{\chi x } ^{(0)} \right)^2  
\frac{\left (\varepsilon_{\chi} - \mu \right )
\varepsilon_{\chi} ^{(m)}}{T}
\, f_{0} ^{\prime} (\varepsilon_\chi)
\nn &
=- \frac{\chi \,e\, \tau\, J^3\,v_z\,\alpha_J^6}{8\,\pi^3\,T}
\int\limits_{-\infty}^{\infty}
dk_z\int\limits_0^{\infty}
dk_{\perp}
\, k_{\perp}^{6J-2}\;\frac{\varepsilon_{\chi}-\mu}{\varepsilon_{\chi}^4}
\,\,f_{0} ^{\prime } (\varepsilon_\chi)
\int\limits_0^{2 \, \pi} d\phi 
\left[ B_{\chi x} \cos^3\phi + B_{\chi y} \cos^2\phi \, \sin\phi \right ]
 = 0\,,
\end{align}
\begin{align}
I^{\ell, 1}_{2, xx}
& \equiv - \,2\,
\tau
\int\frac{  d^3{\mathbf k} }{(2\, \pi ) ^{3} } \,
v _{\chi x}^{(0)} \, v _{\chi x} ^{(m)} \,
\frac{(\varepsilon_{\chi}-\mu) ^{2}}{T}
\, \,f_{0} ^{\prime} (\varepsilon_\chi)
\nn
& =
\frac{\chi \, e \, \tau\, J^2 \, v_z\, \alpha_J^4 } {8\, \pi^3\, T}
\int\limits_{-\infty}^{\infty}dk_z
\int\limits_0^{\infty} dk_\perp \, k_{\perp}^{4J-2}\,
\frac{\left(\varepsilon_{\chi}-\mu\right)^2}{\varepsilon_{\chi}^5}
\,f_{0} ^{\prime } (\varepsilon_\chi)
\nn & \hspace{ 3 cm} \times
\int\limits_0^{2 \, \pi} d\phi \left [
B_{\chi x}
\left \lbrace \left(\varepsilon_{\chi}^2-J \, v_z^2 \, k_z^2\right)
\cos\phi \cos(2\phi)-J\,v_z^2 \,k_z^2 \cos\phi \right \rbrace
+ B_{\chi y}
\left(\varepsilon_{\chi}^2-J \, v_z^2 \, k_z^2\right)\cos\phi\sin(2\phi)
\right ]
\nn & = 0\,,
\end{align}
and
\begin{align}
I^{\ell, 1}_{ 3, xx} & \equiv -\tau\,\int\frac{  d^3{\mathbf k} }{(2 \,\pi ) ^{3} }
\left( v _{\chi x } ^{(0)} \right)^2 
\frac{ ( \varepsilon_{\chi}-\mu) ^{2}
\,\varepsilon_{\chi} ^{(m)}} {T}
\,f_{0} ^{\prime \prime} (\varepsilon_\chi)
\nn &
= -\frac{\chi \, e \, \tau \, J^3 \, v_z \, \alpha_J^6} {16 \,\pi^3}
\int\limits_{-\infty}^{\infty} dk_z
\int\limits_0^{\infty} dk_\perp \,
k_{\perp}^{6J-2}\,
\frac{\left(\varepsilon_{\chi}-\mu\right)^2}{\varepsilon_{\chi}^4}
\, f_{0} ^{\prime \prime} (\varepsilon_\chi)
\int\limits_0^{2 \, \pi} d\phi
\left [
B_{\chi x} \cos^3\phi + 
B_{\chi y} \cos^2\phi \, \sin\phi  \right ]
= 0 \,.
\end{align}

\subsubsection{Part two}

Using the partitioned integrals defined in Eqs.~\eqref{l2} and \eqref{l3}, we have $ \ell_{xx}^{2, m, \chi} = I^{ \ell, 2}_{1, xx } + I^{ \ell, 2}_{ 2, xx } + I^{ \ell, 2}_{ 3, xx}  $, where
\begin{align}
\label{l9}
I^{ \ell, 2}_{1, xx } & = -\, \,
\tau \int\frac{d^3 \mathbf{ k }}{\left( 2 \,\pi \right)^3} \,
 \Bigg[ \left( v_{\chi x}^{\left( 0 \right)} \right)^2 
 \frac{ \left( \varepsilon_{\chi}^{\left( m \right)} \right)^2}{T} + 4 \, v_{\chi x}^{(0)} \, v_{\chi x}^{(m)} \, \frac{\left( \varepsilon_{\chi} - \mu \right) \, \varepsilon_{\chi}^{ (m) }}{ T } + \left( v_{\chi x}^{( m )} \right)^2 \, \frac{ \left( \varepsilon_{\chi} - \mu \right)^2 }{T} \Bigg] \, f_0^{ \prime } (\varepsilon_{\chi} )
\nn & = -  \frac{ e^2 \, \tau \, v_z \, \alpha_J^{  \frac{2} {J}  } }
{ 128 \, \pi^{ \frac{3} {2} } \, T \, J } \,
 \frac{ \Gamma\left( 2 - \frac{1}{J} \right) }{ \Gamma \big ( \frac{9}{2} - \frac{1}{J} \big ) } 
 \int_{ 0 }^{ \infty } d \varepsilon_{ \chi } 
 \Bigg [ \Big \lbrace 
 \left( 51 \, J^4 - 97 \, J^3 + 86 \, J^2 - 32 \, J + 4 \right) \, \varepsilon_{ \chi }^{ 2 } 
-\left( 42 \, J^4 - 96 \, J^3 + 122 \, J^2 - 56 \, J + 8 \right) \, \mu \, 
\varepsilon_{ \chi } 
\nn 
& \hspace{ 6 cm}
+ \left(9 \, J^4 - 14 \, J^3 + 39 \, J^2 - 24 \, J + 4 \right) \,
 \mu^2 \Big \rbrace  \,B_{ \chi x }^2  \nn
& \hspace{ 4 cm}
+ \Big \lbrace \left( 17 \, J^4 - 79 \, J^3 + 84 \, J^2 - 32 \, J + 4\right)  
\varepsilon_{ \chi }^{ 2} 
- \frac{   14 \, J^5 - 102 \, J^4 + 210 \, J^3 - 178 \, J^2 + 64 \, J - 8  } { J-1 } 
\, \mu \,  \epsilon_{ \chi }  \nn 
& \hspace{ 5 cm}  
+ \left( 3 \, J^4 - 14 \, J^3 + 39 \, J^2 - 24 \, J + 4 \right) \, \mu ^2 
 \Big \rbrace \, B_{ \chi y }^2 
\Bigg ] \, \, \varepsilon_{ \chi }^{-\frac{ 2 }{ J }}\, f_0^{ \prime } (\varepsilon_{\chi} ) 
\nn & = -\frac{ e^2 \, \tau \, v_z \, \alpha_J^{\frac{2}{J} } \, T
\,  \Gamma \big ( 2 - \frac{1}{J} \big ) }
{ 384 \, \pi^{ \frac{3} {2} } \, J \, \mu_{\chi}^{ \frac{ 2 }{ J } }
\,  \Gamma\big( \frac{9}{2} - \frac{1}{J} \big )} 
\, \Big [ \left \lbrace  \left( -51 \, J^4 + 199 \, J^3 - 303 \, J^2 + 159 \, J - 26 \right)  \pi^2 
- \left( 54 \, J^2 - 45 \, J + 9 \right) \, J^2 \, \beta^2 \mu^2 
\right \rbrace B_{ \chi x }^2 
 \nn & \hspace{ 4.5 cm} 
  + \left \lbrace  \left( -17 \, J^4 + 113 \,J^3 - 231 \, J^2 + 133 \, J - 22 \right)  \pi^2 
  - \left(18 \, J^2 - 15 \, J + 3 \right) J^2 \, \beta^2 \, 
  \mu^2 \right \rbrace B_{ \chi y }^2 \Big ]\,,
\end{align}
\begin{align}
\label{l12}
 I^{ \ell, 2}_{ 2, xx }   & = -2 \, \tau \int\frac{d^3 \mathbf{ k }}{\left( 2\pi \right)^3} \, 
\Bigg[ \, v_{\chi x}^{( 0 )} \, v_{\chi x}^{( m )} \, 
\frac{ \left( \varepsilon_{\chi} - \mu \right)^2 \varepsilon_{\chi}^{ (m) } }{T} 
+ \left( v_{ \chi x }^{ (0) } \right)^2 \
 \frac{ \left( \varepsilon_{\chi} - \mu \right) 
 \left( \varepsilon_{\chi}^{ (m) } \right)^2}{T} \Bigg]
 \, f_0^{ \prime \prime } (\varepsilon_{\chi} )
\nn & =
- \, \frac{ e^2 \, \tau \, v_{ z } \, \alpha_J^{  \frac{2} {J}  } }
{ 64 \, \pi^{ \frac{3} {2} } \, T } \, 
 \frac{ \Gamma \big ( 2 - \frac{1}{J} \big) }
 { \Gamma \big ( \frac{9}{2} - \frac{1}{J} \big ) }  
  \int_{ 0 }^{ \infty } d \varepsilon_{ \chi } \,
\Bigg [ \,
\Big \lbrace \left  ( 24 \, J^3 - 32 \, J^2 + 14 \, J - 2 \right  ) 
 \varepsilon_{ \chi }^3
 -  \left ( 30 \, J^3 - 49 \, J^2 + 25 \,J - 4 \right ) 
  \mu \, \varepsilon_\chi^2
  \nn
& \hspace{ 6 cm } 
+ \left ( 6 \, J^3 - 17 \, J^2 + 11 \, J - 2 \right ) 
 \mu^2 \, \varepsilon_\chi \,  \Big \rbrace
 \, B_{ \chi x }^2 
 \nn & \hspace{ 5.5 cm } 
 + \Big \lbrace  \left ( 8 \, J^3 - 20 \, J^2 + 12 \, J - 2 \right ) 
 \varepsilon_{ \chi }^{ 3 - \frac{ 2 }{ J }} 
 - \left ( 10 \, J^3 - 35 \, J^2 + 23 \, J - 4  \right ) 
 \mu \, \varepsilon_{ \chi }^{ 2 }
  \nn  &  \hspace{ 6.75 cm } 
+ \left ( 2 \, J^3 - 15 \, J^2 + 11 \, J - 2 \right  )  \mu^2 \, 
\varepsilon_{ \chi } \, \Big \rbrace \, 
B_{ \chi y }^2 \Bigg ] \,\varepsilon_{ \chi }^{ - \frac{ 2 }{ J }}\,
 f_0^{ \prime \prime } (\varepsilon_{\chi} )\nn
& = -\frac{ e^2 \, \tau \, v_z \, \alpha_J^{ \frac{2} {J} } \, T}
{ 64 \, \pi^{ \frac{3} {2} } 
\, J \, \mu_{\chi}^{ \frac{ 2 }{ J } }} \, 
 \frac{ \Gamma \big ( 2 - \frac{1}{J} \big ) }
 { \Gamma \big ( \frac{9}{2} - \frac{1}{J} \big ) } \,
  \Big [ \left \lbrace
   \left( 24 \, J^4 - 98 \, J^3 + 129 \, J^2 - 63 \, J + 10 \right)   \pi ^2 
   + 3  \left( 6 \, J^2 - 5 \, J + 1 \right) J^2 \, \beta ^2 \mu ^2 \right \rbrace
    B_{ \chi x }^2  \nn 
& \hspace{ 4.75 cm }
 + \left \lbrace
  \left( 8 \, J^4 - 42 \, J^3 + 69 \, J^2 - 37 \, J + 6 \right)  \pi ^2 
+ \left(6 \, J^2 - 5 \, J + 1\right)   J^2 \, \beta^2 \, \mu ^2 \right \rbrace B_{ \chi y }^2
\, \Big] \,,
\end{align}
and
\begin{align}
\label{l15}
 I^{ \ell, 2}_{ 3, xx}  &= -\,\tau
 \int\frac{d^3 \mathbf{ k }}{\left( 2 \,\pi \right)^3} 
  \left( v_{\chi x}^{ (0) } \right)^2 \, \frac{ \left( \varepsilon_{\chi} - \mu \right)^2 
 \left( \varepsilon_{\chi}^{ (m) } \right)^2 }{ 2\,T} \, f_0^{ \prime \prime \prime} (\varepsilon_{\chi} ) \nn 
& = - \frac{ e^2 \, \tau \, v_{ z } \, \alpha_{ J }^{  \frac{2} {J}  } \, J}{ 256 \, \pi^{ \frac{3} {2} } \, T } \, 
 \frac{ \Gamma\left( 2 - \frac{1}{J} \right) }{ \Gamma \left( \frac{9}{2} - \frac{1}{J} \right) }  
 \left( 3 \, B_{ \chi x }^2 + B_{ \chi y }^2 \right) 
  \nn 
& \qquad
 \times  \int_0^{ \infty } d \varepsilon_{ \chi } 
 \left [ \left ( 6 \, J^2 - 5 \, J + 1 \right ) 
 \left( \varepsilon_\chi^4 
 + \mu^2 \, \varepsilon_{ \chi }^{ 2  } \right) 
 -\left ( 12 \, J^2 - 10 \, J + 2 \right )  \mu 
 \,\varepsilon_{ \chi }^3 \right ] 
\varepsilon_{ \chi }^{ - \frac{ 2 }{ J }}\,
  f_0^{ \prime \prime \prime } (\varepsilon_{\chi} )
\nn & =
 \frac{ e^2 \, \tau \, v_z \, \alpha_J^{ \frac{2} {J} } \, T}
 {  128 \, \pi^{ \frac{3} {2} } \, J \, \mu_{\chi}^{ \frac{ 2 }{ J } }}
  \, \frac{ \Gamma \big( 2 - \frac{1}{J} \big ) }
  { \Gamma \big ( \frac{9}{2} - \frac{1}{J} \big ) }
   \left( 3 \, B_{ \chi x }^2 
 + B_{ \chi y }^2 \right) \left[ \left( 12 \, J^4 - 46 \, J^3 + 56 \, J^2 - 26 \, J + 4 \right)  \pi^2 
 + \left(6 J^2-5 J+1\right) J^2 \, \beta^2 \, \mu ^2 \right].
\end{align}

Adding up the three partitions, we obtain
\begin{align}
\label{qommxx}
\ell_{ x x }^{ 2, \, m, \, \chi } &= 
 \frac{ e^2 \, \tau \, v_z \, \alpha_J^{ \frac{2} {J} } \, T}
 { 384 \, J \, \mu_{\chi}^{ \frac{ 2 }{ J } }} \,
  \frac{ \sqrt{ \pi } \,
   \Gamma\big ( 2 - \frac{1}{J} \big ) }
  { \Gamma \big ( \frac{9}{2} - \frac{1}{J} \big ) } 
  \left[ \left ( 15 \, J^4 - 25 \, J^3 + 33 \, J^2 - 15 \, J + 2  \right) 
  B_{ \chi x }^2 + \left ( 5 \, J^4 + \, J^3 - 15 \, J^2 + 11 \, J - 2 \right )   B_{ \chi y }^2 \right]. 
\end{align}

\subsubsection{Part three}

Using the partitioned integrals defined in Eqs.~\eqref{l4} and \eqref{l5}, we have $ \ell_{xx}^{2, ( m , \Omega ), \chi} =
 I^{ \ell, 3}_{ 1,xx}  
+  I^{ \ell, 3}_{ 2, xx} +  I^{ \ell, 3}_{ 3 , xx} $, where
\begin{align}
\label{l19}
 I^{ \ell, 3}_{ 1,xx} & =  -\, 2 \, e \, \tau 
 \int \frac{ d^3 \mathbf { k }}{( 2 \, \pi )^3} 
 \left[ 2  \left ( \boldsymbol{ v }_{\chi}^{ (0) } \cdot \mathbf {  \Omega }_{\chi} \right ) 
 v_{\chi x }^{( 0 )} \, B_{\chi x } - \left( \mathbf { B }_{\chi} \cdot \mathbf { \Omega}_{\chi} \right) \, v_{\chi x }^{( 0 )} \, v_{\chi 
x }^{ (0) } \right] \, \frac{ \left( \varepsilon_{\chi} - \mu \right) \, \varepsilon_{\chi}^{( m )}}{T} \, f_0^{ \prime } (\varepsilon_{\chi} )  
\nn &=
 \frac{ e^2 \, \tau \, v_{ z } \, \alpha_{ J }^{ \frac{2} {J} } \, J }
 { 64 \, \pi^{ \frac{3} {2} } \, T } \,
  \frac{ \Gamma\big ( 2 - \frac{1}{J} \big ) }
  { \Gamma \big ( \frac{9}{2} - \frac{1}{J} \big ) } 
  \, \Big[  \left ( 38 \, J^2 -29 \, J + 5 \right  ) 
   B_{ \chi x }^2 - ( 6 \, J^2 - 5 \, J + 1 ) \, B_{ \chi y }^2 \Big] 
   \int_{ 0 }^{ \infty } d \varepsilon_{ \chi } 
    \left( \varepsilon_{ \chi }^2  -  \mu \, \varepsilon_{ \chi } \right) 
  \varepsilon_{ \chi } ^{ - \frac{ 2 }{ J } }\,
   f_0^{ \prime } (\varepsilon_{\chi} )
\nn & =
 - \frac{ e^2 \, \tau \, v_{ z } \, \alpha_{ J }^{ \frac{2} {J} } \, T }
 { 192 \, \mu^{ \frac{ 2 }{ J } } }
  \, \frac{ \sqrt{ \pi } \, \Gamma\big ( 2 - \frac{1}{J} \big ) }
  { \Gamma \big ( \frac{9}{2} - \frac{1}{J} \big ) } \, 
  \Big[ \left ( 38 \, J^3 - 105 \, J^2 + 63 \, J - 10 \right  )   B_{ \chi x }^2 
  - \left ( 6 \, J^3 - 17 \, J^2 + 11 \, J - 2 \right )  B_{ \chi y }^2 \Big]\,,
\end{align}
\begin{align}
\label{l20}
 I^{ \ell, 3}_{ 2, xx} &=  
- \,2 \, e \, \tau \int \frac{ d^3 \mathbf { k }}{( 2 \, \pi )^3} 
\left [  \left ( \boldsymbol{ v }_{\chi}^{ (0) } \cdot \mathbf {  \Omega }_{\chi} \right)
 v_{\chi x}^{( m )} \, B_{\chi x} 
+ \big( \boldsymbol{ v }_{\chi}^{ ( m ) } \cdot 
\mathbf {  \Omega }_{\chi} \big) \, v_{\chi x }^{( 0 )} \, B_{\chi x } 
- \left( \mathbf { B }_{\chi} \cdot \mathbf { \Omega}_{\chi} \right) 
\, v_{ \chi x }^{ (0) } \, v_{ \chi x }^{ (m) } \right ]
 \, \frac{ \left( \varepsilon_{\chi} - \mu \right)^2 }{T} 
 \, f_0^{ \prime } (\varepsilon_{\chi} )
\nn & = 
\frac{ e^2 \, \tau \, v_{ z } \, \alpha_{ J }^{ \frac{2} {J} } }
{ 64 \, \pi^{ \frac{3} {2} } \, T } \,
 \frac{ \Gamma\big ( 2 - \frac{1}{J} \big ) }
 { \Gamma \big ( \frac{9}{2} - \frac{1}{J} \big ) } 
 \left [ \left ( 8 \, J^3 - 15 \, J^2 + 11 \, J - 2 \right ) B_{ \chi x }^2  
  - \left ( 2 \, J^3 - 15 \, J^2 + 11 \, J - 2 \right )  B_{ \chi y }^2 \right ] 
\nn &  \qquad \times  
\int_{ 0 }^{ \infty } 
d \varepsilon_{ \chi } 
 \left( \varepsilon_{ \chi }^2
- 2\, \mu \, \varepsilon_{ \chi }
 + \mu^2 \right)
\varepsilon_{ \chi }^{ - \frac{ 2 }{ J } } \,  
 f_0^{ \prime } (\varepsilon_{\chi} )\nn
 &=
 - \frac{ e^2 \, \tau \, v_{ z } \, \alpha_{ J }^{ \frac{2} {J} } \, T }
 { 192 \, \mu^{ \frac{ 2 }{ J } } } \, \frac{ \sqrt{ \pi } \, 
 \Gamma\big ( 2 - \frac{1}{J} \big ) }
 { \Gamma \big ( \frac{9}{2} - \frac{1}{J} \big ) } 
 \left [ \left ( 8 \, J^3 - 15 \, J^2 + 11 \, J - 2 \right  ) 
 B_{ \chi x }^2 
 - \left ( 2 \, J^3 - 15 \, J^2 + 11 \, J - 2 \right )  B_{ \chi y }^2 \right ]
\end{align} 
and
 \begin{align}
I^{ \ell, 3}_{ 3 , xx} & = -\, e \, \tau \int \frac{ d^3 \mathbf { k }}{( 2 \, \pi )^3} \, \bigg[ 2\, 
 \left( \boldsymbol{ v }_{\chi}^{ (0) } \cdot \mathbf {  \Omega }_{\chi} \right) \, v_{\chi x}^{ (0) } 
 \, B_{\chi x} - \left( \mathbf {B}_{ \chi } \cdot \mathbf {  \Omega }_{ \chi } \right) 
  \left( v_{ \chi x }^{ (0) } \right)^2 \bigg] \, 
  \frac{ \left( \varepsilon_{\chi} - \mu \right)^2 \,
   \varepsilon_{ \chi }^{ (m) } }{ T } \, f_0^{ \prime \prime} (\varepsilon_{\chi} )\nn
& = 
\frac{ e^2 \, \tau \, v_{ z } \, \alpha_{ J }^{ \frac{2} {J} } \, J }
{ 128 \, \pi^{ \frac{3} {2} } \, T } \, 
\frac{ \Gamma\big ( 2 - \frac{1}{J} \big ) }
{ \Gamma \big ( \frac{9}{2} - \frac{1}{J} \big ) } 
\left[ \left ( 38 \, J^2 - 29 \, J + 5 \right )  B_{ \chi x }^2  
-\left ( 6 \, J^2 - 5 \, J + 1 \right )  B_{ \chi y }^2 \right ]  \nn 
& \qquad \times\int_{ 0 }^{ \infty } d \varepsilon_{ \chi } 
\left( \varepsilon_{ \chi }^3 -2 \, \mu \, 
\varepsilon_{ \chi }^2 
+ \mu^2 \, \varepsilon_{ \chi }  \right) 
 \varepsilon_{ \chi }^{  - \frac{ 2 }{ J } } \,
 f_0^{ \prime \prime} (\varepsilon_{\chi} )
\nn & = 
\frac{ e^2 \, \tau \, v_{ z } \, \alpha_J^{ \frac{2} {J} } \, T}
{ 128 \, \mu^{ \frac{ 2 }{ J } } } \, \frac{ \sqrt{ \pi } \, 
\Gamma\big ( 2 - \frac{1}{J} \big ) }
{ \Gamma \big ( \frac{9}{2} - \frac{1}{J} \big ) } 
\left [ \left ( 38 \, J^3 - 105 \, J^2 + 63 \, J - 10 \right )  B_{ \chi x }^2 
- \left ( 6 \, J^3 - 17 \, J^2 + 11 \, J - 2 \right )  B_{ \chi y }^2 \right ].  
\end{align}

Adding up the three partitions, we get
\begin{align}
\label{mommxx}
\ell_{xx}^{2, ( m , \Omega ), \chi}  
& = \frac{ e^2 \, \tau \, v_{ z } \, \alpha_{ J }^{ \frac{2} {J} } \, T}
{ 384 \, \mu^{ \frac{ 2 }{ J } } } \, \frac{ \sqrt \pi 
\, \Gamma\big ( 2 - \frac{1}{J} \big ) }
{ \Gamma \big ( \frac{9}{2} - \frac{1}{J} \big ) } \left [
\left ( 22 \, J^3 - 75 \, J^2 + 41 \,J - 6 \right ) 
\, B_{ \chi x }^2 -\left  ( 2 \, J^3 + 13 \, J^2 - 11 \, J + 2  \right) B_{ \chi y }^2 \right ].  
\end{align}

\subsubsection{Total}

Collecting the long and cumbersome expressions for the individual parts, the final result is
\begin{align}
\label{l28}
\ell_{ x x }^{ m, \chi } & = \ell_{ x x }^{ 1, m, \chi } + \ell_{ x x }^{ 2, m, \chi } + \ell_{ x x }^{ 2, ( \Omega, m ), \chi } \nn
& = \frac{ e^2 \, \tau \, v_z \, \alpha_J^{ \frac{2} {J} } \, T }{ 384 \, J \, \mu^{ \frac{ 2 }{ J } } } \, \frac{ \sqrt{ \pi } \, \Gamma\big ( 2 - \frac{1}{J} \big ) }{ \Gamma \big ( \frac{9}{2} - \frac{1}{J} \big ) } \left[ ( 37 \, J^4 - 100 \, J^3 + 74 \, J^2 - 21 \, J + 2 ) \, B_{ \chi x }^2 + ( 3 \, J^4 - 12 \, J^3 - 4 \, J^2 + 9 \, J - 2 ) \, B_{ \chi y }^2  \right].
\end{align}
On comparing with Eq.~\eqref{eqomegaOMM}, we find that it satisfies the relation
\begin{align}
  \sigma ^{ m , \chi}_{xx} = \frac{ 3 \, e^2 } {\pi^2 \, T} 
  \,  \ell^{ m , \chi}_{xx}   + \mathcal{O} (\beta ^{-2})\, .  
\end{align}

\section{Computation of the transverse components of the magnetothermal coefficient}
\label{appellyx}

In this section, we show the derivation of the explicit expressions for the transverse components $\ell^\chi_{yx} = \ell^\chi_{xy}$ of the magnetothermal coefficient $\ell^\chi $, shown in Sec.~\ref{secell}, starting from the integrals appearing in Eq.~\eqref{eq_thermal}.

\subsection{Field-independent contribution}

Analogous to $\sigma^{ 0, \chi}_{xy}$ and $\alpha^{ 0, \chi}_{xy}$, the $\mathbf B_\chi$-independent contribution of Eq.~\eqref{eq_thermoel} vanishes, as shown below:
\begin{align}
\ell^{ 0, \chi}_{xy} = -
\, \tau  \int \frac{d^3 {\mathbf k} } {(2\, \pi )^3 }  \,  v_{\chi x }^{(0)} \,
 v_{\chi y }^{(0)}   
\,  \frac{ ( \varepsilon_\chi - \mu   )^2   }{T} \, 
 f_{0} ^{\prime} ( \varepsilon_\chi ) = 0\, .
\end{align}

\subsection{Contribution solely from BC (no OMM)}

The contribution to $\ell_{xy}^\chi$ solely from the BC, captured by the term $\ell^{\Omega , \chi}_{xy}  \equiv 
\ell^{1,\Omega , \chi}_{xy}
+ \ell^{2,\Omega , \chi}_{xy}  $ of Eq.~\eqref{eqalphatot}, is given by the sum of

\begin{align}
\label{ll2}
\ell^{ 1, \Omega , \chi}_{xy}
& =- \,e\, \tau
\int\frac{d^3 {\mathbf k}}   {(2\,\pi)^ 3}
\left [ \left ({\boldsymbol  v}_{\chi} ^{(0)} \cdot \mathbf{\Omega } _{\chi} \right )
\left (v_{\chi x }^{(0)}\, B _{\chi y} +  B_{\chi x}  \, v_{\chi y} ^{(0)} \right )
- \left( {\mathbf B} _{\chi} \cdot \mathbf{\Omega } _{\chi}  \right)
v_{\chi x}^{(0)} \, v_{\chi y} ^{(0)} \right ]
\frac{\left( \varepsilon_{\chi}-\mu\right)^{2}}  {T}
\, f_0^\prime (\varepsilon_{\chi})
\nn &
=\frac{\chi \, e \, \tau \, J^3 \, v_z\,\alpha^4_J }
{16\, \pi^3 \, T}
\int\limits_{-\infty}^{\infty}  dk_z
\int \limits_0^{\infty} dk_\perp \,k_{\perp}^{4J-2} \,
\frac{\left(\varepsilon_{\chi}-\mu\right)^2 } {\varepsilon_{\chi}^5}\,
f_0^\prime (\varepsilon_{\chi} )
\nn & \hspace{ 3 cm}\times
\int \limits_0^{2 \, \pi}  d\phi
\left [ B_{\chi x}
\left(\alpha^2_J \, k_{\perp}^{2J} \sin^3 \phi + v_z^2 \, k_z^2 \sin\phi\right)
 + B_{\chi y} \left(\alpha^2_J \, k_{\perp}^{2J} \cos^3\phi + v_z^2 \, k_z^2 \cos\phi\right) 
 \right ]
\nn & = 0\,,
\end{align}
and
\begin{align}
\ell^{ 2, \Omega , \chi}_{xy} & =  -\, e ^2 \, \tau   \int \frac{d^3 {\mathbf k} }
{(2 \,\pi )^3 }  \, Q_{\chi x }  \, Q_{\chi y }   \,  \frac{ ( \varepsilon_\chi - \mu   )^2   }
{T}
\, f_{0} ^{\prime} ( \varepsilon_\chi ) 
\nn &=
- \frac{e ^2 \, \tau \, v_z \, J\, \alpha_{J}^{ \frac{2} {J} } } {64\, \pi^2 \, T } 
 \left ( 13 \, J^2 - 7 \, J +1 \right ) \, 
\frac{
\Gamma \big (2 - \frac{1}{J} \big ) }
{\Gamma \big ( \frac{9}{2} - \frac{1}{J} \big )}  \, B_{\chi x} \,  B_{\chi y}  
 \int_{0} ^{  \infty}  d \varepsilon_\chi \,  \varepsilon_\chi ^{-\frac{2} {J}} 
 \, (\varepsilon_\chi -\mu )^2   f_{0} ^{\prime } ( \varepsilon_\chi ) 
\nn &=  \frac{ \sqrt \pi \, e ^2 \, \tau \, v_z \, J\, \alpha_{J}^{ \frac{2} {J} } \, T}
{ 3 \times 64 \, \mu ^{ \frac{2} {J}} } 
\, \frac{\Gamma \big (2 - \frac{1}{J} \big) }
 {\Gamma \big ( \frac{9}{2} - \frac{1}{J} \big )} 
 \left (13 \,J^{2} - 7 \,J +1 \right) 
  \, B_{\chi x} \,  B_{\chi y} \,.  
\end{align}
Comparing with Eq.~\eqref{eqsigmaxyBC}, we find that
\begin{align}
\sigma ^{ \Omega , \chi}_{xy} = \frac{ 3 \,e^2 }{\pi^2 \, T}  
\,   \ell^{ 2, \Omega , \chi}_{xy}   + \order{ \beta ^{-2} } .  
\end{align}

\subsection{Contribution with the integrand proportional to nonzero powers of OMM}

The contribution to $\ell_{xy}^\chi$, which vanishes if the OMM is omitted, is captured by the term $
\ell^{m, \chi}_{xy}  =  \ell^{1, m , \chi}_{xy} + \ell^{2, m , \chi}_{xy} 
+ \ell^{2, (\Omega, m), \chi}_{xy} $ of Eq.~\eqref{eqelltot}.

\subsubsection{Part one}

We find that the part contributed by a linear-in-OMM term in the integrand is given by
\begin{align}
 \ell^{1, m , \chi}_{xy} 
=  I^{\ell, 1}_{1, xy} + I^{\ell,1}_{2, xy} + I^{\ell,1 }_{3, xy} \,,
\end{align}
where
\begin{align}
 I^{\ell, 1}_{1, xy} & \equiv 
-\,2\,\tau
\int\frac{  d^3{\mathbf k} } {(2\, \pi ) ^{3} }\,v _{\chi x } ^{(0)} 
\, v _{\chi y} ^{(0)}   \,\frac{ \left (\varepsilon_{\chi} - \mu \right )
\varepsilon_{\chi} ^{(m)}} {T}
\,
f_0^\prime ( \varepsilon_{\chi} ) \nn
 & =
-\frac{\chi \, e\, \tau\, J^3 \, v_z\, \alpha_J^6 }
{8 \, \pi^3 \, T}
\int\limits_{-\infty}^{\infty}dk_z
\int\limits_0^{\infty}  dk_{\perp} \, k_{\perp}^{6J-2}\,
\frac{\varepsilon_{\chi}-\mu}{\varepsilon_{\chi}^4}
\, f_0^\prime ( \varepsilon_{\chi} )
\int\limits_0^{2\pi} d\phi
\left[ B_{\chi x} \cos^2\phi \,\sin\phi
+ B_{\chi y} \sin^2\phi \, \cos\phi \right ] = 0 \,,
\end{align}
\begin{align}
I^\ell_{2, xy} & \equiv -\,\tau
\int\frac{  d^3{\mathbf k} }{(2 \,\pi ) ^{3} } 
\left [ v _{\chi x}^{(0)} \,v _{\chi y} ^{(m)} 
+ v _{\chi x} ^{(m)} \, v _{\chi y}^{(0)}\right ]
\frac{(\varepsilon_{\chi}-\mu) ^{2}}{T} \,f_0^\prime ( \varepsilon_{\chi} )
\nn &
=-\frac{\chi \, e\, \tau\, J^2\, v_z\, \alpha_J^4}
{16\,\pi^3\,T}
\int\limits_{-\infty}^{\infty}dk_z
\int\limits_0^{\infty} dk_\perp \,
k_{\perp}^{4J-2}\, \frac{\left(\varepsilon_{\chi}-\mu\right)^2}
{\varepsilon_{\chi}^5}\, f_0^\prime ( \varepsilon_{\chi} )
\nn & \hspace{2 cm} \times
\int\limits_0^{2\pi} d\phi 
\left[
B_{\chi x} \left  \lbrace\left(J \,v_z^2 \, k_z^2 -\varepsilon_{\chi}^2\right)\sin(3\phi)
+J \, v_z^2 \, k_z^2 \sin\phi \right \rbrace
+ B_{\chi y} \left  \lbrace \left(\varepsilon_{\chi}^2-J \,v_z^2 \, k_z^2 \right)\cos(3\phi)
+J \, v_z^2 \, k_z^2 \cos\phi \right \rbrace
\right ]
\nn & =0 \,,
\end{align}
\begin{align}
I^\ell_{3, xy} & \equiv
-\tau\,\int\frac{  d^3{\mathbf k} } {(2 \, \pi ) ^{3} }
v _{\chi x }^{(0)} \, v _{\chi y} ^{(0)}\,
\frac{\left ( \varepsilon_{\chi}-\mu \right )^{2}
\varepsilon_{\chi}^{(m)}}{T} \,
f_0^{\prime \prime}(\varepsilon_{\chi})
\nn &
= -\frac{\chi \, e\, \tau\, J^3 \, v_z\, \alpha_J^6} 
{ 32\, \pi^3  \,T }
\int\limits_{-\infty}^{\infty}
dk_z\int\limits_0^{\infty} dk_{\perp}  \, k_{\perp}^{6J-2}
\, \frac{\left(\varepsilon_{\chi}-\mu\right)^2} {\varepsilon_{\chi}^4}
\, f_0^{\prime \prime} ( \varepsilon_{\chi} )
\int\limits_0^{2\pi} d\phi 
\left [ B_{\chi x}
\cos\phi \, \sin(2\phi) + B_{\chi y} \sin\phi \, \sin(2\phi) \right ] = 0\,.
\end{align}

\subsubsection{Part two}

Using the partitioned integrals defined in Eqs.~\eqref{l2} and \eqref{l3}, we have $ \ell_{xy}^{2, m, \chi} = 
I^{ \ell, 2}_{1, xy } + I^{ \ell, 2}_{ 2, xy } + I^{ \ell, 2}_{ 3, xy}  $, where
\begin{align}
\label{lyx1}
I^{ \ell, 2}_{1, xy } &= -\,
\tau \int\frac{d^3 \mathbf{ k }}{\left( 2\,\pi \right)^3} \, 
\left [ v_{\chi y}^{\left( 0 \right)} \, v_{\chi x}^{\left( 0 \right)} \,
 \frac{ \left (  \varepsilon_{\chi}^{\left( m \right)} \right)^2}{T} 
 + 2 \left \lbrace v_{ \chi y }^{(0)} \, v_{\chi x }^{(m)} 
+ v_{\chi y }^{(m)} \, v_{\chi x }^{(0)} \right \rbrace
\frac{\left( \varepsilon_{\chi} - \mu \right) 
 \varepsilon_{\chi}^{(m)}}{T} + v_{ \chi y }^{( m )} \, v_{ \chi x }^{( m )} 
 \, \frac{ \left( \varepsilon_{\chi} - \mu \right)^2 }{T} \right ] 
\, f_0^{ \prime } (\varepsilon_{\chi} )
\nn & = -\frac{ e^2 \, \tau \, v_z \, \alpha_J^{ \frac{2} {J} } \, J }
{ 64 \, \pi^{ \frac{3} {2} } \, T } \, 
\frac{ \Gamma\big ( 2 - \frac{1}{J} \big ) }
{ \Gamma \big ( \frac{9}{2} - \frac{1}{J} \big ) } 
\, B_{ \chi x } \, B_{ \chi y } 
\int_{ 0 }^{ \infty } d \varepsilon_{ \chi } \left[
\left  ( 17 \, J^2 - 9 \, J + 1 \right ) 
 \varepsilon_{ \chi }^2 + ( 4 \, J - 14 \, J^2 ) \, \mu \, \varepsilon_{ \chi }
 +  
+ 3 \, J^2 \, \mu^2 \right ]
\varepsilon_{ \chi }^{ - \frac{ 2 }{J} } \, 
f_0^{ \prime } (\varepsilon_{\chi} )
\nn & = 
-\frac{ e^2 \, \tau \, v_z \, \alpha_J^{ \frac{2} {J} } \, T }{ 192 \, \pi^{ \frac{3} {2} } \, J \, \mu^{ \frac{ 2 }{ J } } } \, \frac{ \Gamma\big ( 2 - \frac{1}{J} \big ) }{ \Gamma \big ( \frac{9}{2} - \frac{1}{J} \big ) } \, \left[ ( -17 \, J^4 + 43 \, J^3 - 36 \, J^2 + 13 \, J - 2 ) \, \pi^2 -3 \, ( 6 \, J^2 - 5 \, J + 1 ) \, J^2 \, \beta^2 \, \mu^2 \right] \, B_{ \chi x } \, B_{ \chi y } \,,
\end{align}
\begin{align}
\label{lyx4}
I^{ \ell, 2}_{ 2, xy } &= -\,
\tau \int\frac{d^3 \mathbf{ k }}{\left( 2\, \pi \right)^3} \,
 \left[ \left \lbrace v_{\chi y }^{( 0 )} \, v_{\chi x }^{( m )} 
 + v_{\chi y }^{( m )} \, v_{\chi x }^{ (0) } \right \rbrace
  \frac{ \left( \varepsilon_{\chi} - \mu \right)^2 \varepsilon_{\chi}^{ (m) } }{T} 
  + 2 \, v_{\chi y }^{ (0) } \, v_{\chi x }^{ (0) }
  \, \frac{ \left( \varepsilon_{\chi} - \mu \right) \left( \varepsilon_{\chi}^{ (m) } \right)^2}{T} \right] 
  f_0^{ \prime \prime } (\varepsilon_{\chi} )
\nn & = -\frac{ e^2 \, \tau \, v_z \, \alpha_J^{ \frac{2} {J} } \, J 
\, \Gamma\big ( 2 - \frac{1}{J} \big )
}
{ 32 \, \pi^{ \frac{3} {2} } \, T\, \Gamma \big ( \frac{9}{2} - \frac{1}{J} \big ) }  
\, B_{ \chi x } \, B_{ \chi y } 
 \int_{ 0 }^{ \infty } d \varepsilon_{ \chi } \left [ 
\left ( 8 \, J^2 - 6 \, J + 1 \right ) 
 \varepsilon_{ \chi }^2 + \left ( -10 \, J^2 + 7 \, J - 1  \right )
  \mu \, \varepsilon_{ \chi } + 
+ \left ( 2 \, J^2 - J  \right ) 
 \mu^2 \right ] 
 \varepsilon_{ \chi }^{ 1 - \frac{ 2 }{ J } }\, 
 f_0^{ \prime \prime} (\varepsilon_{\chi} )
\nn & = -\frac{ e^2 \, \tau \, v_z \, \alpha_J^{ \frac{2} {J} } \, T }{ 32 \, \pi^{ \frac{3} {2} } \, J \, \mu^{ \frac{ 2 }{ J } } } \, \frac{ \Gamma\big ( 2 - \frac{1}{J} \big ) }
{ \Gamma \big ( \frac{9}{2} - \frac{1}{J} \big ) } 
 \left[ \left ( 8 \, J^4 - 28 \, J^3 + 30 \, J^2 - 13 \, J + 2 \right )  \pi^2
  + \left  ( 6 \, J^2 - 5 \, J + 1 \right ) 
   J^2 \, \beta^2 \, \mu^2 \right] \, B_{ \chi x } \, B_{ \chi y }\,,
\end{align}
and
\begin{align}
\label{lyx7}
 I^{ \ell, 2}_{ 3, xy}   &= - \,\tau 
 \int\frac{d^3 \mathbf{ k }}{\left( 2\, \pi \right)^3} \, 
 \frac{ v_{\chi  y }^{ (0) } \, v_{\chi x }^{ (0) }} {2}  \, 
 \frac{ \left( \varepsilon_{\chi} - \mu \right)^2 \left( \varepsilon_{\chi}^{ (m) } \right)^2 }{T} \, 
 f_0^{ \prime \prime \prime } (\varepsilon_{\chi} )
\nn & = -\frac{ e^2 \, \tau \, v_z \, \alpha_J^{ \frac{2} {J} } \, J }
{ 128 \, \pi^{ \frac{3} {2} } \, T } 
\, 
\frac{ \Gamma\big ( 2 - \frac{1}{J} \big ) \,  B_{ \chi x } \, B_{ \chi y } 
}
{
 \Gamma \big ( \frac{9}{2} - \frac{1}{J} \big ) } 
\int_{ 0 }^{ \infty } d \varepsilon_{ \chi } 
\,\frac{ 
\left ( 6 \, J^2 - 5 \, J + 1 \right )  \varepsilon_{ \chi }^2 
-\left ( 12 \, J^2 - 10 \, J + 2 \right  )  \mu \, \varepsilon_{ \chi }+  
+ \left ( 6 \, J^2 - 5 \, J + 1 \right )  \mu^2  }
{
\varepsilon_{ \chi }^{ \frac{ 2 }{ J } -2 }
}
\,f_0^{ \prime \prime \prime } (\varepsilon_{\chi} )
\nn & = \frac{ e^2 \, \tau \, v_z \, \alpha_J^{ \frac{2} {J} } \, T }
{ 64 \, \pi^{ \frac{3} {2} } 
\, J \, \mu^{ \frac{ 2 }{ J } } } \, \frac{ \Gamma\big ( 2 - \frac{1}{J} \big ) }{ \Gamma \big ( \frac{9}{2} - \frac{1}{J} \big ) } 
 \left[ \left ( 12 \, J^4 - 46 \, J^3 + 56 \, J^2 - 26 \, J + 4 \right)  \pi^2
 + \left ( 6 \, J^2 - 5 \, J + 1 \right ) J^2 \, \beta^2 \, \mu^2 \right] 
  B_{ \chi x } \, B_{ \chi y } \,.
\end{align}

Adding up the three partitions, we obtain
\begin{align}
\label{qommyx}
\ell_{ x y }^{ 2, \, m, \, \chi } 
& = \frac{ e^2 \, \tau \, v_z \, \alpha_J^{ \frac{2} {J} } \, T }
{ 192 \, J \, \mu^{ \frac{ 2 }{ J } } } \,
 \frac{ \sqrt{ \pi } \, \Gamma\big ( 2 - \frac{1}{J} \big ) }
{ \Gamma \big ( \frac{9}{2} - \frac{1}{J} \big ) }
\left ( 5 \, J^4 - 13 \, J^3 + 24 \, J^2 - 13 \, J + 2 \right ) 
 B_{ \chi x } \, B_{ \chi y } \,.
\end{align}

\subsubsection{Part three}

Using the partitioned integrals defined in Eqs.~\eqref{l4} and \eqref{l5}, we have $ \ell_{xy}^{2, ( m , \Omega ), \chi} =
 I^{ \ell, 3}_{ 1,xy}  
+  I^{ \ell, 3}_{ 2, xx} +  I^{ \ell, 3}_{ 3 , xy} $, where
\begin{align}
\label{lyx11}
I^{ \ell, 3}_{ 1,xy}  & = - 2 \, e \, \tau 
\int \frac{ d^3 \mathbf { k }}{( 2 \, \pi )^3} 
\left[ \left ( \boldsymbol{ v }_{\chi}^{ (0) } \cdot \mathbf {  \Omega }_{\chi} \right ) 
 \left \lbrace v_{\chi y}^{( 0 )} \, B_{\chi x } + B_{\chi y } \, 
 v_{\chi x }^{ ( 0 ) }  \right \rbrace
 - \left( \mathbf{ B } _{\chi} \cdot \mathbf { \Omega}_{\chi} \right) 
 \, v_{\chi y }^{( 0 )} \, v_{\chi x }^{ (0) } \right] 
  \frac{ \left( \varepsilon_{\chi} - \mu \right)  
 \varepsilon_{\chi}^{( m )}}{T} \, 
  f_0^{ \prime } (\varepsilon_{\chi} )  
\nn & = \frac{ e^2 \, \tau \, v_z \, \alpha_J^{ \frac{2} {J} } \, J }
{ 32 \, \pi^{ 3/2 } \, T } \, \frac{ \Gamma\big ( 2 - \frac{1}{J} \big ) }
{ \Gamma \big ( \frac{9}{2} - \frac{1}{J} \big ) } \, B_{ \chi x } \, 
B_{ \chi y }   \int_{ 0 }^{ \infty } d \varepsilon_{ \chi } 
\left [ \left  ( 22 \, J^2 - 17 \, J + 3 \right )  \varepsilon_{ \chi }
+ \left  ( -22 \, J^2 + 17 \, J - 3 \right ) 
 \mu  \right ]  
 \varepsilon_{ \chi }^{ 1 - \frac{ 2 }{ J } } \,f_0^{ \prime } (\varepsilon_{\chi} )
\nn &
= -\frac{ e^2 \, \tau \, v_z \, \alpha_J^{ \frac{2} {J} } \, T }
{ 96 \, \mu^{ \frac{ 2 }{ J } } } \, 
\frac{ \sqrt{ \pi } \, \Gamma\big ( 2 - \frac{1}{J} \big ) }
{ \Gamma \big ( \frac{9}{2} - \frac{1}{J} \big ) } 
\left ( 22 \, J^3 - 61 \, J^2 + 37 \, J - 6 \right )  B_{ \chi x } \, B_{ \chi y }\,,
\end{align}
\begin{align}
\label{lyx12}
I^{ \ell, 3}_{ 2 , xy} &=
 - \, e \, \tau \int \frac{ d^3 \mathbf { k }}{( 2 \, \pi )^3} \,
  \Big[ \left(  \boldsymbol{ v }_{\chi}^{ (0) } \cdot \mathbf {  \Omega }_{\chi} \right ) 
  \left \lbrace  v_{\chi y}^{( m )} \, B_{\chi x} 
  + B_{\chi y} \, v_{\chi x}^{ ( m ) }  \right \rbrace 
  +  \left ( \boldsymbol{ v }_{\chi}^{ ( m ) } \cdot \mathbf {  \Omega }_{\chi} \right)
   \left \lbrace v_{\chi y}^{( 0 )} \, B_{\chi x} 
  + B_{\chi y} \, v_{\chi x}^{ (0) }  \right \rbrace
\nn & \hspace{ 3 cm }
-  \left( \mathbf{ B } _{\chi} \cdot \mathbf { \Omega}_{\chi} \right) 
\left \lbrace v_{ \chi y }^{ (0) } \, v_{ \chi x }^{ (m) } 
+ v_{ \chi y }^{ (m) } \, v_{ \chi x }^{ (0) } \right \rbrace 
\Big] \, \frac{ \left( \varepsilon_{\chi} - \mu \right)^2 }{T}
 \, f_0^{ \prime } (\varepsilon_{\chi} )
\nn & = \frac{ e^2 \, \tau \, v_z \, \alpha_J^{ \frac{2} {J} } \, J }
{ 32 \, \pi^{ 3/2 } \, T } \, \frac{ \Gamma\big ( 2 - \frac{1}{J} \big ) }
{ \Gamma \big ( \frac{9}{2} - \frac{1}{J} \big ) } ( 5 \, J^3 - 15 \, J^2 + 11 \, J - 2 ) \, B_{ \chi x } \, B_{ \chi y } \, \int_{ 0 }^{ \infty } d \varepsilon_{ \chi }  \left( \varepsilon_{ \chi }^2 
-2\, \mu \, \varepsilon_{ \chi }
+ \mu^2  \right) \varepsilon_{ \chi }^{ - \frac{ 2 }{ J } }\,
 f_0^{ \prime } (\varepsilon_{\chi} )
\nn &  =  -\frac{ e^2 \, \tau \, v_z \, \alpha_J^{ \frac{2} {J} } \, T }
{ 96 \, \mu^{ \frac{ 2 }{ J } } } \, \frac{ \sqrt{ \pi } \,
 \Gamma\big ( 2 - \frac{1}{J} \big ) }
 { \Gamma \big ( \frac{9}{2} - \frac{1}{J} \big ) } 
 \left  ( 5 \, J^3 - 15 \, J^2 + 11 \, J - 2 \right ) 
 B_{ \chi x } \, B_{ \chi y }\,,
\end{align}
and
\begin{align}
\label{lyx18}
I^{ \ell, 3}_{ 3 , xy}  & = - e \, \tau 
\int \frac{ d^3 \mathbf { k }}{( 2 \, \pi )^3} \,
 \left [ \left( \boldsymbol{ v }_{\chi}^{ (0) } \cdot \mathbf {  \Omega }_{\chi} \right) 
  \left \lbrace  v_{\chi y }^{ (0) }
  \, B_{\chi x } + B_{\chi y } \, v_{\chi x }^{ (0) } \right \rbrace
 - \left( \mathbf {B}_{ \chi } \cdot \mathbf {  \Omega }_{ \chi } \right) 
 v_{ \chi y }^{ (0) } \, v_{ \chi x }^{ (0) } \right ] 
 \frac{ \left( \varepsilon_{\chi} - \mu \right)^2 \, \varepsilon_{ \chi }^{ (m) } }{ T } 
 \, f_0^{ \prime \prime } (\varepsilon_{\chi} )\nn
 & = \frac{ e^2 \, \tau \, v_z \, \alpha_J^{ \frac{2} {J} } \, J }
 { 64 \, \pi^{ 3/2 } \, T } \, 
 \frac{ \Gamma\big ( 2 - \frac{1}{J} \big ) }
 { \Gamma \big ( \frac{9}{2} - \frac{1}{J} \big ) } 
 \left ( 22 \, J^2 - 17 \, J + 3 \right )  B_{ \chi x } \, B_{ \chi y } 
 \int_{ 0 }^{ \infty } d \varepsilon_{ \chi }
  \left( \varepsilon_{ \chi }^2 
 - 2 \, \mu \, \varepsilon_{ \chi } 
 + \mu^2 \right)
 \varepsilon_{ \chi }^{ 1 - \frac{ 2 }{ J } } \, f_0^{ \prime \prime } (\varepsilon_{\chi} )
\nn &  = \frac{ e^2 \, \tau \, v_z \, \alpha_J^{ \frac{2} {J} } \, T }
{ 64 \, \mu^{ \frac{ 2 }{ J } } } \, 
\frac{ \sqrt{ \pi } \, \Gamma\big ( 2 - \frac{1}{J} \big ) }
{ \Gamma \big ( \frac{9}{2} - \frac{1}{J} \big ) } 
\left ( 22 \, J^3 - 61 \, J^2 + 37 \, J - 6 \right ) B_{ \chi x } \, B_{ \chi y }\,.
\end{align}

Adding up the three partitions, we get
\begin{align}
\label{mommxx}
\ell_{xy}^{2, ( m , \Omega ), \chi}  
& =  \frac{ e^2 \, \tau \, v_z \, \alpha_J^{ \frac{2} {J} } \, T }
{ 192 \, \mu^{ \frac{ 2 }{ J } } } \, 
\frac{ \sqrt{ \pi } \, \Gamma\big ( 2 - \frac{1}{J} \big ) }
{ \Gamma \big ( \frac{9}{2} - \frac{1}{J} \big ) } 
\left ( 12 \, J^3 - 31 \, J^2 + 15 \, J - 2 \right) 
 B_{ \chi x } \, B_{ \chi y } \,.  
\end{align}

\subsubsection{Total}

Collecting the long and cumbersome expressions for the individual parts, the final result is
\begin{align}
\label{l28}
\ell_{ x y}^{ m, \chi } & = 
\ell_{ x y }^{ 1, m, \chi } + \ell_{ x y }^{ 2, m, \chi } + \ell_{ x y }^{ 2, ( \Omega, m ), \chi } \nn
& = 
\frac{ e^2 \, \tau \, v_z \, \alpha_J^{ \frac{2} {J} } \, T }{ 192 \, J \, \mu^{ \frac{ 2 }{ J } } } \,
 \frac{ \sqrt{ \pi } \, \Gamma\big ( 2 - \frac{1}{J} \big ) }
 { \Gamma \big ( \frac{9}{2} - \frac{1}{J} \big ) } 
 \left ( 17 \, J^4 - 44 \, J^3 + 39 \, J^2 - 15 \, J + 2 \right ) 
  B_{ \chi x } \, B_{ \chi y } \,.
\end{align}
On comparing with Eq.~\eqref{eqomegaOMM_xy}, we find that it satisfies the relation
\begin{align}
  \sigma ^{ m , \chi}_{xy} = \frac{ 3 \, e^2 } {\pi^2 \, T} 
  \,  \ell^{ m , \chi}_{xy}   + \mathcal{O} (\beta ^{-2})\, .  
\end{align}


 
\bibliography{ref_ruiz}

\begin{thebibliography}{78}%
\makeatletter
\providecommand \@ifxundefined [1]{%
 \@ifx{#1\undefined}
}%
\providecommand \@ifnum [1]{%
 \ifnum #1\expandafter \@firstoftwo
 \else \expandafter \@secondoftwo
 \fi
}%
\providecommand \@ifx [1]{%
 \ifx #1\expandafter \@firstoftwo
 \else \expandafter \@secondoftwo
 \fi
}%
\providecommand \natexlab [1]{#1}%
\providecommand \enquote  [1]{``#1''}%
\providecommand \bibnamefont  [1]{#1}%
\providecommand \bibfnamefont [1]{#1}%
\providecommand \citenamefont [1]{#1}%
\providecommand \href@noop [0]{\@secondoftwo}%
\providecommand \href [0]{\begingroup \@sanitize@url \@href}%
\providecommand \@href[1]{\@@startlink{#1}\@@href}%
\providecommand \@@href[1]{\endgroup#1\@@endlink}%
\providecommand \@sanitize@url [0]{\catcode `\\12\catcode `\$12\catcode
  `\&12\catcode `\#12\catcode `\^12\catcode `\_12\catcode `\%12\relax}%
\providecommand \@@startlink[1]{}%
\providecommand \@@endlink[0]{}%
\providecommand \url  [0]{\begingroup\@sanitize@url \@url }%
\providecommand \@url [1]{\endgroup\@href {#1}{\urlprefix }}%
\providecommand \urlprefix  [0]{URL }%
\providecommand \Eprint [0]{\href }%
\providecommand \doibase [0]{https://doi.org/}%
\providecommand \selectlanguage [0]{\@gobble}%
\providecommand \bibinfo  [0]{\@secondoftwo}%
\providecommand \bibfield  [0]{\@secondoftwo}%
\providecommand \translation [1]{[#1]}%
\providecommand \BibitemOpen [0]{}%
\providecommand \bibitemStop [0]{}%
\providecommand \bibitemNoStop [0]{.\EOS\space}%
\providecommand \EOS [0]{\spacefactor3000\relax}%
\providecommand \BibitemShut  [1]{\csname bibitem#1\endcsname}%
\let\auto@bib@innerbib\@empty
\bibitem [{\citenamefont {Burkov}\ and\ \citenamefont
  {Balents}(2011)}]{burkov11_weyl}%
  \BibitemOpen
  \bibfield  {author} {\bibinfo {author} {\bibfnamefont {A.~A.}\ \bibnamefont
  {Burkov}}\ and\ \bibinfo {author} {\bibfnamefont {L.}~\bibnamefont
  {Balents}},\ }\bibfield  {title} {\bibinfo {title} {Weyl semimetal in a
  topological insulator multilayer},\ }\href
  {https://doi.org/10.1103/PhysRevLett.107.127205} {\bibfield  {journal}
  {\bibinfo  {journal} {Phys. Rev. Lett.}\ }\textbf {\bibinfo {volume} {107}},\
  \bibinfo {pages} {127205} (\bibinfo {year} {2011})}\BibitemShut {NoStop}%
\bibitem [{\citenamefont {Yan}\ and\ \citenamefont
  {Felser}(2017)}]{yan17_topological}%
  \BibitemOpen
  \bibfield  {author} {\bibinfo {author} {\bibfnamefont {B.}~\bibnamefont
  {Yan}}\ and\ \bibinfo {author} {\bibfnamefont {C.}~\bibnamefont {Felser}},\
  }\bibfield  {title} {\bibinfo {title} {Topological materials: {W}eyl
  semimetals},\ }\href
  {https://doi.org/10.1146/annurev-conmatphys-031016-025458} {\bibfield
  {journal} {\bibinfo  {journal} {Annual Review of Condensed Matter Physics}\
  }\textbf {\bibinfo {volume} {8}},\ \bibinfo {pages} {337} (\bibinfo {year}
  {2017})}\BibitemShut {NoStop}%
\bibitem [{\citenamefont {Bradlyn}\ \emph {et~al.}(2016)\citenamefont
  {Bradlyn}, \citenamefont {Cano}, \citenamefont {Wang}, \citenamefont
  {Vergniory}, \citenamefont {Felser}, \citenamefont {Cava},\ and\
  \citenamefont {Bernevig}}]{bernevig}%
  \BibitemOpen
  \bibfield  {author} {\bibinfo {author} {\bibfnamefont {B.}~\bibnamefont
  {Bradlyn}}, \bibinfo {author} {\bibfnamefont {J.}~\bibnamefont {Cano}},
  \bibinfo {author} {\bibfnamefont {Z.}~\bibnamefont {Wang}}, \bibinfo {author}
  {\bibfnamefont {M.~G.}\ \bibnamefont {Vergniory}}, \bibinfo {author}
  {\bibfnamefont {C.}~\bibnamefont {Felser}}, \bibinfo {author} {\bibfnamefont
  {R.~J.}\ \bibnamefont {Cava}},\ and\ \bibinfo {author} {\bibfnamefont
  {B.~A.}\ \bibnamefont {Bernevig}},\ }\bibfield  {title} {\bibinfo {title}
  {{Beyond Dirac and Weyl fermions: Unconventional quasiparticles in
  conventional crystals}},\ }\href
  {https://science.sciencemag.org/content/353/6299/aaf5037} {\bibfield
  {journal} {\bibinfo  {journal} {Science}\ }\textbf {\bibinfo {volume} {353}}
  (\bibinfo {year} {2016})}\BibitemShut {NoStop}%
\bibitem [{\citenamefont {Fang}\ \emph {et~al.}(2012)\citenamefont {Fang},
  \citenamefont {Gilbert}, \citenamefont {Dai},\ and\ \citenamefont
  {Bernevig}}]{bernevig2}%
  \BibitemOpen
  \bibfield  {author} {\bibinfo {author} {\bibfnamefont {C.}~\bibnamefont
  {Fang}}, \bibinfo {author} {\bibfnamefont {M.~J.}\ \bibnamefont {Gilbert}},
  \bibinfo {author} {\bibfnamefont {X.}~\bibnamefont {Dai}},\ and\ \bibinfo
  {author} {\bibfnamefont {B.~A.}\ \bibnamefont {Bernevig}},\ }\bibfield
  {title} {\bibinfo {title} {Multi-{W}eyl topological semimetals stabilized by
  point group symmetry},\ }\href
  {https://doi.org/10.1103/PhysRevLett.108.266802} {\bibfield  {journal}
  {\bibinfo  {journal} {Phys. Rev. Lett.}\ }\textbf {\bibinfo {volume} {108}},\
  \bibinfo {pages} {266802} (\bibinfo {year} {2012})}\BibitemShut {NoStop}%
\bibitem [{\citenamefont {Dantas}\ \emph {et~al.}(2018)\citenamefont {Dantas},
  \citenamefont {Pena-Benitez}, \citenamefont {Roy},\ and\ \citenamefont
  {Sur{\'o}wka}}]{dantas18_magnetotransport}%
  \BibitemOpen
  \bibfield  {author} {\bibinfo {author} {\bibfnamefont {R.}~\bibnamefont
  {Dantas}}, \bibinfo {author} {\bibfnamefont {F.}~\bibnamefont
  {Pena-Benitez}}, \bibinfo {author} {\bibfnamefont {B.}~\bibnamefont {Roy}},\
  and\ \bibinfo {author} {\bibfnamefont {P.}~\bibnamefont {Sur{\'o}wka}},\
  }\bibfield  {title} {\bibinfo {title} {Magnetotransport in multi-{W}eyl
  semimetals: A kinetic theory approach},\ }\href
  {https://doi.org/10.1007/JHEP12(2018)069} {\bibfield  {journal} {\bibinfo
  {journal} {Journal of High Energy Physics}\ }\textbf {\bibinfo {volume}
  {2018}},\ \bibinfo {pages} {1} (\bibinfo {year} {2018})}\BibitemShut
  {NoStop}%
\bibitem [{\citenamefont {Nielsen}\ and\ \citenamefont
  {Ninomiya}(1981)}]{nielsen}%
  \BibitemOpen
  \bibfield  {author} {\bibinfo {author} {\bibfnamefont {H.}~\bibnamefont
  {Nielsen}}\ and\ \bibinfo {author} {\bibfnamefont {M.}~\bibnamefont
  {Ninomiya}},\ }\bibfield  {title} {\bibinfo {title} {A no-go theorem for
  regularizing chiral fermions},\ }\href
  {https://doi.org/https://doi.org/10.1016/0370-2693(81)91026-1} {\bibfield
  {journal} {\bibinfo  {journal} {Physics Letters B}\ }\textbf {\bibinfo
  {volume} {105}},\ \bibinfo {pages} {219} (\bibinfo {year}
  {1981})}\BibitemShut {NoStop}%
\bibitem [{\citenamefont {Huang}\ \emph {et~al.}(2015)\citenamefont {Huang},
  \citenamefont {Zhao}, \citenamefont {Long}, \citenamefont {Wang},
  \citenamefont {Chen}, \citenamefont {Yang}, \citenamefont {Liang},
  \citenamefont {Xue}, \citenamefont {Weng}, \citenamefont {Fang},
  \citenamefont {Dai},\ and\ \citenamefont {Chen}}]{huang15_observation}%
  \BibitemOpen
  \bibfield  {author} {\bibinfo {author} {\bibfnamefont {X.}~\bibnamefont
  {Huang}}, \bibinfo {author} {\bibfnamefont {L.}~\bibnamefont {Zhao}},
  \bibinfo {author} {\bibfnamefont {Y.}~\bibnamefont {Long}}, \bibinfo {author}
  {\bibfnamefont {P.}~\bibnamefont {Wang}}, \bibinfo {author} {\bibfnamefont
  {D.}~\bibnamefont {Chen}}, \bibinfo {author} {\bibfnamefont {Z.}~\bibnamefont
  {Yang}}, \bibinfo {author} {\bibfnamefont {H.}~\bibnamefont {Liang}},
  \bibinfo {author} {\bibfnamefont {M.}~\bibnamefont {Xue}}, \bibinfo {author}
  {\bibfnamefont {H.}~\bibnamefont {Weng}}, \bibinfo {author} {\bibfnamefont
  {Z.}~\bibnamefont {Fang}}, \bibinfo {author} {\bibfnamefont {X.}~\bibnamefont
  {Dai}},\ and\ \bibinfo {author} {\bibfnamefont {G.}~\bibnamefont {Chen}},\
  }\bibfield  {title} {\bibinfo {title} {Observation of the
  chiral-anomaly-induced negative magnetoresistance in 3d {W}eyl semimetal
  {T}a{A}s},\ }\href {https://doi.org/10.1103/PhysRevX.5.031023} {\bibfield
  {journal} {\bibinfo  {journal} {Phys. Rev. X}\ }\textbf {\bibinfo {volume}
  {5}},\ \bibinfo {pages} {031023} (\bibinfo {year} {2015})}\BibitemShut
  {NoStop}%
\bibitem [{\citenamefont {Lv}\ \emph {et~al.}(2015)\citenamefont {Lv},
  \citenamefont {Weng}, \citenamefont {Fu}, \citenamefont {Wang}, \citenamefont
  {Miao}, \citenamefont {Ma}, \citenamefont {Richard}, \citenamefont {Huang},
  \citenamefont {Zhao}, \citenamefont {Chen}, \citenamefont {Fang},
  \citenamefont {Dai}, \citenamefont {Qian},\ and\ \citenamefont
  {Ding}}]{lv_Weyl}%
  \BibitemOpen
  \bibfield  {author} {\bibinfo {author} {\bibfnamefont {B.~Q.}\ \bibnamefont
  {Lv}}, \bibinfo {author} {\bibfnamefont {H.~M.}\ \bibnamefont {Weng}},
  \bibinfo {author} {\bibfnamefont {B.~B.}\ \bibnamefont {Fu}}, \bibinfo
  {author} {\bibfnamefont {X.~P.}\ \bibnamefont {Wang}}, \bibinfo {author}
  {\bibfnamefont {H.}~\bibnamefont {Miao}}, \bibinfo {author} {\bibfnamefont
  {J.}~\bibnamefont {Ma}}, \bibinfo {author} {\bibfnamefont {P.}~\bibnamefont
  {Richard}}, \bibinfo {author} {\bibfnamefont {X.~C.}\ \bibnamefont {Huang}},
  \bibinfo {author} {\bibfnamefont {L.~X.}\ \bibnamefont {Zhao}}, \bibinfo
  {author} {\bibfnamefont {G.~F.}\ \bibnamefont {Chen}}, \bibinfo {author}
  {\bibfnamefont {Z.}~\bibnamefont {Fang}}, \bibinfo {author} {\bibfnamefont
  {X.}~\bibnamefont {Dai}}, \bibinfo {author} {\bibfnamefont {T.}~\bibnamefont
  {Qian}},\ and\ \bibinfo {author} {\bibfnamefont {H.}~\bibnamefont {Ding}},\
  }\bibfield  {title} {\bibinfo {title} {{Experimental Discovery of Weyl
  Semimetal TaAs}},\ }\href {https://doi.org/10.1103/PhysRevX.5.031013}
  {\bibfield  {journal} {\bibinfo  {journal} {Phys. Rev. X}\ }\textbf {\bibinfo
  {volume} {5}},\ \bibinfo {pages} {031013} (\bibinfo {year}
  {2015})}\BibitemShut {NoStop}%
\bibitem [{\citenamefont {Xu}\ \emph {et~al.}(2015)\citenamefont {Xu},
  \citenamefont {Belopolski}, \citenamefont {Alidoust}, \citenamefont
  {Neupane}, \citenamefont {Bian}, \citenamefont {Zhang}, \citenamefont
  {Sankar}, \citenamefont {Chang}, \citenamefont {Yuan}, \citenamefont {Lee},
  \citenamefont {Huang}, \citenamefont {Zheng}, \citenamefont {Ma},
  \citenamefont {Sanchez}, \citenamefont {Wang}, \citenamefont {Bansil},
  \citenamefont {Chou}, \citenamefont {Shibayev}, \citenamefont {Lin},
  \citenamefont {Jia},\ and\ \citenamefont {Hasan}}]{yang_Weyl}%
  \BibitemOpen
  \bibfield  {author} {\bibinfo {author} {\bibfnamefont {S.-Y.}\ \bibnamefont
  {Xu}}, \bibinfo {author} {\bibfnamefont {I.}~\bibnamefont {Belopolski}},
  \bibinfo {author} {\bibfnamefont {N.}~\bibnamefont {Alidoust}}, \bibinfo
  {author} {\bibfnamefont {M.}~\bibnamefont {Neupane}}, \bibinfo {author}
  {\bibfnamefont {G.}~\bibnamefont {Bian}}, \bibinfo {author} {\bibfnamefont
  {C.}~\bibnamefont {Zhang}}, \bibinfo {author} {\bibfnamefont
  {R.}~\bibnamefont {Sankar}}, \bibinfo {author} {\bibfnamefont
  {G.}~\bibnamefont {Chang}}, \bibinfo {author} {\bibfnamefont
  {Z.}~\bibnamefont {Yuan}}, \bibinfo {author} {\bibfnamefont {C.-C.}\
  \bibnamefont {Lee}}, \bibinfo {author} {\bibfnamefont {S.-M.}\ \bibnamefont
  {Huang}}, \bibinfo {author} {\bibfnamefont {H.}~\bibnamefont {Zheng}},
  \bibinfo {author} {\bibfnamefont {J.}~\bibnamefont {Ma}}, \bibinfo {author}
  {\bibfnamefont {D.~S.}\ \bibnamefont {Sanchez}}, \bibinfo {author}
  {\bibfnamefont {B.}~\bibnamefont {Wang}}, \bibinfo {author} {\bibfnamefont
  {A.}~\bibnamefont {Bansil}}, \bibinfo {author} {\bibfnamefont
  {F.}~\bibnamefont {Chou}}, \bibinfo {author} {\bibfnamefont {P.~P.}\
  \bibnamefont {Shibayev}}, \bibinfo {author} {\bibfnamefont {H.}~\bibnamefont
  {Lin}}, \bibinfo {author} {\bibfnamefont {S.}~\bibnamefont {Jia}},\ and\
  \bibinfo {author} {\bibfnamefont {M.~Z.}\ \bibnamefont {Hasan}},\ }\bibfield
  {title} {\bibinfo {title} {{Discovery of a Weyl fermion semimetal and
  topological Fermi arcs}},\ }\href {https://doi.org/10.1126/science.aaa9297}
  {\bibfield  {journal} {\bibinfo  {journal} {Science}\ }\textbf {\bibinfo
  {volume} {349}},\ \bibinfo {pages} {613} (\bibinfo {year}
  {2015})}\BibitemShut {NoStop}%
\bibitem [{\citenamefont {Ruan}\ \emph {et~al.}(2016)\citenamefont {Ruan},
  \citenamefont {Jian}, \citenamefont {Yao}, \citenamefont {Zhang},
  \citenamefont {Zhang},\ and\ \citenamefont {Xing}}]{ruan_Weyl}%
  \BibitemOpen
  \bibfield  {author} {\bibinfo {author} {\bibfnamefont {J.}~\bibnamefont
  {Ruan}}, \bibinfo {author} {\bibfnamefont {S.-K.}\ \bibnamefont {Jian}},
  \bibinfo {author} {\bibfnamefont {H.}~\bibnamefont {Yao}}, \bibinfo {author}
  {\bibfnamefont {H.}~\bibnamefont {Zhang}}, \bibinfo {author} {\bibfnamefont
  {S.-C.}\ \bibnamefont {Zhang}},\ and\ \bibinfo {author} {\bibfnamefont
  {D.}~\bibnamefont {Xing}},\ }\bibfield  {title} {\bibinfo {title}
  {{Symmetry-protected ideal Weyl semimetal in HgTe-class materials}},\ }\href
  {https://doi.org/10.1038/ncomms11136} {\bibfield  {journal} {\bibinfo
  {journal} {Nature Communications}\ }\textbf {\bibinfo {volume} {7}},\
  \bibinfo {pages} {11136} (\bibinfo {year} {2016})}\BibitemShut {NoStop}%
\bibitem [{\citenamefont {Xu}\ \emph {et~al.}(2011)\citenamefont {Xu},
  \citenamefont {Weng}, \citenamefont {Wang}, \citenamefont {Dai},\ and\
  \citenamefont {Fang}}]{Gang2011}%
  \BibitemOpen
  \bibfield  {author} {\bibinfo {author} {\bibfnamefont {G.}~\bibnamefont
  {Xu}}, \bibinfo {author} {\bibfnamefont {H.}~\bibnamefont {Weng}}, \bibinfo
  {author} {\bibfnamefont {Z.}~\bibnamefont {Wang}}, \bibinfo {author}
  {\bibfnamefont {X.}~\bibnamefont {Dai}},\ and\ \bibinfo {author}
  {\bibfnamefont {Z.}~\bibnamefont {Fang}},\ }\bibfield  {title} {\bibinfo
  {title} {Chern semimetal and the quantized anomalous {H}all effect in
  {H}g{C}r$_{2}${S}e$_{4}$},\ }\href
  {https://doi.org/10.1103/PhysRevLett.107.186806} {\bibfield  {journal}
  {\bibinfo  {journal} {Phys. Rev. Lett.}\ }\textbf {\bibinfo {volume} {107}},\
  \bibinfo {pages} {186806} (\bibinfo {year} {2011})}\BibitemShut {NoStop}%
\bibitem [{\citenamefont {Huang}\ \emph {et~al.}(2016)\citenamefont {Huang},
  \citenamefont {Xu}, \citenamefont {Belopolski}, \citenamefont {Lee},
  \citenamefont {Chang}, \citenamefont {Chang}, \citenamefont {Wang},
  \citenamefont {Alidoust}, \citenamefont {Bian}, \citenamefont {Neupane},
  \citenamefont {Sanchez}, \citenamefont {Zheng}, \citenamefont {Jeng},
  \citenamefont {Bansil}, \citenamefont {Neupert}, \citenamefont {Lin},\ and\
  \citenamefont {Hasan}}]{hasan_mweyl16}%
  \BibitemOpen
  \bibfield  {author} {\bibinfo {author} {\bibfnamefont {S.-M.}\ \bibnamefont
  {Huang}}, \bibinfo {author} {\bibfnamefont {S.-Y.}\ \bibnamefont {Xu}},
  \bibinfo {author} {\bibfnamefont {I.}~\bibnamefont {Belopolski}}, \bibinfo
  {author} {\bibfnamefont {C.-C.}\ \bibnamefont {Lee}}, \bibinfo {author}
  {\bibfnamefont {G.}~\bibnamefont {Chang}}, \bibinfo {author} {\bibfnamefont
  {T.-R.}\ \bibnamefont {Chang}}, \bibinfo {author} {\bibfnamefont
  {B.}~\bibnamefont {Wang}}, \bibinfo {author} {\bibfnamefont {N.}~\bibnamefont
  {Alidoust}}, \bibinfo {author} {\bibfnamefont {G.}~\bibnamefont {Bian}},
  \bibinfo {author} {\bibfnamefont {M.}~\bibnamefont {Neupane}}, \bibinfo
  {author} {\bibfnamefont {D.}~\bibnamefont {Sanchez}}, \bibinfo {author}
  {\bibfnamefont {H.}~\bibnamefont {Zheng}}, \bibinfo {author} {\bibfnamefont
  {H.-T.}\ \bibnamefont {Jeng}}, \bibinfo {author} {\bibfnamefont
  {A.}~\bibnamefont {Bansil}}, \bibinfo {author} {\bibfnamefont
  {T.}~\bibnamefont {Neupert}}, \bibinfo {author} {\bibfnamefont
  {H.}~\bibnamefont {Lin}},\ and\ \bibinfo {author} {\bibfnamefont {M.~Z.}\
  \bibnamefont {Hasan}},\ }\bibfield  {title} {\bibinfo {title} {New type of
  {W}eyl semimetal with quadratic double weyl fermions},\ }\href
  {https://doi.org/10.1073/pnas.1514581113} {\bibfield  {journal} {\bibinfo
  {journal} {Proceedings of the National Academy of Sciences}\ }\textbf
  {\bibinfo {volume} {113}},\ \bibinfo {pages} {1180} (\bibinfo {year}
  {2016})}\BibitemShut {NoStop}%
\bibitem [{\citenamefont {{Singh}}\ \emph {et~al.}(2018)\citenamefont
  {{Singh}}, \citenamefont {{Chang}}, \citenamefont {{Chang}}, \citenamefont
  {{Huang}}, \citenamefont {{Su}}, \citenamefont {{Lin}}, \citenamefont
  {{Lin}},\ and\ \citenamefont {{Bansil}}}]{singh18_tunable}%
  \BibitemOpen
  \bibfield  {author} {\bibinfo {author} {\bibfnamefont {B.}~\bibnamefont
  {{Singh}}}, \bibinfo {author} {\bibfnamefont {G.}~\bibnamefont {{Chang}}},
  \bibinfo {author} {\bibfnamefont {T.-R.}\ \bibnamefont {{Chang}}}, \bibinfo
  {author} {\bibfnamefont {S.-M.}\ \bibnamefont {{Huang}}}, \bibinfo {author}
  {\bibfnamefont {C.}~\bibnamefont {{Su}}}, \bibinfo {author} {\bibfnamefont
  {M.-C.}\ \bibnamefont {{Lin}}}, \bibinfo {author} {\bibfnamefont
  {H.}~\bibnamefont {{Lin}}},\ and\ \bibinfo {author} {\bibfnamefont
  {A.}~\bibnamefont {{Bansil}}},\ }\bibfield  {title} {\bibinfo {title}
  {Tunable double-{W}eyl fermion semimetal state in the {S}r{S}i$_{2}$
  materials class},\ }\href {https://doi.org/10.1038/s41598-018-28644-y}
  {\bibfield  {journal} {\bibinfo  {journal} {Scientific Reports}\ }\textbf
  {\bibinfo {volume} {8}},\ \bibinfo {eid} {10540} (\bibinfo {year}
  {2018})}\BibitemShut {NoStop}%
\bibitem [{\citenamefont {Liu}\ and\ \citenamefont
  {Zunger}(2017)}]{liu2017predicted}%
  \BibitemOpen
  \bibfield  {author} {\bibinfo {author} {\bibfnamefont {Q.}~\bibnamefont
  {Liu}}\ and\ \bibinfo {author} {\bibfnamefont {A.}~\bibnamefont {Zunger}},\
  }\bibfield  {title} {\bibinfo {title} {{Predicted realization of cubic
  {Dirac} fermion in quasi-one-dimensional transition-metal
  monochalcogenides}},\ }\href {https://doi.org/10.1103/PhysRevX.7.021019}
  {\bibfield  {journal} {\bibinfo  {journal} {Phys. Rev. X}\ }\textbf {\bibinfo
  {volume} {7}},\ \bibinfo {pages} {021019} (\bibinfo {year}
  {2017})}\BibitemShut {NoStop}%
\bibitem [{\citenamefont {Son}\ and\ \citenamefont
  {Spivak}(2013)}]{son13_chiral}%
  \BibitemOpen
  \bibfield  {author} {\bibinfo {author} {\bibfnamefont {D.~T.}\ \bibnamefont
  {Son}}\ and\ \bibinfo {author} {\bibfnamefont {B.~Z.}\ \bibnamefont
  {Spivak}},\ }\bibfield  {title} {\bibinfo {title} {{Chiral anomaly and
  classical negative magnetoresistance of Weyl metals}},\ }\href
  {https://doi.org/10.1103/PhysRevB.88.104412} {\bibfield  {journal} {\bibinfo
  {journal} {Phys. Rev. B}\ }\textbf {\bibinfo {volume} {88}},\ \bibinfo
  {pages} {104412} (\bibinfo {year} {2013})}\BibitemShut {NoStop}%
\bibitem [{\citenamefont {Burkov}(2017)}]{burkov17_giant}%
  \BibitemOpen
  \bibfield  {author} {\bibinfo {author} {\bibfnamefont {A.~A.}\ \bibnamefont
  {Burkov}},\ }\bibfield  {title} {\bibinfo {title} {Giant planar {H}all effect
  in topological metals},\ }\href {https://doi.org/10.1103/PhysRevB.96.041110}
  {\bibfield  {journal} {\bibinfo  {journal} {Phys. Rev. B}\ }\textbf {\bibinfo
  {volume} {96}},\ \bibinfo {pages} {041110} (\bibinfo {year}
  {2017})}\BibitemShut {NoStop}%
\bibitem [{\citenamefont {Li}\ \emph {et~al.}(2017)\citenamefont {Li},
  \citenamefont {Wang}, \citenamefont {Li}, \citenamefont {Yang}, \citenamefont
  {Shen}, \citenamefont {Sheng}, \citenamefont {Li}, \citenamefont {Lu},
  \citenamefont {Zheng},\ and\ \citenamefont {Xu}}]{li_nmr17}%
  \BibitemOpen
  \bibfield  {author} {\bibinfo {author} {\bibfnamefont {Y.}~\bibnamefont
  {Li}}, \bibinfo {author} {\bibfnamefont {Z.}~\bibnamefont {Wang}}, \bibinfo
  {author} {\bibfnamefont {P.}~\bibnamefont {Li}}, \bibinfo {author}
  {\bibfnamefont {X.}~\bibnamefont {Yang}}, \bibinfo {author} {\bibfnamefont
  {Z.}~\bibnamefont {Shen}}, \bibinfo {author} {\bibfnamefont {F.}~\bibnamefont
  {Sheng}}, \bibinfo {author} {\bibfnamefont {X.}~\bibnamefont {Li}}, \bibinfo
  {author} {\bibfnamefont {Y.}~\bibnamefont {Lu}}, \bibinfo {author}
  {\bibfnamefont {Y.}~\bibnamefont {Zheng}},\ and\ \bibinfo {author}
  {\bibfnamefont {Z.-A.}\ \bibnamefont {Xu}},\ }\bibfield  {title} {\bibinfo
  {title} {{Negative magnetoresistance in {W}eyl semimetals NbAs and NbP:
  Intrinsic chiral anomaly and extrinsic effects}},\ }\href
  {https://doi.org/10.1007/s11467-016-0636-8} {\bibfield  {journal} {\bibinfo
  {journal} {Frontiers of Physics}\ }\textbf {\bibinfo {volume} {12}},\
  \bibinfo {pages} {127205} (\bibinfo {year} {2017})}\BibitemShut {NoStop}%
\bibitem [{\citenamefont {Nandy}\ \emph {et~al.}(2017)\citenamefont {Nandy},
  \citenamefont {Sharma}, \citenamefont {Taraphder},\ and\ \citenamefont
  {Tewari}}]{nandy_2017_chiral}%
  \BibitemOpen
  \bibfield  {author} {\bibinfo {author} {\bibfnamefont {S.}~\bibnamefont
  {Nandy}}, \bibinfo {author} {\bibfnamefont {G.}~\bibnamefont {Sharma}},
  \bibinfo {author} {\bibfnamefont {A.}~\bibnamefont {Taraphder}},\ and\
  \bibinfo {author} {\bibfnamefont {S.}~\bibnamefont {Tewari}},\ }\bibfield
  {title} {\bibinfo {title} {Chiral anomaly as the origin of the planar {H}all
  effect in {W}eyl semimetals},\ }\href
  {https://doi.org/10.1103/PhysRevLett.119.176804} {\bibfield  {journal}
  {\bibinfo  {journal} {Phys. Rev. Lett.}\ }\textbf {\bibinfo {volume} {119}},\
  \bibinfo {pages} {176804} (\bibinfo {year} {2017})}\BibitemShut {NoStop}%
\bibitem [{\citenamefont {Nandy}\ \emph {et~al.}(2018)\citenamefont {Nandy},
  \citenamefont {Taraphder},\ and\ \citenamefont {Tewari}}]{nandy18_Berry}%
  \BibitemOpen
  \bibfield  {author} {\bibinfo {author} {\bibfnamefont {S.}~\bibnamefont
  {Nandy}}, \bibinfo {author} {\bibfnamefont {A.}~\bibnamefont {Taraphder}},\
  and\ \bibinfo {author} {\bibfnamefont {S.}~\bibnamefont {Tewari}},\
  }\bibfield  {title} {\bibinfo {title} {Berry phase theory of planar {H}all
  effect in topological insulators},\ }\href
  {https://doi.org/10.1038/s41598-018-33258-5} {\bibfield  {journal} {\bibinfo
  {journal} {Scientific Reports}\ }\textbf {\bibinfo {volume} {8}},\ \bibinfo
  {pages} {14983} (\bibinfo {year} {2018})}\BibitemShut {NoStop}%
\bibitem [{\citenamefont {Nag}\ and\ \citenamefont {Nandy}(2020)}]{Nag_2020}%
  \BibitemOpen
  \bibfield  {author} {\bibinfo {author} {\bibfnamefont {T.}~\bibnamefont
  {Nag}}\ and\ \bibinfo {author} {\bibfnamefont {S.}~\bibnamefont {Nandy}},\
  }\bibfield  {title} {\bibinfo {title} {Magneto-transport phenomena of
  type-{I} multi-{W}eyl semimetals in co-planar setups},\ }\href
  {https://doi.org/10.1088/1361-648x/abc310} {\bibfield  {journal} {\bibinfo
  {journal} {Journal of Physics: Condensed Matter}\ }\textbf {\bibinfo {volume}
  {33}},\ \bibinfo {pages} {075504} (\bibinfo {year} {2020})}\BibitemShut
  {NoStop}%
\bibitem [{\citenamefont {{Yadav}}\ \emph {et~al.}(2022)\citenamefont
  {{Yadav}}, \citenamefont {{Fazzini}},\ and\ \citenamefont
  {{Mandal}}}]{ips-serena}%
  \BibitemOpen
  \bibfield  {author} {\bibinfo {author} {\bibfnamefont {S.}~\bibnamefont
  {{Yadav}}}, \bibinfo {author} {\bibfnamefont {S.}~\bibnamefont {{Fazzini}}},\
  and\ \bibinfo {author} {\bibfnamefont {I.}~\bibnamefont {{Mandal}}},\
  }\bibfield  {title} {\bibinfo {title} {{Magneto-transport signatures in
  periodically-driven Weyl and multi-Weyl semimetals}},\ }\href
  {https://doi.org/10.1016/j.physe.2022.115444} {\bibfield  {journal} {\bibinfo
   {journal} {Physica E Low-Dimensional Systems and Nanostructures}\ }\textbf
  {\bibinfo {volume} {144}},\ \bibinfo {eid} {115444} (\bibinfo {year}
  {2022})}\BibitemShut {NoStop}%
\bibitem [{\citenamefont {Nielsen}\ and\ \citenamefont
  {Ninomiya}(1983)}]{chiral_ABJ}%
  \BibitemOpen
  \bibfield  {author} {\bibinfo {author} {\bibfnamefont {H.}~\bibnamefont
  {Nielsen}}\ and\ \bibinfo {author} {\bibfnamefont {M.}~\bibnamefont
  {Ninomiya}},\ }\bibfield  {title} {\bibinfo {title} {{The Adler-Bell-Jackiw
  anomaly and Weyl fermions in a crystal}},\ }\href
  {https://doi.org/https://doi.org/10.1016/0370-2693(83)91529-0} {\bibfield
  {journal} {\bibinfo  {journal} {Physics Letters B}\ }\textbf {\bibinfo
  {volume} {130}},\ \bibinfo {pages} {389} (\bibinfo {year}
  {1983})}\BibitemShut {NoStop}%
\bibitem [{\citenamefont {Huang}\ \emph {et~al.}(2017)\citenamefont {Huang},
  \citenamefont {Zhou},\ and\ \citenamefont {Shen}}]{chiral_ano_mWSM}%
  \BibitemOpen
  \bibfield  {author} {\bibinfo {author} {\bibfnamefont {Z.-M.}\ \bibnamefont
  {Huang}}, \bibinfo {author} {\bibfnamefont {J.}~\bibnamefont {Zhou}},\ and\
  \bibinfo {author} {\bibfnamefont {S.-Q.}\ \bibnamefont {Shen}},\ }\bibfield
  {title} {\bibinfo {title} {{Topological responses from chiral anomaly in
  multi-Weyl semimetals}},\ }\href {https://doi.org/10.1103/PhysRevB.96.085201}
  {\bibfield  {journal} {\bibinfo  {journal} {Phys. Rev. B}\ }\textbf {\bibinfo
  {volume} {96}},\ \bibinfo {pages} {085201} (\bibinfo {year}
  {2017})}\BibitemShut {NoStop}%
\bibitem [{\citenamefont {Sharma}\ \emph {et~al.}(2016)\citenamefont {Sharma},
  \citenamefont {Goswami},\ and\ \citenamefont {Tewari}}]{girish1}%
  \BibitemOpen
  \bibfield  {author} {\bibinfo {author} {\bibfnamefont {G.}~\bibnamefont
  {Sharma}}, \bibinfo {author} {\bibfnamefont {P.}~\bibnamefont {Goswami}},\
  and\ \bibinfo {author} {\bibfnamefont {S.}~\bibnamefont {Tewari}},\
  }\bibfield  {title} {\bibinfo {title} {Nernst and magnetothermal conductivity
  in a lattice model of {W}eyl fermions},\ }\href
  {https://doi.org/10.1103/PhysRevB.93.035116} {\bibfield  {journal} {\bibinfo
  {journal} {Phys. Rev. B}\ }\textbf {\bibinfo {volume} {93}},\ \bibinfo
  {pages} {035116} (\bibinfo {year} {2016})}\BibitemShut {NoStop}%
\bibitem [{\citenamefont {Zhang}\ \emph {et~al.}(2016)\citenamefont {Zhang},
  \citenamefont {Lu},\ and\ \citenamefont {Shen}}]{zhang16_linear}%
  \BibitemOpen
  \bibfield  {author} {\bibinfo {author} {\bibfnamefont {S.-B.}\ \bibnamefont
  {Zhang}}, \bibinfo {author} {\bibfnamefont {H.-Z.}\ \bibnamefont {Lu}},\ and\
  \bibinfo {author} {\bibfnamefont {S.-Q.}\ \bibnamefont {Shen}},\ }\bibfield
  {title} {\bibinfo {title} {Linear magnetoconductivity in an intrinsic
  topological weyl semimetal},\ }\href
  {https://doi.org/10.1088/1367-2630/18/5/053039} {\bibfield  {journal}
  {\bibinfo  {journal} {New Journal of Physics}\ }\textbf {\bibinfo {volume}
  {18}},\ \bibinfo {pages} {053039} (\bibinfo {year} {2016})}\BibitemShut
  {NoStop}%
\bibitem [{\citenamefont {Chen}\ and\ \citenamefont
  {Fiete}(2016)}]{chen16_thermoelectric}%
  \BibitemOpen
  \bibfield  {author} {\bibinfo {author} {\bibfnamefont {Q.}~\bibnamefont
  {Chen}}\ and\ \bibinfo {author} {\bibfnamefont {G.~A.}\ \bibnamefont
  {Fiete}},\ }\bibfield  {title} {\bibinfo {title} {Thermoelectric transport in
  double-weyl semimetals},\ }\href {https://doi.org/10.1103/PhysRevB.93.155125}
  {\bibfield  {journal} {\bibinfo  {journal} {Phys. Rev. B}\ }\textbf {\bibinfo
  {volume} {93}},\ \bibinfo {pages} {155125} (\bibinfo {year}
  {2016})}\BibitemShut {NoStop}%
\bibitem [{\citenamefont {Das}\ and\ \citenamefont
  {Agarwal}(2019{\natexlab{a}})}]{das19_linear}%
  \BibitemOpen
  \bibfield  {author} {\bibinfo {author} {\bibfnamefont {K.}~\bibnamefont
  {Das}}\ and\ \bibinfo {author} {\bibfnamefont {A.}~\bibnamefont {Agarwal}},\
  }\bibfield  {title} {\bibinfo {title} {Linear magnetochiral transport in
  tilted type-{I} and type-{II} {W}eyl semimetals},\ }\href
  {https://doi.org/10.1103/PhysRevB.99.085405} {\bibfield  {journal} {\bibinfo
  {journal} {Phys. Rev. B}\ }\textbf {\bibinfo {volume} {99}},\ \bibinfo
  {pages} {085405} (\bibinfo {year} {2019}{\natexlab{a}})}\BibitemShut
  {NoStop}%
\bibitem [{\citenamefont {Das}\ and\ \citenamefont
  {Agarwal}(2019{\natexlab{b}})}]{das19_linear2}%
  \BibitemOpen
  \bibfield  {author} {\bibinfo {author} {\bibfnamefont {K.}~\bibnamefont
  {Das}}\ and\ \bibinfo {author} {\bibfnamefont {A.}~\bibnamefont {Agarwal}},\
  }\bibfield  {title} {\bibinfo {title} {{Berry curvature induced thermopower
  in type-I and type-II Weyl semimetals}},\ }\href
  {https://doi.org/10.1103/PhysRevB.100.085406} {\bibfield  {journal} {\bibinfo
   {journal} {Phys. Rev. B}\ }\textbf {\bibinfo {volume} {100}},\ \bibinfo
  {pages} {085406} (\bibinfo {year} {2019}{\natexlab{b}})}\BibitemShut
  {NoStop}%
\bibitem [{\citenamefont {Das}\ and\ \citenamefont
  {Agarwal}(2020)}]{das20_thermal}%
  \BibitemOpen
  \bibfield  {author} {\bibinfo {author} {\bibfnamefont {K.}~\bibnamefont
  {Das}}\ and\ \bibinfo {author} {\bibfnamefont {A.}~\bibnamefont {Agarwal}},\
  }\bibfield  {title} {\bibinfo {title} {Thermal and gravitational chiral
  anomaly induced magneto-transport in {W}eyl semimetals},\ }\href
  {https://doi.org/10.1103/PhysRevResearch.2.013088} {\bibfield  {journal}
  {\bibinfo  {journal} {Phys. Rev. Res.}\ }\textbf {\bibinfo {volume} {2}},\
  \bibinfo {pages} {013088} (\bibinfo {year} {2020})}\BibitemShut {NoStop}%
\bibitem [{\citenamefont {Das}\ \emph {et~al.}(2022)\citenamefont {Das},
  \citenamefont {Das},\ and\ \citenamefont {Agarwal}}]{das22_nonlinear}%
  \BibitemOpen
  \bibfield  {author} {\bibinfo {author} {\bibfnamefont {S.}~\bibnamefont
  {Das}}, \bibinfo {author} {\bibfnamefont {K.}~\bibnamefont {Das}},\ and\
  \bibinfo {author} {\bibfnamefont {A.}~\bibnamefont {Agarwal}},\ }\bibfield
  {title} {\bibinfo {title} {{Nonlinear magnetoconductivity in Weyl and
  multi-Weyl semimetals in quantizing magnetic field}},\ }\href
  {https://doi.org/10.1103/PhysRevB.105.235408} {\bibfield  {journal} {\bibinfo
   {journal} {Phys. Rev. B}\ }\textbf {\bibinfo {volume} {105}},\ \bibinfo
  {pages} {235408} (\bibinfo {year} {2022})}\BibitemShut {NoStop}%
\bibitem [{\citenamefont {Pal}\ \emph {et~al.}(2022{\natexlab{a}})\citenamefont
  {Pal}, \citenamefont {Dey},\ and\ \citenamefont {Ghosh}}]{pal22a_berry}%
  \BibitemOpen
  \bibfield  {author} {\bibinfo {author} {\bibfnamefont {O.}~\bibnamefont
  {Pal}}, \bibinfo {author} {\bibfnamefont {B.}~\bibnamefont {Dey}},\ and\
  \bibinfo {author} {\bibfnamefont {T.~K.}\ \bibnamefont {Ghosh}},\ }\bibfield
  {title} {\bibinfo {title} {Berry curvature induced magnetotransport in 3d
  noncentrosymmetric metals},\ }\href
  {https://doi.org/10.1088/1361-648X/ac2fd4} {\bibfield  {journal} {\bibinfo
  {journal} {Journal of Physics: Condensed Matter}\ }\textbf {\bibinfo {volume}
  {34}},\ \bibinfo {pages} {025702} (\bibinfo {year}
  {2022}{\natexlab{a}})}\BibitemShut {NoStop}%
\bibitem [{\citenamefont {Pal}\ \emph {et~al.}(2022{\natexlab{b}})\citenamefont
  {Pal}, \citenamefont {Dey},\ and\ \citenamefont {Ghosh}}]{pal22b_berry}%
  \BibitemOpen
  \bibfield  {author} {\bibinfo {author} {\bibfnamefont {O.}~\bibnamefont
  {Pal}}, \bibinfo {author} {\bibfnamefont {B.}~\bibnamefont {Dey}},\ and\
  \bibinfo {author} {\bibfnamefont {T.~K.}\ \bibnamefont {Ghosh}},\ }\bibfield
  {title} {\bibinfo {title} {Berry curvature induced anisotropic
  magnetotransport in a quadratic triple-component fermionic system},\ }\href
  {https://doi.org/10.1088/1361-648X/ac4cee} {\bibfield  {journal} {\bibinfo
  {journal} {Journal of Physics: Condensed Matter}\ }\textbf {\bibinfo {volume}
  {34}},\ \bibinfo {pages} {155702} (\bibinfo {year}
  {2022}{\natexlab{b}})}\BibitemShut {NoStop}%
\bibitem [{\citenamefont {Fu}\ and\ \citenamefont
  {Wang}(2022)}]{fu22_thermoelectric}%
  \BibitemOpen
  \bibfield  {author} {\bibinfo {author} {\bibfnamefont {L.~X.}\ \bibnamefont
  {Fu}}\ and\ \bibinfo {author} {\bibfnamefont {C.~M.}\ \bibnamefont {Wang}},\
  }\bibfield  {title} {\bibinfo {title} {{Thermoelectric transport of
  multi-Weyl semimetals in the quantum limit}},\ }\href
  {https://doi.org/10.1103/PhysRevB.105.035201} {\bibfield  {journal} {\bibinfo
   {journal} {Phys. Rev. B}\ }\textbf {\bibinfo {volume} {105}},\ \bibinfo
  {pages} {035201} (\bibinfo {year} {2022})}\BibitemShut {NoStop}%
\bibitem [{\citenamefont {{Araki}}(2020)}]{araki20_magnetic}%
  \BibitemOpen
  \bibfield  {author} {\bibinfo {author} {\bibfnamefont {Y.}~\bibnamefont
  {{Araki}}},\ }\bibfield  {title} {\bibinfo {title} {Magnetic textures and
  dynamics in magnetic {W}eyl semimetals},\ }\href
  {https://doi.org/10.1002/andp.201900287} {\bibfield  {journal} {\bibinfo
  {journal} {Annalen der Physik}\ }\textbf {\bibinfo {volume} {532}},\ \bibinfo
  {pages} {1900287} (\bibinfo {year} {2020})}\BibitemShut {NoStop}%
\bibitem [{\citenamefont {Mizuta}\ and\ \citenamefont
  {Ishii}(2014)}]{mizuta14_contribution}%
  \BibitemOpen
  \bibfield  {author} {\bibinfo {author} {\bibfnamefont {Y.~P.}\ \bibnamefont
  {Mizuta}}\ and\ \bibinfo {author} {\bibfnamefont {F.}~\bibnamefont {Ishii}},\
  }\bibfield  {title} {\bibinfo {title} {Contribution of {B}erry curvature to
  thermoelectric effects},\ }\href {https://doi.org/10.7566/JPSCP.3.017035}
  {\bibfield  {journal} {\bibinfo  {journal} {Proceedings of the International
  Conference on Strongly Correlated Electron Systems (SCES2013)}\ }\textbf
  {\bibinfo {volume} {3}},\ \bibinfo {pages} {017035} (\bibinfo {year}
  {2014})}\BibitemShut {NoStop}%
\bibitem [{\citenamefont {Knoll}\ \emph {et~al.}(2020)\citenamefont {Knoll},
  \citenamefont {Timm},\ and\ \citenamefont {Meng}}]{timm_omm}%
  \BibitemOpen
  \bibfield  {author} {\bibinfo {author} {\bibfnamefont {A.}~\bibnamefont
  {Knoll}}, \bibinfo {author} {\bibfnamefont {C.}~\bibnamefont {Timm}},\ and\
  \bibinfo {author} {\bibfnamefont {T.}~\bibnamefont {Meng}},\ }\bibfield
  {title} {\bibinfo {title} {{Negative longitudinal magnetoconductance at weak
  fields in Weyl semimetals}},\ }\href
  {https://doi.org/10.1103/PhysRevB.101.201402} {\bibfield  {journal} {\bibinfo
   {journal} {Phys. Rev. B}\ }\textbf {\bibinfo {volume} {101}},\ \bibinfo
  {pages} {201402} (\bibinfo {year} {2020})}\BibitemShut {NoStop}%
\bibitem [{\citenamefont {Medel~Onofre}\ and\ \citenamefont
  {Mart\'{\i}n-Ruiz}(2023)}]{onofre}%
  \BibitemOpen
  \bibfield  {author} {\bibinfo {author} {\bibfnamefont {L.}~\bibnamefont
  {Medel~Onofre}}\ and\ \bibinfo {author} {\bibfnamefont {A.}~\bibnamefont
  {Mart\'{\i}n-Ruiz}},\ }\bibfield  {title} {\bibinfo {title} {Planar hall
  effect in {W}eyl semimetals induced by pseudoelectromagnetic fields},\ }\href
  {https://doi.org/10.1103/PhysRevB.108.155132} {\bibfield  {journal} {\bibinfo
   {journal} {Phys. Rev. B}\ }\textbf {\bibinfo {volume} {108}},\ \bibinfo
  {pages} {155132} (\bibinfo {year} {2023})}\BibitemShut {NoStop}%
\bibitem [{\citenamefont {Ghosh}\ and\ \citenamefont
  {Mandal}(2024{\natexlab{a}})}]{ips-rahul-ph}%
  \BibitemOpen
  \bibfield  {author} {\bibinfo {author} {\bibfnamefont {R.}~\bibnamefont
  {Ghosh}}\ and\ \bibinfo {author} {\bibfnamefont {I.}~\bibnamefont {Mandal}},\
  }\bibfield  {title} {\bibinfo {title} {{Electric and thermoelectric response
  for Weyl and multi-Weyl semimetals in planar Hall configurations including
  the effects of strain}},\ }\href
  {https://doi.org/https://doi.org/10.1016/j.physe.2024.115914} {\bibfield
  {journal} {\bibinfo  {journal} {Physica E: Low-dimensional Systems and
  Nanostructures}\ }\textbf {\bibinfo {volume} {159}},\ \bibinfo {pages}
  {115914} (\bibinfo {year} {2024}{\natexlab{a}})}\BibitemShut {NoStop}%
\bibitem [{\citenamefont {Ghosh}\ and\ \citenamefont
  {Mandal}(2024{\natexlab{b}})}]{ips-rahul-tilt}%
  \BibitemOpen
  \bibfield  {author} {\bibinfo {author} {\bibfnamefont {R.}~\bibnamefont
  {Ghosh}}\ and\ \bibinfo {author} {\bibfnamefont {I.}~\bibnamefont {Mandal}},\
  }\bibfield  {title} {\bibinfo {title} {{Direction-dependent conductivity in
  planar Hall set-ups with tilted Weyl/multi-Weyl semimetals}},\ }\href
  {https://doi.org/10.1088/1361-648X/ad38fa} {\bibfield  {journal} {\bibinfo
  {journal} {Journal of Physics Condensed Matter}\ }\textbf {\bibinfo {volume}
  {36}},\ \bibinfo {pages} {275501} (\bibinfo {year}
  {2024}{\natexlab{b}})}\BibitemShut {NoStop}%
\bibitem [{\citenamefont {Li}\ \emph {et~al.}(2023)\citenamefont {Li},
  \citenamefont {Cao}, \citenamefont {Cui}, \citenamefont {Yu},\ and\
  \citenamefont {Yao}}]{nodal_ph}%
  \BibitemOpen
  \bibfield  {author} {\bibinfo {author} {\bibfnamefont {L.}~\bibnamefont
  {Li}}, \bibinfo {author} {\bibfnamefont {J.}~\bibnamefont {Cao}}, \bibinfo
  {author} {\bibfnamefont {C.}~\bibnamefont {Cui}}, \bibinfo {author}
  {\bibfnamefont {Z.-M.}\ \bibnamefont {Yu}},\ and\ \bibinfo {author}
  {\bibfnamefont {Y.}~\bibnamefont {Yao}},\ }\bibfield  {title} {\bibinfo
  {title} {{Planar Hall effect in topological Weyl and nodal-line
  semimetals}},\ }\href {https://doi.org/10.1103/PhysRevB.108.085120}
  {\bibfield  {journal} {\bibinfo  {journal} {Phys. Rev. B}\ }\textbf {\bibinfo
  {volume} {108}},\ \bibinfo {pages} {085120} (\bibinfo {year}
  {2023})}\BibitemShut {NoStop}%
\bibitem [{\citenamefont {Guinea}\ \emph
  {et~al.}(2010{\natexlab{a}})\citenamefont {Guinea}, \citenamefont
  {Katsnelson},\ and\ \citenamefont {Geim}}]{guinea10_energy}%
  \BibitemOpen
  \bibfield  {author} {\bibinfo {author} {\bibfnamefont {F.}~\bibnamefont
  {Guinea}}, \bibinfo {author} {\bibfnamefont {M.~I.}\ \bibnamefont
  {Katsnelson}},\ and\ \bibinfo {author} {\bibfnamefont {A.}~\bibnamefont
  {Geim}},\ }\bibfield  {title} {\bibinfo {title} {Energy gaps and a zero-field
  quantum {H}all effect in graphene by strain engineering},\ }\href
  {https://doi.org/10.1038/nphys1420} {\bibfield  {journal} {\bibinfo
  {journal} {Nature Physics}\ }\textbf {\bibinfo {volume} {6}},\ \bibinfo
  {pages} {30} (\bibinfo {year} {2010}{\natexlab{a}})}\BibitemShut {NoStop}%
\bibitem [{\citenamefont {Guinea}\ \emph
  {et~al.}(2010{\natexlab{b}})\citenamefont {Guinea}, \citenamefont {Geim},
  \citenamefont {Katsnelson},\ and\ \citenamefont
  {Novoselov}}]{guinea10_generating}%
  \BibitemOpen
  \bibfield  {author} {\bibinfo {author} {\bibfnamefont {F.}~\bibnamefont
  {Guinea}}, \bibinfo {author} {\bibfnamefont {A.~K.}\ \bibnamefont {Geim}},
  \bibinfo {author} {\bibfnamefont {M.~I.}\ \bibnamefont {Katsnelson}},\ and\
  \bibinfo {author} {\bibfnamefont {K.~S.}\ \bibnamefont {Novoselov}},\
  }\bibfield  {title} {\bibinfo {title} {Generating quantizing pseudomagnetic
  fields by bending graphene ribbons},\ }\href
  {https://doi.org/10.1103/PhysRevB.81.035408} {\bibfield  {journal} {\bibinfo
  {journal} {Phys. Rev. B}\ }\textbf {\bibinfo {volume} {81}},\ \bibinfo
  {pages} {035408} (\bibinfo {year} {2010}{\natexlab{b}})}\BibitemShut
  {NoStop}%
\bibitem [{\citenamefont {Low}\ and\ \citenamefont
  {Guinea}(2010)}]{low10_strain}%
  \BibitemOpen
  \bibfield  {author} {\bibinfo {author} {\bibfnamefont {T.}~\bibnamefont
  {Low}}\ and\ \bibinfo {author} {\bibfnamefont {F.}~\bibnamefont {Guinea}},\
  }\bibfield  {title} {\bibinfo {title} {Strain-induced pseudomagnetic field
  for novel graphene electronics},\ }\href {https://doi.org/10.1021/nl1018063}
  {\bibfield  {journal} {\bibinfo  {journal} {Nano letters}\ }\textbf {\bibinfo
  {volume} {10}},\ \bibinfo {pages} {3551} (\bibinfo {year}
  {2010})}\BibitemShut {NoStop}%
\bibitem [{\citenamefont {Cortijo}\ \emph {et~al.}(2015)\citenamefont
  {Cortijo}, \citenamefont {Ferreir\'os}, \citenamefont {Landsteiner},\ and\
  \citenamefont {Vozmediano}}]{landsteiner_gaguge}%
  \BibitemOpen
  \bibfield  {author} {\bibinfo {author} {\bibfnamefont {A.}~\bibnamefont
  {Cortijo}}, \bibinfo {author} {\bibfnamefont {Y.}~\bibnamefont
  {Ferreir\'os}}, \bibinfo {author} {\bibfnamefont {K.}~\bibnamefont
  {Landsteiner}},\ and\ \bibinfo {author} {\bibfnamefont {M.~A.~H.}\
  \bibnamefont {Vozmediano}},\ }\bibfield  {title} {\bibinfo {title} {Elastic
  gauge fields in {W}eyl semimetals},\ }\href
  {https://doi.org/10.1103/PhysRevLett.115.177202} {\bibfield  {journal}
  {\bibinfo  {journal} {Phys. Rev. Lett.}\ }\textbf {\bibinfo {volume} {115}},\
  \bibinfo {pages} {177202} (\bibinfo {year} {2015})}\BibitemShut {NoStop}%
\bibitem [{\citenamefont {Liu}\ \emph {et~al.}(2013)\citenamefont {Liu},
  \citenamefont {Ye},\ and\ \citenamefont {Qi}}]{liu_gauge}%
  \BibitemOpen
  \bibfield  {author} {\bibinfo {author} {\bibfnamefont {C.-X.}\ \bibnamefont
  {Liu}}, \bibinfo {author} {\bibfnamefont {P.}~\bibnamefont {Ye}},\ and\
  \bibinfo {author} {\bibfnamefont {X.-L.}\ \bibnamefont {Qi}},\ }\bibfield
  {title} {\bibinfo {title} {{Chiral gauge field and axial anomaly in a Weyl
  semimetal}},\ }\href {https://doi.org/10.1103/PhysRevB.87.235306} {\bibfield
  {journal} {\bibinfo  {journal} {Phys. Rev. B}\ }\textbf {\bibinfo {volume}
  {87}},\ \bibinfo {pages} {235306} (\bibinfo {year} {2013})}\BibitemShut
  {NoStop}%
\bibitem [{\citenamefont {Pikulin}\ \emph {et~al.}(2016)\citenamefont
  {Pikulin}, \citenamefont {Chen},\ and\ \citenamefont
  {Franz}}]{pikulin_gauge}%
  \BibitemOpen
  \bibfield  {author} {\bibinfo {author} {\bibfnamefont {D.~I.}\ \bibnamefont
  {Pikulin}}, \bibinfo {author} {\bibfnamefont {A.}~\bibnamefont {Chen}},\ and\
  \bibinfo {author} {\bibfnamefont {M.}~\bibnamefont {Franz}},\ }\bibfield
  {title} {\bibinfo {title} {Chiral anomaly from strain-induced gauge fields in
  {D}irac and {W}eyl semimetals},\ }\href
  {https://doi.org/10.1103/PhysRevX.6.041021} {\bibfield  {journal} {\bibinfo
  {journal} {Phys. Rev. X}\ }\textbf {\bibinfo {volume} {6}},\ \bibinfo {pages}
  {041021} (\bibinfo {year} {2016})}\BibitemShut {NoStop}%
\bibitem [{\citenamefont {Arjona}\ and\ \citenamefont
  {Vozmediano}(2018)}]{arjona18_rotational}%
  \BibitemOpen
  \bibfield  {author} {\bibinfo {author} {\bibfnamefont {V.}~\bibnamefont
  {Arjona}}\ and\ \bibinfo {author} {\bibfnamefont {M.~A.}\ \bibnamefont
  {Vozmediano}},\ }\bibfield  {title} {\bibinfo {title} {{Rotational strain in
  Weyl semimetals: A continuum approach}},\ }\href
  {https://doi.org/10.1103/PhysRevB.97.201404} {\bibfield  {journal} {\bibinfo
  {journal} {Physical Review B}\ }\textbf {\bibinfo {volume} {97}},\ \bibinfo
  {pages} {201404} (\bibinfo {year} {2018})}\BibitemShut {NoStop}%
\bibitem [{\citenamefont {Ghosh}\ \emph {et~al.}(2020)\citenamefont {Ghosh},
  \citenamefont {Sinha}, \citenamefont {Nandy},\ and\ \citenamefont
  {Taraphder}}]{ghosh20_chirality}%
  \BibitemOpen
  \bibfield  {author} {\bibinfo {author} {\bibfnamefont {S.}~\bibnamefont
  {Ghosh}}, \bibinfo {author} {\bibfnamefont {D.}~\bibnamefont {Sinha}},
  \bibinfo {author} {\bibfnamefont {S.}~\bibnamefont {Nandy}},\ and\ \bibinfo
  {author} {\bibfnamefont {A.}~\bibnamefont {Taraphder}},\ }\bibfield  {title}
  {\bibinfo {title} {Chirality-dependent planar {H}all effect in inhomogeneous
  {W}eyl semimetals},\ }\href {https://doi.org/10.1103/PhysRevB.102.121105}
  {\bibfield  {journal} {\bibinfo  {journal} {Phys. Rev. B}\ }\textbf {\bibinfo
  {volume} {102}},\ \bibinfo {pages} {121105} (\bibinfo {year}
  {2020})}\BibitemShut {NoStop}%
\bibitem [{\citenamefont {Ahmad}\ \emph {et~al.}(2023)\citenamefont {Ahmad},
  \citenamefont {Raman}, \citenamefont {Tewari},\ and\ \citenamefont
  {Sharma}}]{girish2023}%
  \BibitemOpen
  \bibfield  {author} {\bibinfo {author} {\bibfnamefont {A.}~\bibnamefont
  {Ahmad}}, \bibinfo {author} {\bibfnamefont {K.~V.}\ \bibnamefont {Raman}},
  \bibinfo {author} {\bibfnamefont {S.}~\bibnamefont {Tewari}},\ and\ \bibinfo
  {author} {\bibfnamefont {G.}~\bibnamefont {Sharma}},\ }\bibfield  {title}
  {\bibinfo {title} {{Longitudinal magnetoconductance and the planar Hall
  conductance in inhomogeneous Weyl semimetals}},\ }\href
  {https://doi.org/10.1103/PhysRevB.107.144206} {\bibfield  {journal} {\bibinfo
   {journal} {Phys. Rev. B}\ }\textbf {\bibinfo {volume} {107}},\ \bibinfo
  {pages} {144206} (\bibinfo {year} {2023})}\BibitemShut {NoStop}%
\bibitem [{\citenamefont {Kamboj}\ \emph {et~al.}(2019)\citenamefont {Kamboj},
  \citenamefont {Rana}, \citenamefont {Sirohi}, \citenamefont {Vasdev},
  \citenamefont {Mandal}, \citenamefont {Marik}, \citenamefont {Singh},
  \citenamefont {Das},\ and\ \citenamefont {Sheet}}]{exp_gauge}%
  \BibitemOpen
  \bibfield  {author} {\bibinfo {author} {\bibfnamefont {S.}~\bibnamefont
  {Kamboj}}, \bibinfo {author} {\bibfnamefont {P.~S.}\ \bibnamefont {Rana}},
  \bibinfo {author} {\bibfnamefont {A.}~\bibnamefont {Sirohi}}, \bibinfo
  {author} {\bibfnamefont {A.}~\bibnamefont {Vasdev}}, \bibinfo {author}
  {\bibfnamefont {M.}~\bibnamefont {Mandal}}, \bibinfo {author} {\bibfnamefont
  {S.}~\bibnamefont {Marik}}, \bibinfo {author} {\bibfnamefont {R.~P.}\
  \bibnamefont {Singh}}, \bibinfo {author} {\bibfnamefont {T.}~\bibnamefont
  {Das}},\ and\ \bibinfo {author} {\bibfnamefont {G.}~\bibnamefont {Sheet}},\
  }\bibfield  {title} {\bibinfo {title} {{Generation of strain-induced
  pseudo-magnetic field in a doped type-II Weyl semimetal}},\ }\href
  {https://doi.org/10.1103/PhysRevB.100.115105} {\bibfield  {journal} {\bibinfo
   {journal} {Phys. Rev. B}\ }\textbf {\bibinfo {volume} {100}},\ \bibinfo
  {pages} {115105} (\bibinfo {year} {2019})}\BibitemShut {NoStop}%
\bibitem [{\citenamefont {Xiao}\ \emph {et~al.}(2010)\citenamefont {Xiao},
  \citenamefont {Chang},\ and\ \citenamefont {Niu}}]{xiao_review}%
  \BibitemOpen
  \bibfield  {author} {\bibinfo {author} {\bibfnamefont {D.}~\bibnamefont
  {Xiao}}, \bibinfo {author} {\bibfnamefont {M.-C.}\ \bibnamefont {Chang}},\
  and\ \bibinfo {author} {\bibfnamefont {Q.}~\bibnamefont {Niu}},\ }\bibfield
  {title} {\bibinfo {title} {Berry phase effects on electronic properties},\
  }\href {https://doi.org/10.1103/RevModPhys.82.1959} {\bibfield  {journal}
  {\bibinfo  {journal} {Rev. Mod. Phys.}\ }\textbf {\bibinfo {volume} {82}},\
  \bibinfo {pages} {1959} (\bibinfo {year} {2010})}\BibitemShut {NoStop}%
\bibitem [{\citenamefont {Sundaram}\ and\ \citenamefont
  {Niu}(1999)}]{sundaram99_wavepacket}%
  \BibitemOpen
  \bibfield  {author} {\bibinfo {author} {\bibfnamefont {G.}~\bibnamefont
  {Sundaram}}\ and\ \bibinfo {author} {\bibfnamefont {Q.}~\bibnamefont {Niu}},\
  }\bibfield  {title} {\bibinfo {title} {{Wave-packet dynamics in slowly
  perturbed crystals: Gradient corrections and Berry-phase effects}},\ }\href
  {https://doi.org/10.1103/PhysRevB.59.14915} {\bibfield  {journal} {\bibinfo
  {journal} {Phys. Rev. B}\ }\textbf {\bibinfo {volume} {59}},\ \bibinfo
  {pages} {14915} (\bibinfo {year} {1999})}\BibitemShut {NoStop}%
\bibitem [{\citenamefont {Xiao}\ \emph {et~al.}(2007)\citenamefont {Xiao},
  \citenamefont {Yao},\ and\ \citenamefont {Niu}}]{xiao07_valley}%
  \BibitemOpen
  \bibfield  {author} {\bibinfo {author} {\bibfnamefont {D.}~\bibnamefont
  {Xiao}}, \bibinfo {author} {\bibfnamefont {W.}~\bibnamefont {Yao}},\ and\
  \bibinfo {author} {\bibfnamefont {Q.}~\bibnamefont {Niu}},\ }\bibfield
  {title} {\bibinfo {title} {Valley-contrasting physics in graphene: Magnetic
  moment and topological transport},\ }\href
  {https://doi.org/10.1103/PhysRevLett.99.236809} {\bibfield  {journal}
  {\bibinfo  {journal} {Phys. Rev. Lett.}\ }\textbf {\bibinfo {volume} {99}},\
  \bibinfo {pages} {236809} (\bibinfo {year} {2007})}\BibitemShut {NoStop}%
\bibitem [{\citenamefont {K\"onye}\ and\ \citenamefont
  {Ogata}(2021)}]{konye21_microscopic}%
  \BibitemOpen
  \bibfield  {author} {\bibinfo {author} {\bibfnamefont {V.}~\bibnamefont
  {K\"onye}}\ and\ \bibinfo {author} {\bibfnamefont {M.}~\bibnamefont
  {Ogata}},\ }\bibfield  {title} {\bibinfo {title} {Microscopic theory of
  magnetoconductivity at low magnetic fields in terms of {Berry} curvature and
  orbital magnetic moment},\ }\href
  {https://doi.org/10.1103/PhysRevResearch.3.033076} {\bibfield  {journal}
  {\bibinfo  {journal} {Phys. Rev. Res.}\ }\textbf {\bibinfo {volume} {3}},\
  \bibinfo {pages} {033076} (\bibinfo {year} {2021})}\BibitemShut {NoStop}%
\bibitem [{\citenamefont {Ashcroft}\ and\ \citenamefont
  {Mermin}(2011)}]{mermin}%
  \BibitemOpen
  \bibfield  {author} {\bibinfo {author} {\bibfnamefont {N.}~\bibnamefont
  {Ashcroft}}\ and\ \bibinfo {author} {\bibfnamefont {N.}~\bibnamefont
  {Mermin}},\ }\href {https://books.google.de/books?id=x\_s\_YAAACAAJ} {\emph
  {\bibinfo {title} {Solid State Physics}}}\ (\bibinfo  {publisher} {Cengage
  Learning},\ \bibinfo {year} {2011})\BibitemShut {NoStop}%
\bibitem [{\citenamefont {{Mandal}}\ and\ \citenamefont
  {{Saha}}(2024)}]{ips-kush-review}%
  \BibitemOpen
  \bibfield  {author} {\bibinfo {author} {\bibfnamefont {I.}~\bibnamefont
  {{Mandal}}}\ and\ \bibinfo {author} {\bibfnamefont {K.}~\bibnamefont
  {{Saha}}},\ }\bibfield  {title} {\bibinfo {title} {{Thermoelectric response
  in nodal-point semimetals}},\ }\href {https://doi.org/10.1002/andp.202400016}
  {\bibfield  {journal} {\bibinfo  {journal} {Annalen der Physik}\ }\textbf
  {\bibinfo {volume} {536}},\ \bibinfo {pages} {2400016} (\bibinfo {year}
  {2024})}\BibitemShut {NoStop}%
\bibitem [{\citenamefont {Nag}\ \emph {et~al.}(2020)\citenamefont {Nag},
  \citenamefont {Menon},\ and\ \citenamefont {Basu}}]{Nag_floquet_2020}%
  \BibitemOpen
  \bibfield  {author} {\bibinfo {author} {\bibfnamefont {T.}~\bibnamefont
  {Nag}}, \bibinfo {author} {\bibfnamefont {A.}~\bibnamefont {Menon}},\ and\
  \bibinfo {author} {\bibfnamefont {B.}~\bibnamefont {Basu}},\ }\bibfield
  {title} {\bibinfo {title} {{Thermoelectric transport properties of Floquet
  multi-Weyl semimetals}},\ }\href
  {https://doi.org/10.1103/PhysRevB.102.014307} {\bibfield  {journal} {\bibinfo
   {journal} {Phys. Rev. B}\ }\textbf {\bibinfo {volume} {102}},\ \bibinfo
  {pages} {014307} (\bibinfo {year} {2020})}\BibitemShut {NoStop}%
\bibitem [{\citenamefont {Watzman}\ \emph {et~al.}(2018)\citenamefont
  {Watzman}, \citenamefont {McCormick}, \citenamefont {Shekhar}, \citenamefont
  {Wu}, \citenamefont {Sun}, \citenamefont {Prakash}, \citenamefont {Felser},
  \citenamefont {Trivedi},\ and\ \citenamefont {Heremans}}]{watzman18_dirac}%
  \BibitemOpen
  \bibfield  {author} {\bibinfo {author} {\bibfnamefont {S.~J.}\ \bibnamefont
  {Watzman}}, \bibinfo {author} {\bibfnamefont {T.~M.}\ \bibnamefont
  {McCormick}}, \bibinfo {author} {\bibfnamefont {C.}~\bibnamefont {Shekhar}},
  \bibinfo {author} {\bibfnamefont {S.-C.}\ \bibnamefont {Wu}}, \bibinfo
  {author} {\bibfnamefont {Y.}~\bibnamefont {Sun}}, \bibinfo {author}
  {\bibfnamefont {A.}~\bibnamefont {Prakash}}, \bibinfo {author} {\bibfnamefont
  {C.}~\bibnamefont {Felser}}, \bibinfo {author} {\bibfnamefont
  {N.}~\bibnamefont {Trivedi}},\ and\ \bibinfo {author} {\bibfnamefont {J.~P.}\
  \bibnamefont {Heremans}},\ }\bibfield  {title} {\bibinfo {title} {{Dirac
  dispersion generates unusually large Nernst effect in Weyl semimetals}},\
  }\href {https://doi.org/10.1103/PhysRevB.97.161404} {\bibfield  {journal}
  {\bibinfo  {journal} {Phys. Rev. B}\ }\textbf {\bibinfo {volume} {97}},\
  \bibinfo {pages} {161404} (\bibinfo {year} {2018})}\BibitemShut {NoStop}%
\bibitem [{\citenamefont {Xiao}\ \emph {et~al.}(2006)\citenamefont {Xiao},
  \citenamefont {Yao}, \citenamefont {Fang},\ and\ \citenamefont
  {Niu}}]{prl_niu}%
  \BibitemOpen
  \bibfield  {author} {\bibinfo {author} {\bibfnamefont {D.}~\bibnamefont
  {Xiao}}, \bibinfo {author} {\bibfnamefont {Y.}~\bibnamefont {Yao}}, \bibinfo
  {author} {\bibfnamefont {Z.}~\bibnamefont {Fang}},\ and\ \bibinfo {author}
  {\bibfnamefont {Q.}~\bibnamefont {Niu}},\ }\bibfield  {title} {\bibinfo
  {title} {Berry-phase effect in anomalous thermoelectric transport},\ }\href
  {https://doi.org/10.1103/PhysRevLett.97.026603} {\bibfield  {journal}
  {\bibinfo  {journal} {Phys. Rev. Lett.}\ }\textbf {\bibinfo {volume} {97}},\
  \bibinfo {pages} {026603} (\bibinfo {year} {2006})}\BibitemShut {NoStop}%
\bibitem [{\citenamefont {{Li}}\ \emph {et~al.}(2016)\citenamefont {{Li}},
  \citenamefont {{Kharzeev}}, \citenamefont {{Zhang}}, \citenamefont {{Huang}},
  \citenamefont {{Pletikosi{\'c}}}, \citenamefont {{Fedorov}}, \citenamefont
  {{Zhong}}, \citenamefont {{Schneeloch}}, \citenamefont {{Gu}},\ and\
  \citenamefont {{Valla}}}]{li_2016}%
  \BibitemOpen
  \bibfield  {author} {\bibinfo {author} {\bibfnamefont {Q.}~\bibnamefont
  {{Li}}}, \bibinfo {author} {\bibfnamefont {D.~E.}\ \bibnamefont
  {{Kharzeev}}}, \bibinfo {author} {\bibfnamefont {C.}~\bibnamefont {{Zhang}}},
  \bibinfo {author} {\bibfnamefont {Y.}~\bibnamefont {{Huang}}}, \bibinfo
  {author} {\bibfnamefont {I.}~\bibnamefont {{Pletikosi{\'c}}}}, \bibinfo
  {author} {\bibfnamefont {A.~V.}\ \bibnamefont {{Fedorov}}}, \bibinfo {author}
  {\bibfnamefont {R.~D.}\ \bibnamefont {{Zhong}}}, \bibinfo {author}
  {\bibfnamefont {J.~A.}\ \bibnamefont {{Schneeloch}}}, \bibinfo {author}
  {\bibfnamefont {G.~D.}\ \bibnamefont {{Gu}}},\ and\ \bibinfo {author}
  {\bibfnamefont {T.}~\bibnamefont {{Valla}}},\ }\bibfield  {title} {\bibinfo
  {title} {{Chiral magnetic effect in ZrTe$_{5}$}},\ }\href
  {https://doi.org/10.1038/nphys3648} {\bibfield  {journal} {\bibinfo
  {journal} {Nature Physics}\ }\textbf {\bibinfo {volume} {12}},\ \bibinfo
  {pages} {550} (\bibinfo {year} {2016})}\BibitemShut {NoStop}%
\bibitem [{\citenamefont {{Zhang}}\ \emph {et~al.}(2016)\citenamefont
  {{Zhang}}, \citenamefont {{Xu}}, \citenamefont {{Belopolski}}, \citenamefont
  {{Yuan}}, \citenamefont {{Lin}}, \citenamefont {{Tong}}, \citenamefont
  {{Bian}}, \citenamefont {{Alidoust}}, \citenamefont {{Lee}}, \citenamefont
  {{Huang}}, \citenamefont {{Chang}}, \citenamefont {{Chang}}, \citenamefont
  {{Hsu}}, \citenamefont {{Jeng}}, \citenamefont {{Neupane}}, \citenamefont
  {{Sanchez}}, \citenamefont {{Zheng}}, \citenamefont {{Wang}}, \citenamefont
  {{Lin}}, \citenamefont {{Zhang}}, \citenamefont {{Lu}}, \citenamefont
  {{Shen}}, \citenamefont {{Neupert}}, \citenamefont {{Zahid Hasan}},\ and\
  \citenamefont {{Jia}}}]{cheng-long}%
  \BibitemOpen
  \bibfield  {author} {\bibinfo {author} {\bibfnamefont {C.-L.}\ \bibnamefont
  {{Zhang}}}, \bibinfo {author} {\bibfnamefont {S.-Y.}\ \bibnamefont {{Xu}}},
  \bibinfo {author} {\bibfnamefont {I.}~\bibnamefont {{Belopolski}}}, \bibinfo
  {author} {\bibfnamefont {Z.}~\bibnamefont {{Yuan}}}, \bibinfo {author}
  {\bibfnamefont {Z.}~\bibnamefont {{Lin}}}, \bibinfo {author} {\bibfnamefont
  {B.}~\bibnamefont {{Tong}}}, \bibinfo {author} {\bibfnamefont
  {G.}~\bibnamefont {{Bian}}}, \bibinfo {author} {\bibfnamefont
  {N.}~\bibnamefont {{Alidoust}}}, \bibinfo {author} {\bibfnamefont {C.-C.}\
  \bibnamefont {{Lee}}}, \bibinfo {author} {\bibfnamefont {S.-M.}\ \bibnamefont
  {{Huang}}}, \bibinfo {author} {\bibfnamefont {T.-R.}\ \bibnamefont
  {{Chang}}}, \bibinfo {author} {\bibfnamefont {G.}~\bibnamefont {{Chang}}},
  \bibinfo {author} {\bibfnamefont {C.-H.}\ \bibnamefont {{Hsu}}}, \bibinfo
  {author} {\bibfnamefont {H.-T.}\ \bibnamefont {{Jeng}}}, \bibinfo {author}
  {\bibfnamefont {M.}~\bibnamefont {{Neupane}}}, \bibinfo {author}
  {\bibfnamefont {D.~S.}\ \bibnamefont {{Sanchez}}}, \bibinfo {author}
  {\bibfnamefont {H.}~\bibnamefont {{Zheng}}}, \bibinfo {author} {\bibfnamefont
  {J.}~\bibnamefont {{Wang}}}, \bibinfo {author} {\bibfnamefont
  {H.}~\bibnamefont {{Lin}}}, \bibinfo {author} {\bibfnamefont
  {C.}~\bibnamefont {{Zhang}}}, \bibinfo {author} {\bibfnamefont {H.-Z.}\
  \bibnamefont {{Lu}}}, \bibinfo {author} {\bibfnamefont {S.-Q.}\ \bibnamefont
  {{Shen}}}, \bibinfo {author} {\bibfnamefont {T.}~\bibnamefont {{Neupert}}},
  \bibinfo {author} {\bibfnamefont {M.}~\bibnamefont {{Zahid Hasan}}},\ and\
  \bibinfo {author} {\bibfnamefont {S.}~\bibnamefont {{Jia}}},\ }\bibfield
  {title} {\bibinfo {title} {{Signatures of the Adler-Bell-Jackiw chiral
  anomaly in a Weyl fermion semimetal}},\ }\href
  {https://doi.org/10.1038/ncomms10735} {\bibfield  {journal} {\bibinfo
  {journal} {Nature Communications}\ }\textbf {\bibinfo {volume} {7}},\
  \bibinfo {eid} {10735} (\bibinfo {year} {2016})}\BibitemShut {NoStop}%
\bibitem [{\citenamefont {Shama}\ \emph {et~al.}(2020)\citenamefont {Shama},
  \citenamefont {Gopal},\ and\ \citenamefont {Singh}}]{shama}%
  \BibitemOpen
  \bibfield  {author} {\bibinfo {author} {\bibnamefont {Shama}}, \bibinfo
  {author} {\bibfnamefont {R.}~\bibnamefont {Gopal}},\ and\ \bibinfo {author}
  {\bibfnamefont {Y.}~\bibnamefont {Singh}},\ }\bibfield  {title} {\bibinfo
  {title} {{Observation of planar Hall effect in the ferromagnetic Weyl
  semimetal Co$_3$Sn$_2$S$_2$}},\ }\href
  {https://doi.org/https://doi.org/10.1016/j.jmmm.2020.166547} {\bibfield
  {journal} {\bibinfo  {journal} {Journal of Magnetism and Magnetic Materials}\
  }\textbf {\bibinfo {volume} {502}},\ \bibinfo {pages} {166547} (\bibinfo
  {year} {2020})}\BibitemShut {NoStop}%
\bibitem [{\citenamefont {Tanwar}\ \emph {et~al.}(2023)\citenamefont {Tanwar},
  \citenamefont {Ahmad}, \citenamefont {Alam}, \citenamefont {Yao},
  \citenamefont {Tafti},\ and\ \citenamefont {Matusiak}}]{marcin}%
  \BibitemOpen
  \bibfield  {author} {\bibinfo {author} {\bibfnamefont {P.~K.}\ \bibnamefont
  {Tanwar}}, \bibinfo {author} {\bibfnamefont {M.}~\bibnamefont {Ahmad}},
  \bibinfo {author} {\bibfnamefont {M.~S.}\ \bibnamefont {Alam}}, \bibinfo
  {author} {\bibfnamefont {X.}~\bibnamefont {Yao}}, \bibinfo {author}
  {\bibfnamefont {F.}~\bibnamefont {Tafti}},\ and\ \bibinfo {author}
  {\bibfnamefont {M.}~\bibnamefont {Matusiak}},\ }\bibfield  {title} {\bibinfo
  {title} {{Gravitational anomaly in the ferrimagnetic topological Weyl
  semimetal NdAlSi}},\ }\href {https://doi.org/10.1103/PhysRevB.108.L161106}
  {\bibfield  {journal} {\bibinfo  {journal} {Phys. Rev. B}\ }\textbf {\bibinfo
  {volume} {108}},\ \bibinfo {pages} {L161106} (\bibinfo {year}
  {2023})}\BibitemShut {NoStop}%
\bibitem [{\citenamefont {Diaz}\ \emph {et~al.}(2021)\citenamefont {Diaz},
  \citenamefont {Putzke}, \citenamefont {Huang}, \citenamefont {Estry},
  \citenamefont {Analytis}, \citenamefont {Sabsovich}, \citenamefont {Grushin},
  \citenamefont {Ilan},\ and\ \citenamefont {Moll}}]{diaz_2022}%
  \BibitemOpen
  \bibfield  {author} {\bibinfo {author} {\bibfnamefont {J.}~\bibnamefont
  {Diaz}}, \bibinfo {author} {\bibfnamefont {C.}~\bibnamefont {Putzke}},
  \bibinfo {author} {\bibfnamefont {X.}~\bibnamefont {Huang}}, \bibinfo
  {author} {\bibfnamefont {A.}~\bibnamefont {Estry}}, \bibinfo {author}
  {\bibfnamefont {J.~G.}\ \bibnamefont {Analytis}}, \bibinfo {author}
  {\bibfnamefont {D.}~\bibnamefont {Sabsovich}}, \bibinfo {author}
  {\bibfnamefont {A.~G.}\ \bibnamefont {Grushin}}, \bibinfo {author}
  {\bibfnamefont {R.}~\bibnamefont {Ilan}},\ and\ \bibinfo {author}
  {\bibfnamefont {P.~J.~W.}\ \bibnamefont {Moll}},\ }\bibfield  {title}
  {\bibinfo {title} {{Bending strain in 3D topological semi-metals}},\ }\href
  {https://doi.org/10.1088/1361-6463/ac357f} {\bibfield  {journal} {\bibinfo
  {journal} {Journal of Physics D: Applied Physics}\ }\textbf {\bibinfo
  {volume} {55}},\ \bibinfo {pages} {084001} (\bibinfo {year}
  {2021})}\BibitemShut {NoStop}%
\bibitem [{\citenamefont {Ghosh}\ \emph {et~al.}(2024)\citenamefont {Ghosh},
  \citenamefont {Haidar},\ and\ \citenamefont {Mandal}}]{ips-rsw}%
  \BibitemOpen
  \bibfield  {author} {\bibinfo {author} {\bibfnamefont {R.}~\bibnamefont
  {Ghosh}}, \bibinfo {author} {\bibfnamefont {F.}~\bibnamefont {Haidar}},\ and\
  \bibinfo {author} {\bibfnamefont {I.}~\bibnamefont {Mandal}},\ }\bibfield
  {title} {\bibinfo {title} {{Linear response in planar Hall and thermal Hall
  setups for Rarita-Schwinger-Weyl semimetals}},\ }\href
  {https://arxiv.org/abs/2408.01422} {\bibfield  {journal} {\bibinfo  {journal}
  {arXiv e-prints}\ } (\bibinfo {year} {2024})},\ \Eprint
  {https://arxiv.org/abs/2408.01422} {arXiv:2408.01422 [cond-mat.mes-hall]}
  \BibitemShut {NoStop}%
\bibitem [{\citenamefont {Mandal}\ and\ \citenamefont {Saha}(2020)}]{ips-kush}%
  \BibitemOpen
  \bibfield  {author} {\bibinfo {author} {\bibfnamefont {I.}~\bibnamefont
  {Mandal}}\ and\ \bibinfo {author} {\bibfnamefont {K.}~\bibnamefont {Saha}},\
  }\bibfield  {title} {\bibinfo {title} {Thermopower in an anisotropic
  two-dimensional {W}eyl semimetal},\ }\href
  {https://doi.org/10.1103/PhysRevB.101.045101} {\bibfield  {journal} {\bibinfo
   {journal} {Phys. Rev. B}\ }\textbf {\bibinfo {volume} {101}},\ \bibinfo
  {pages} {045101} (\bibinfo {year} {2020})}\BibitemShut {NoStop}%
\bibitem [{\citenamefont {St\aa{}lhammar}\ \emph {et~al.}(2020)\citenamefont
  {St\aa{}lhammar}, \citenamefont {Larana-Aragon}, \citenamefont {Knolle},\
  and\ \citenamefont {Bergholtz}}]{staalhammar20_magneto}%
  \BibitemOpen
  \bibfield  {author} {\bibinfo {author} {\bibfnamefont {M.}~\bibnamefont
  {St\aa{}lhammar}}, \bibinfo {author} {\bibfnamefont {J.}~\bibnamefont
  {Larana-Aragon}}, \bibinfo {author} {\bibfnamefont {J.}~\bibnamefont
  {Knolle}},\ and\ \bibinfo {author} {\bibfnamefont {E.~J.}\ \bibnamefont
  {Bergholtz}},\ }\bibfield  {title} {\bibinfo {title} {Magneto-optical
  conductivity in generic {W}eyl semimetals},\ }\href
  {https://doi.org/10.1103/PhysRevB.102.235134} {\bibfield  {journal} {\bibinfo
   {journal} {Phys. Rev. B}\ }\textbf {\bibinfo {volume} {102}},\ \bibinfo
  {pages} {235134} (\bibinfo {year} {2020})}\BibitemShut {NoStop}%
\bibitem [{\citenamefont {Yadav}\ \emph {et~al.}(2023)\citenamefont {Yadav},
  \citenamefont {Sekh},\ and\ \citenamefont {Mandal}}]{yadav23_magneto}%
  \BibitemOpen
  \bibfield  {author} {\bibinfo {author} {\bibfnamefont {S.}~\bibnamefont
  {Yadav}}, \bibinfo {author} {\bibfnamefont {S.}~\bibnamefont {Sekh}},\ and\
  \bibinfo {author} {\bibfnamefont {I.}~\bibnamefont {Mandal}},\ }\bibfield
  {title} {\bibinfo {title} {Magneto-optical conductivity in the type-{I} and
  type-{II} phases of {W}eyl/multi-{W}eyl semimetals},\ }\href
  {https://doi.org/10.1016/j.physb.2023.414765} {\bibfield  {journal} {\bibinfo
   {journal} {Physica B: Condensed Matter}\ }\textbf {\bibinfo {volume}
  {656}},\ \bibinfo {pages} {414765} (\bibinfo {year} {2023})}\BibitemShut
  {NoStop}%
\bibitem [{\citenamefont {{Mandal}}\ and\ \citenamefont
  {{Gemsheim}}(2019)}]{ips-seb}%
  \BibitemOpen
  \bibfield  {author} {\bibinfo {author} {\bibfnamefont {I.}~\bibnamefont
  {{Mandal}}}\ and\ \bibinfo {author} {\bibfnamefont {S.}~\bibnamefont
  {{Gemsheim}}},\ }\bibfield  {title} {\bibinfo {title} {{Emergence of
  topological {M}ott insulators in proximity of quadratic band touching
  points}},\ }\href {https://doi.org/10.5488/CMP.22.13701} {\bibfield
  {journal} {\bibinfo  {journal} {Condensed Matter Physics}\ }\textbf {\bibinfo
  {volume} {22}},\ \bibinfo {pages} {13701} (\bibinfo {year}
  {2019})}\BibitemShut {NoStop}%
\bibitem [{\citenamefont {Mandal}(2020)}]{ips_cpge}%
  \BibitemOpen
  \bibfield  {author} {\bibinfo {author} {\bibfnamefont {I.}~\bibnamefont
  {Mandal}},\ }\bibfield  {title} {\bibinfo {title} {Effect of interactions on
  the quantization of the chiral photocurrent for double-{W}eyl semimetals},\
  }\href {https://www.mdpi.com/2073-8994/12/6/919} {\bibfield  {journal}
  {\bibinfo  {journal} {Symmetry}\ }\textbf {\bibinfo {volume} {12}} (\bibinfo
  {year} {2020})}\BibitemShut {NoStop}%
\bibitem [{\citenamefont {{Mandal}}(2021)}]{ips-biref}%
  \BibitemOpen
  \bibfield  {author} {\bibinfo {author} {\bibfnamefont {I.}~\bibnamefont
  {{Mandal}}},\ }\bibfield  {title} {\bibinfo {title} {{Robust marginal {F}ermi
  liquid in birefringent semimetals}},\ }\href
  {https://doi.org/10.1016/j.physleta.2021.127707} {\bibfield  {journal}
  {\bibinfo  {journal} {Physics Letters A}\ }\textbf {\bibinfo {volume}
  {418}},\ \bibinfo {eid} {127707} (\bibinfo {year} {2021})}\BibitemShut
  {NoStop}%
\bibitem [{\citenamefont {{Mandal}}\ and\ \citenamefont
  {{Ziegler}}(2021)}]{ips-klaus}%
  \BibitemOpen
  \bibfield  {author} {\bibinfo {author} {\bibfnamefont {I.}~\bibnamefont
  {{Mandal}}}\ and\ \bibinfo {author} {\bibfnamefont {K.}~\bibnamefont
  {{Ziegler}}},\ }\bibfield  {title} {\bibinfo {title} {{Robust quantum
  transport at particle-hole symmetry}},\ }\href
  {https://doi.org/10.1209/0295-5075/ac1a25} {\bibfield  {journal} {\bibinfo
  {journal} {EPL (Europhysics Letters)}\ }\textbf {\bibinfo {volume} {135}},\
  \bibinfo {eid} {17001} (\bibinfo {year} {2021})}\BibitemShut {NoStop}%
\bibitem [{\citenamefont {Nandkishore}\ and\ \citenamefont
  {Parameswaran}(2017)}]{rahul-sid}%
  \BibitemOpen
  \bibfield  {author} {\bibinfo {author} {\bibfnamefont {R.~M.}\ \bibnamefont
  {Nandkishore}}\ and\ \bibinfo {author} {\bibfnamefont {S.~A.}\ \bibnamefont
  {Parameswaran}},\ }\bibfield  {title} {\bibinfo {title} {Disorder-driven
  destruction of a non-{F}ermi liquid semimetal studied by renormalization
  group analysis},\ }\href {https://doi.org/10.1103/PhysRevB.95.205106}
  {\bibfield  {journal} {\bibinfo  {journal} {Phys. Rev. B}\ }\textbf {\bibinfo
  {volume} {95}},\ \bibinfo {pages} {205106} (\bibinfo {year}
  {2017})}\BibitemShut {NoStop}%
\bibitem [{\citenamefont {Mandal}\ and\ \citenamefont
  {Nandkishore}(2018)}]{ips-rahul-qbt}%
  \BibitemOpen
  \bibfield  {author} {\bibinfo {author} {\bibfnamefont {I.}~\bibnamefont
  {Mandal}}\ and\ \bibinfo {author} {\bibfnamefont {R.~M.}\ \bibnamefont
  {Nandkishore}},\ }\bibfield  {title} {\bibinfo {title} {{Interplay of Coulomb
  interactions and disorder in three-dimensional quadratic band crossings
  without time-reversal symmetry and with unequal masses for conduction and
  valence bands}},\ }\href {https://doi.org/10.1103/PhysRevB.97.125121}
  {\bibfield  {journal} {\bibinfo  {journal} {Phys. Rev. B}\ }\textbf {\bibinfo
  {volume} {97}},\ \bibinfo {pages} {125121} (\bibinfo {year}
  {2018})}\BibitemShut {NoStop}%
\bibitem [{\citenamefont {Mandal}(2018)}]{ips-qbt-sc}%
  \BibitemOpen
  \bibfield  {author} {\bibinfo {author} {\bibfnamefont {I.}~\bibnamefont
  {Mandal}},\ }\bibfield  {title} {\bibinfo {title} {Fate of superconductivity
  in three-dimensional disordered {L}uttinger semimetals},\ }\href
  {https://doi.org/https://doi.org/10.1016/j.aop.2018.03.004} {\bibfield
  {journal} {\bibinfo  {journal} {Annals of Physics}\ }\textbf {\bibinfo
  {volume} {392}},\ \bibinfo {pages} {179 } (\bibinfo {year}
  {2018})}\BibitemShut {NoStop}%
\bibitem [{\citenamefont {Mandal}\ and\ \citenamefont
  {Freire}(2024)}]{ips-hermann-review}%
  \BibitemOpen
  \bibfield  {author} {\bibinfo {author} {\bibfnamefont {I.}~\bibnamefont
  {Mandal}}\ and\ \bibinfo {author} {\bibfnamefont {H.}~\bibnamefont
  {Freire}},\ }\bibfield  {title} {\bibinfo {title} {{Transport properties in
  non-Fermi liquid phases of nodal-point semimetals}},\ }\href
  {https://doi.org/10.1088/1361-648X/ad665e} {\bibfield  {journal} {\bibinfo
  {journal} {Journal of Physics: Condensed Matter}\ }\textbf {\bibinfo {volume}
  {36}},\ \bibinfo {pages} {443002} (\bibinfo {year} {2024})}\BibitemShut
  {NoStop}%
\bibitem [{\citenamefont {{Bera}}\ and\ \citenamefont
  {{Mandal}}(2021)}]{ips-sandip}%
  \BibitemOpen
  \bibfield  {author} {\bibinfo {author} {\bibfnamefont {S.}~\bibnamefont
  {{Bera}}}\ and\ \bibinfo {author} {\bibfnamefont {I.}~\bibnamefont
  {{Mandal}}},\ }\bibfield  {title} {\bibinfo {title} {{Floquet scattering of
  quadratic band-touching semimetals through a time-periodic potential well}},\
  }\href {https://doi.org/10.1088/1361-648X/ac020a} {\bibfield  {journal}
  {\bibinfo  {journal} {Journal of Physics Condensed Matter}\ }\textbf
  {\bibinfo {volume} {33}},\ \bibinfo {eid} {295502} (\bibinfo {year}
  {2021})}\BibitemShut {NoStop}%
\bibitem [{\citenamefont {{Bera}}\ \emph {et~al.}(2023)\citenamefont {{Bera}},
  \citenamefont {{Sekh}},\ and\ \citenamefont {{Mandal}}}]{ips-sandip-sajid}%
  \BibitemOpen
  \bibfield  {author} {\bibinfo {author} {\bibfnamefont {S.}~\bibnamefont
  {{Bera}}}, \bibinfo {author} {\bibfnamefont {S.}~\bibnamefont {{Sekh}}},\
  and\ \bibinfo {author} {\bibfnamefont {I.}~\bibnamefont {{Mandal}}},\
  }\bibfield  {title} {\bibinfo {title} {{Floquet transmission in
  Weyl/multi-Weyl and nodal-line semimetals through a time-periodic potential
  well}},\ }\href {https://doi.org/10.1002/andp.202200460} {\bibfield
  {journal} {\bibinfo  {journal} {Ann. Phys. (Berlin)}\ }\textbf {\bibinfo
  {volume} {535}},\ \bibinfo {pages} {2200460} (\bibinfo {year}
  {2023})}\BibitemShut {NoStop}%
\end{thebibliography}%


\end{document}